\documentclass{aastex631}

\usepackage{newtxtext,newtxmath,amsmath,float,xspace,threeparttable,bm,amsmath,booktabs}
\usepackage{rotating}

\newcommand{\todo}{\ifmmode \text{\color{red}\Huge{\(\bullet\)}} \else {\color{red}{\Huge$\bullet$}}\fi}
\newcommand{\tido}{\ifmmode {{\color{red}\bullet}} \else {\color{red}$\bullet$}\fi}

\newcommand{\E        }[1]{\ifmmode 10^{#1} \else $10^{#1}$\fi}
\newcommand{\tE        }[1]{\ifmmode \times10^{#1} \else $\times10^{#1}$\fi}
\newcommand{\til}{\ifmmode \sim \else $\sim$\fi}
\renewcommand{\~} {\ifmmode \sim \else $\sim$\fi}

\newcommand{\logNH }{\ifmmode \log (N_{\rm H}/{\rm cm}^{-2}) \else $\log (N_{\rm H}/{\rm cm}^{-2})$\fi}

\newcommand{\Mbh   }{\ifmmode M_{\rm BH} \else $M_{\rm BH}$\fi}

\newcommand{\Catrip}{\ifmmode \left[{\rm Ca}\,\textsc{ii}\right\,\lambda8498, 8542, 8662 \else Ca\,\textsc{ii} $\,\lambda8498, 8542, 8662$\fi}

\newcommand{\pc}	{\ifmmode {\rm pc} \else pc\fi}
\newcommand{\ld}	{\ifmmode {\rm l.d.} \else l.d.\fi}
\newcommand{\kms}	{\ifmmode {\rm km\,s}^{-1} \else km\,s$^{-1}$\fi}
\newcommand{\cc}	{\ifmmode {\rm cm}^{-3}    \else cm$^{-3}$\fi}
\newcommand{\cmii}	{\ifmmode {\rm cm}^{-2}    \else cm$^{-2}$\fi}
\newcommand{\ergs}	{\ifmmode {\rm erg\,s}^{-1} \else erg s$^{-1}$\fi}
\newcommand{\ergcms}	{\ifmmode {\rm erg\,cm}^{-2}\,{\rm s}^{-1} \else erg\,cm$^{-2}$\,s$^{-1}$\fi}
\newcommand{\ergcmsA}	{\ifmmode {\rm erg\,cm}^{-2}\,{\rm s}^{-1}\,{\rm\AA}^{-1}
\else erg\,cm$^{-2}$\,s$^{-1}$\,\AA$^{-1}$\fi}
\newcommand{  \ergcmsHz  }{\ifmmode{\rm erg\,cm}^{-2}\,{\rm s}^{-1}\,{\rm Hz}^{-1}
                       \else ergs\,cm$^{-2}$\,s$^{-1}$\,Hz$^{-1}$\fi}
\newcommand{\kev}	{\ifmmode {\rm keV} \else keV\fi}

\newcommand{\mic}	{\ifmmode {\rm \mu m} \else $\mu$m\fi}
\newcommand{\vFWHM}	{\ifmmode v_{\mbox{\tiny FWHM}} \else $v_{\mbox{\tiny FWHM}}$\fi}
\newcommand{\vBLR}	{\ifmmode v_{\mbox{\tiny BLR}} \else $v_{\mbox{\tiny BLR}}$\fi}
\newcommand{\sigBLR}	{\ifmmode \sigma_{\mbox{\tiny BLR}} \else $\sigma_{\mbox{\tiny BLR}}$\fi}
\newcommand{\vNLR}	{\ifmmode v_{\mbox{\tiny NLR}} \else $v_{\mbox{\tiny NLR}}$\fi}
\newcommand{\tauBLR}	{\ifmmode \tau_{\mbox{\tiny BLR}} \else $\tau_{\mbox{\tiny BLR}}$\fi}

\newcommand{\Hubble}	{\ifmmode {\rm km\,s}^{-1}\,{\rm Mpc}^{-1} \else km\,s$^{-1}$\,Mpc$^{-1}$\fi}
\newcommand{\NDunit}	{\ifmmode {\rm Mpc}^{-3} \else Mpc$^{-3}$\fi}
\newcommand{\LFunit}	{\ifmmode {\rm Mpc}^{-3}\,{\rm mag}^{-1} \else Mpc$^{-3}$\,mag$^{-1}$\fi}
\newcommand{\MFunit}	{\ifmmode {\rm Mpc}^{-3}\,{\rm dex}^{-1} \else Mpc$^{-3}$\,dex$^{-1}$\fi}

\newcommand{\Msun}{\ifmmode M_{\odot} \else $M_{\odot}$\fi}
\newcommand{\Lsun}{\ifmmode L_{\odot} \else $L_{\odot}$\fi}
\newcommand{\Zsun}{\ifmmode Z_{\odot} \else $Z_{\odot}$\fi}
\newcommand{\mpyr}{\ifmmode \Msun\,{\rm yr}^{-1} \else $\Msun\,{\rm yr}^{-1}$\fi}

\newcommand{\qnote}{\ifmmode q_{0} \else $q_{0}$\fi}
\newcommand{\Hnote}{\ifmmode H_{0} \else $H_{0}$\fi}
\newcommand{\hnote}{\ifmmode h_{0} \else $h_{0}$\fi}
\newcommand{\anote}{\ifmmode a_{0} \else $a_{0}$\fi}

\newcommand{  \Halpha   }{\ifmmode {\rm H}\alpha \else H$\alpha$\fi}
\newcommand{  \ha   	}{\ifmmode {\rm H}\alpha \else H$\alpha$\fi}
\newcommand{  \Hbeta    }{\ifmmode {\rm H}\beta \else H$\beta$\fi}
\newcommand{  \hb    	}{\ifmmode {\rm H}\beta \else H$\beta$\fi}
\newcommand{  \Hgamma   }{\ifmmode {\rm H}\gamma \else H$\gamma$\fi}
\newcommand{  \Hdelta   }{\ifmmode {\rm H}\delta \else H$\delta$\fi}
\newcommand{  \Lya      }{\ifmmode {\rm Ly}\alpha \else Ly$\alpha$\fi}
\newcommand{  \Lyb      }{\ifmmode {\rm Ly}\beta \else Ly$\beta$\fi}
\newcommand{  \Pa       }{\ifmmode {\rm P}\alpha \else P$\alpha$\fi}
\newcommand{  \Pb       }{\ifmmode {\rm P}\beta \else P$\beta$\fi}
\newcommand{  \Paa       }{\ifmmode {\rm P}\alpha \else P$\alpha$\fi}
\newcommand{  \Pab       }{\ifmmode {\rm Pa}\beta \else Pa$\beta$\fi}
\newcommand{  \Bra      }{\ifmmode {\rm Br}\alpha \else Br$\alpha$\fi}
\newcommand{  \Brg      }{\ifmmode {\rm Br}\gamma \else Br$\gamma$\fi}
\newcommand{  \hii      }{\ifmmode {\rm H}\,\textsc{ii} \else H\,\textsc{ii}\fi}
\newcommand{  \hei      }{\ifmmode {\rm He}\,\textsc{i} \else He\,\textsc{i}\fi}
\newcommand{  \heii     }{\ifmmode {\rm He}\,\textsc{ii} \else He\,\textsc{ii}\fi}
\newcommand{  \HeIIuv   }{\ifmmode {\rm He}\,\textsc{ii}\,\lambda1640 \else He\,\textsc{ii}\,$\lambda1640$\fi}
\newcommand{  \HeIIop   }{\ifmmode {\rm He}\,\textsc{ii}\,\lambda4686 \else He\,\textsc{ii}\,$\lambda4686$\fi}
\newcommand{  \cii      }{\ifmmode {\rm C}\,\textsc{ii}  \else C\,\textsc{ii}\fi}
\newcommand{  \ciii     }{\ifmmode {\rm C}\,\textsc{iii}\right] \else C\,\textsc{iii}]\fi}
\newcommand{  \CIII     }{\ifmmode {\rm C}\,\textsc{iii}\right]\,\lambda1909 \else C\,\textsc{iii}]\,$\lambda1909$\fi}
\newcommand{  \civ      }{\ifmmode {\rm C}\,\textsc{iv}  \else C\,\textsc{iv}\fi}
\newcommand{  \CIV      }{\ifmmode {\rm C}\,\textsc{iv}\,\lambda1549 \else C\,\textsc{iv}\,$\lambda1549$\fi}
\newcommand{  \nii      }{\ifmmode [{\rm N}\,\textsc{ii}]  \else [N\,\textsc{ii}]\fi}
\newcommand{  \niii     }{\ifmmode {\rm N}\,\textsc{iii} \else N\,\textsc{iii}\fi}
\newcommand{  \niv      }{\ifmmode {\rm N}\,\textsc{iv}  \else N\,\textsc{iv}\fi}
\newcommand{  \NIVuv    }{\ifmmode {\rm N}\,\textsc{iv}\,\lambda1486 \else N\,\textsc{iv}\,$\lambda1486$\fi}
\newcommand{  \nv       }{\ifmmode {\rm N}\,\textsc{v}   \else N\,\textsc{v}\fi}
\newcommand{\oi}{\ifmmode \left[{\rm O}\,\textsc{i}\right] \else [O\,{\sc i}]\fi}
\newcommand{\OI}{\ifmmode \left[{\rm O}\,\textsc{i}\right]\,\lambda6300 \else [O\,{\sc i}]$\,\lambda6300$\fi}

\newcommand{\oii}{\ifmmode \left[{\rm O}\,\textsc{ii}\right] \else [O\,{\sc ii}]\fi}
\newcommand{\OII}{\ifmmode \left[{\rm O}\,\textsc{ii}\right]\,\lambda3727 \else [O\,{\sc ii}]\,$\lambda3727$\fi}
\newcommand{\oiii}{\ifmmode \left[{\rm O}\,\textsc{iii}\right] \else [O\,{\sc iii}]\fi}
\newcommand{\OIII}{\ifmmode \left[{\rm O}\,\textsc{iii}\right]\,\lambda5007 \else [O\,{\sc iii}]\,$\lambda5007$\fi}

\newcommand{\NII}{\ifmmode \left[{\rm N}\,\textsc{ii}\right]\,\lambda6583 \else [N\,{\sc ii}]$\,\lambda6583$\fi}

\newcommand{\NeIII}{\ifmmode \left[{\rm Ne}\,\textsc{iii}\right]\,\lambda3968 \else [Ne\,{\sc iii}]$\,\lambda3968$\fi}

\newcommand{\NeV}{\ifmmode \left[{\rm Ne}\,\textsc{v}\right]\,\lambda3426 \else [Ne\,{\sc v}]$\,\lambda3426$\fi}

\newcommand{\HeII}{\ifmmode {\rm He}\,\textsc{ii}\,\lambda4686 \else He\,{\sc ii}$\,\lambda4686$\fi}

\newcommand{\sii}{\ifmmode \left[{\rm S}\,\textsc{ii}\right] \else [S\,{\sc ii}]\fi}

\newcommand{\SII}{\ifmmode \left[{\rm S}\,\textsc{ii}\right]\,\lambda6717,6731 \else [S\,{\sc ii}]$\,\lambda6717,6731$\fi}

\newcommand{  \OIIIuv   }{\ifmmode {\rm O}\,\textsc{iii}\,\lambda1663 \else O\,\textsc{iii}\,$\lambda1663$\fi}
\newcommand{  \oiv      }{\ifmmode {\rm O}\,\textsc{iv}  \else O\,\textsc{iv}\fi}
\newcommand{  \OIVuv    }{\ifmmode {\rm O}\,\textsc{iv}\,\lambda1402  \else O\,\textsc{iv}\,$\lambda1402$\fi}
\newcommand{  \OIVIR    }{\ifmmode {\rm O}\,\textsc{iv}\,25.9\,\mu {\rm m} \else O\,\textsc{iv}\,$25.9\,\mu$m\fi}
\newcommand{  \ovi      }{\ifmmode {\rm O}\,\textsc{vi}   \else O\,\textsc{vi}\fi}
\newcommand{  \Ovi      }{\ifmmode {\rm O}\,\textsc{vi}\,\lambda1035 \else O\,\textsc{vi}\,$\lambda1035$\fi}
\newcommand{  \nei      }{\ifmmode {\rm Ne}\,\textsc{i}   \else Ne\,\textsc{i}\fi}
\newcommand{  \neii     }{\ifmmode {\rm Ne}\,\textsc{ii}  \else Ne\,\textsc{ii}\fi}
\newcommand{  \NeiiIR   }{\ifmmode {\rm Ne}\,\textsc{ii}\,12.8\,\mu {\rm m} \else Ne\,\textsc{ii}\,$12.8\,\mu$m\fi}
\newcommand{  \neiii    }{\ifmmode {\rm Ne}\,\textsc{iii} \else Ne\,\textsc{iii}\fi}
\newcommand{  \neiv     }{\ifmmode {\rm Ne}\,\textsc{iv}  \else Ne\,\textsc{iv}\fi}
\newcommand{  \nev      }{\ifmmode {\rm Ne}\,\textsc{v}   \else Ne\,\textsc{v}\fi}
\newcommand{  \NevIR    }{\ifmmode {\rm Ne}\,\textsc{v}\,24.3\,\mu {\rm m} \else Ne\,\textsc{v}\,$24.3\,\mu$m\fi}
\newcommand{  \nevi     }{\ifmmode {\rm Ne}\,\textsc{vi}  \else Ne\,\textsc{vi}\fi}
\newcommand{  \mgi      }{\ifmmode {\rm Mg}\,\textsc{i}   \else Mg\,\textsc{i}\fi}
\newcommand{  \mgii     }{\ifmmode {\rm Mg}\,\textsc{ii}  \else Mg\,\textsc{ii}\fi}
\newcommand{  \MgII     }{\ifmmode {\rm Mg}\,\textsc{ii}\,\lambda2798 \else Mg\,\textsc{ii}\,$\lambda2798$\fi}

\newcommand{  \siii     }{\ifmmode {\rm S}\,\textsc{iii} \else S\,\textsc{iii}\fi}
\newcommand{  \siv      }{\ifmmode {\rm S}\,\textsc{iv}  \else S\,\textsc{iv}\fi}
\newcommand{  \sili     }{\ifmmode {\rm Si}\,\textsc{i}   \else Si\,\textsc{i}\fi}
\newcommand{  \silii    }{\ifmmode {\rm Si}\,\textsc{ii}  \else Si\,\textsc{ii}\fi}
\newcommand{  \Siliv    }{\ifmmode {\rm Si}\,\textsc{iv}  \else Si\,\textsc{iv}\fi}
\newcommand{  \SilIVuv  }{\ifmmode {\rm Si}\,\textsc{iv}\,\lambda1400  \else Si\,\textsc{iv}\,$\lambda1400$\fi}

\newcommand{  \caii     }{\ifmmode {\rm Ca}\,\textsc{ii}   \else Ca\,\textsc{ii}\fi}
 \newcommand{\Mgb}{\ifmmode \left{\rm Mg}\,\textsc{i}\right\,\lambda5175 \else Mg\,{\sc i}\,$\lambda5175$\fi}
\newcommand{\Cahk}{\ifmmode \left[{\rm Ca H+K}\,\textsc{ii}\right\,\lambda3935,3968 \else Ca H+K$\,\lambda3935,3968$\fi}
\newcommand{  \feii     }{\ifmmode {\rm Fe}\,\textsc{ii}  \else Fe\,\textsc{ii}\fi}
\newcommand{  \feiii    }{\ifmmode {\rm Fe}\,\textsc{iii} \else Fe\,\textsc{iii}\fi}

\newcommand{ \Lhb   }{\ifmmode L\left(\hb\right) \else $L\left(\hb\right)$\fi}
\newcommand{ \fwhb  }{\ifmmode {\rm FWHM}\left(\hb\right) \else FWHM(\hb)\fi}
\newcommand{ \Lha   }{\ifmmode L\left(\ha\right) \else $L\left(\ha\right)$\fi}
\newcommand{ \fwha  }{\ifmmode {\rm FWHM}\left(\ha\right) \else FWHM(\ha)\fi}
\newcommand{ \Lmg   }{\ifmmode L\left(\mgii\right) \else $L\left(\mgii\right)$\fi}
\newcommand{ \fwmg  }{\ifmmode {\rm FWHM}\left(\mgii\right) \else FWHM(\mgii)\fi}
\newcommand{ \Lciv  }{\ifmmode L\left(\civ\right) \else $L\left(\civ\right)$\fi}
\newcommand{ \fwciv }{\ifmmode {\rm FWHM}\left(\civ\right) \else FWHM(\civ)\fi}
\newcommand{ \fwhm  }{\ifmmode {\rm FWHM} \else FWHM\fi} 
\newcommand{ \voff  }{\ifmmode v_{\rm off} \else $v_{\rm off}$\fi} 

\newcommand{ \mumg  }{\ifmmode \mu\left(\mgii\right) \else $\mu\left(\mgii\right)$\fi}
\newcommand{ \fmg   }{\ifmmode f\left(\mgii\right) \else $f\left(\mgii\right)$\fi}
\newcommand{ \muciv }{\ifmmode \mu\left(\civ\right) \else $\mu\left(\civ\right)$\fi}
\newcommand{ \fciv  }{\ifmmode f\left(\civ\right) \else $f\left(\civ\right)$\fi}

\newcommand{  \auvo     }{\ifmmode \alpha_{\nu,{\rm UVO}} \else $\alpha_{\nu,{\rm UVO}}$\fi}
\newcommand{  \Ledd     }{\ifmmode L_{\rm Edd} \else $L_{\rm Edd}$\fi}
\newcommand{  \lamLlam  }{\ifmmode \lambda L_{\lambda} \else $\lambda L_{\lambda}$\fi}
\newcommand{  \lLl      }{\ifmmode \lambda L_{\lambda} \else $\lambda L_{\lambda}$\fi}
\newcommand{  \nuLnu    }{\ifmmode \nu L_{\nu} \else $\nu L_{\nu}$\fi}
\newcommand{  \nLn      }{\ifmmode \nu L_{\nu} \else $\nu L_{\nu}$\fi}
\newcommand{  \Luv      }{\ifmmode L_{1450} \else $L_{1450}$\fi}
\newcommand{  \Lop      }{\ifmmode L_{5100} \else $L_{5100}$\fi}
\newcommand{  \lLop     }{\ifmmode \log\left(\Lop/\ergs\right) \else $\log\left(\Lop/\ergs\right)$\fi}
\newcommand{  \Lthree   }{\ifmmode L_{3000} \else $L_{3000}$\fi}
\newcommand{  \lLthree  }{\ifmmode \log\left(\Lthree/\ergs\right) \else $\log\left(\Lthree/\ergs\right)$\fi}

\newcommand{\Fthree}{\ifmmode F_{3000} \else $F_{3000}$\fi}
\newcommand{\fuv}{\ifmmode f_{\lambda}\left(1450{\rm \AA}\right) \else $f_{\lambda}\left(1450 {\rm \AA}\right)$\fi}
\newcommand{\fthree}{\ifmmode f_{\lambda}\left(3000{\rm \AA}\right) \else $f_{\lambda}\left(3000{\rm \AA}\right)$\fi}
\newcommand{\fH}{\ifmmode f_{\lambda}\left(1.65\micron\right) \else
$f_{\lambda}\left(1.65\micron\right)$\fi}

\newcommand{\fbol}{\ifmmode f_{\rm bol} \else $f_{\rm bol}$\fi}
\newcommand{\fbolwv}{\ifmmode f_{\rm bol}\left(\lambda\right) \else $f_{\rm bol}\left(\lambda\right)$\fi}
\newcommand{\fbolopt}{\ifmmode f_{\rm bol}\left(5100{\rm \AA}\right) \else $f_{\rm bol}\left(5100{\rm \AA}\right)$\fi}
\newcommand{\fbolthree}{\ifmmode f_{\rm bol}\left(3000{\rm \AA}\right) \else $f_{\rm bol}\left(3000{\rm \AA}\right)$\fi}
\newcommand{\fboluv}{\ifmmode f_{\rm bol}\left(1450{\rm \AA}\right) \else $f_{\rm bol}\left(1450{\rm \AA}\right)$\fi}

\newcommand{  \mbh      }{\ifmmode M_{\rm BH} \else $M_{\rm BH}$\fi}
\newcommand{  \lmbh     }{\ifmmode \log\left(\mbh/\Msun\right) \else $\log\left(\mbh/\Msun\right)$\fi} 
\newcommand{  \lledd    }{\ifmmode L_{\rm bol}/L_{\rm Edd} \else $L_{\rm bol}/L_{\rm Edd}$\fi}
\newcommand{  \Lbol     }{\ifmmode L_{\rm bol} \else $L_{\rm bol}$\fi}
\newcommand{  \lbol     }{\ifmmode L_{\rm bol} \else $L_{\rm bol}$\fi}
\newcommand{  \lLbol    }{\ifmmode \log\left(\Lbol/\ergs\right) \else $\log\left(\Lbol/\ergs\right)$\fi} 
\newcommand{  \Lagn     }{\ifmmode L_{\rm AGN} \else $L_{\rm AGN}$\fi}
\newcommand{  \lagn     }{\ifmmode L_{\rm AGN} \else $L_{\rm AGN}$\fi}

\newcommand{  \tgrow     }{\ifmmode t_{\rm growth} \else $t_{\rm growth}$\fi}
\newcommand{  \tUni      }{\ifmmode t_{\rm Universe} \else $t_{\rm Universe}$\fi}

\newcommand{  \Mindot	}{\ifmmode \dot{M}_{\rm infall} \else $\dot{M}_{\rm infall}$\fi}
\newcommand{  \Mbhdot	}{\ifmmode \dot{M}_{\rm BH} \else $\dot{M}_{\rm BH}$\fi}
\newcommand{  \Maddot	}{\ifmmode \dot{M}_{\rm AD} \else $\dot{M}_{\rm AD}$\fi}

\newcommand{  \as	}{\ifmmode a_{\rm *} 		\else $a_{\rm *}$\fi}
\newcommand{  \avec	}{\ifmmode \vec{a}_{\rm *} 	\else $\vec{a}_{\rm *}$\fi}
\newcommand{  \re	}{\ifmmode \eta      	\else $\eta$\fi}

\newcommand{  \mseed    }{\ifmmode M_{\rm seed} \else $M_{\rm seed}$\fi}
\newcommand{  \mbul     }{\ifmmode M_{\rm Bulge} \else $M_{\rm Bulge}$\fi} 
\newcommand{  \mstar    }{\ifmmode M_{*} \else $M_{*}$\fi} 
\newcommand{  \mgal     }{\ifmmode M_{*} \else $M_{*}$\fi} 
\newcommand{  \mhost    }{\ifmmode M_{\rm Host} \else $M_{\rm Host}$\fi}
\newcommand{  \mm       }{\ifmmode M_{*}/M_{\rm BH} \else $M_{*}/M_{\rm BH}$\fi}
\newcommand{  \mmsmall  }{\ifmmode M_{\rm BH}/M_{*} \else $M_{\rm BH}/M_{*}$\fi}
\newcommand{  \mmlarge  }{\ifmmode M_{*}/M_{\rm BH} \else $M_{*}/M_{\rm BH}$\fi}
\newcommand{  \mmwp     }{\ifmmode \left(M_{*}/M_{\rm BH}\right) \else $\left(M_{*}/M_{\rm BH}\right)$\fi}
\newcommand{  \ml       }{\ifmmode M_{*}/L_{*} \else $M_{*}/L_{*}$\fi}
\newcommand{  \mlwp     }{\ifmmode \left(M_{*}/L\right) \else $\left(M_{*}/L\right)$\fi}
\newcommand{  \mlk      }{\ifmmode \left(M_{*}/L_{K}\right) \else $\left(M_{*}/L_{K}\right)$\fi}
\newcommand{  \sigs     }{\ifmmode \sigma_{*} \else $\sigma_{*}$\fi}
\newcommand{  \Reff     }{\ifmmode R_{\rm e} \else $R_{\rm e}$\fi}

\def\kmps{\hbox{$\km\s^{-1}\,$}}

\newcommand{\bj}{\ifmmode b_{\rm J} \else $b_{\rm J}$\fi}

\newcommand{\iab}{\ifmmode i_{\rm AB} \else $i_{\rm AB}$\fi}

\newcommand{\jab}{\ifmmode J_{\rm AB} \else $J_{\rm AB}$\fi}
\newcommand{\hab}{\ifmmode H_{\rm AB} \else $H_{\rm AB}$\fi}
\newcommand{\kab}{\ifmmode K_{\rm AB} \else $K_{\rm AB}$\fi}

\newcommand{\jveg}{\ifmmode J_{\rm Vega} \else $J_{\rm Vega}$\fi}
\newcommand{\hveg}{\ifmmode H_{\rm Vega} \else $H_{\rm Vega}$\fi}
\newcommand{\kveg}{\ifmmode K_{\rm Vega} \else $K_{\rm Vega}$\fi}

\def\arcmin{\hbox{$^\prime$}}
\def\arcsec{\hbox{$^{\prime\prime}$}}

\newcommand{  \Chisq    }{\ifmmode \chi^{2} \else $\chi^{2}$}
\newcommand{  \nelec    }{\ifmmode n_{e} \else $n_{e}$\fi}     
\newcommand{  \nh       }{\ifmmode n_{\rm H} \else $n_{\rm H}$\fi}     
\newcommand{  \Ncol     }{\ifmmode N_{col} \else $N_{col}$\fi} 
\newcommand{  \NH       }{\ifmmode N_{\rm H} \else $N_{\rm H}$\fi}     

\def\deg{\hbox{$^\circ$}}
\def\sun{\hbox{$\odot$}}
\def\arcmin{\hbox{$^\prime$}}
\def\arcsec{\hbox{$^{\prime\prime}$}}

\def\ion#1#2{#1$\;${\small\rm\@Roman{#2}}\relax}

\newcommand{\OIIIa}{\ifmmode \left[{\rm O}\,\textsc{iii}\right]\,\lambda4959 \else [O\,{\sc iii}]\,$\lambda4959$\fi}
\newcommand{\NIIa}{\ifmmode \left[{\rm N}\,\textsc{ii}\right]\,\lambda6548 \else [N\,{\sc ii}]\,$\lambda6548$\fi}
\newcommand{\SIIa}{\ifmmode \left[{\rm S}\,\textsc{ii}\right]\,\lambda6716 \else [S\,{\sc ii}]\,$\lambda6716$\fi}
\newcommand{\SIIb}{\ifmmode \left[{\rm S}\,\textsc{ii}\right]\,\lambda6732 \else [S\,{\sc ii}]\,$\lambda6731$\fi}
\newcommand{\NeVa}{\ifmmode \left[{\rm Ne}\,\textsc{v}\right]\,\lambda3346 \else [Ne\,{\sc v}]\,$\lambda3346$\fi}
\newcommand{\NeVb}{\ifmmode \left[{\rm Ne}\,\textsc{v}\right]\,\lambda3426 \else [Ne\,{\sc v}]\,$\lambda3426$\fi}
\newcommand{\NeIIIa}{\ifmmode \left[{\rm Ne}\,\textsc{iii}\right]\,\lambda3869 \else [Ne\,{\sc iii}]\,$\lambda3869$\fi}
\newcommand{\NeIIIb}{\ifmmode \left[{\rm Ne}\,\textsc{iii}\right]\,\lambda3968 \else [Ne\,{\sc iii}]\,$\lambda3968$\fi}

\newcommand{\mgb}{\ifmmode \left{\rm Mg}\,\textsc{i}\right \else Mg\,{\sc i}\fi}

\def\arcmin{{\mbox{$^{\prime}$}}}

\def\arcsec{{\mbox{$^{\prime \prime}$}}}

\def\erg{{\rm\thinspace erg}}

\def\g{{\rm\thinspace g}}

\def\K{{\rm\thinspace K}}

\def\km{{\rm\thinspace km}}
\def\kpc{{\rm\thinspace kpc}}
\def\Lsun{\hbox{$\rm\thinspace L_{\odot}$}}
\def\m{{\rm\thinspace m}}

\def\pc{{\rm\thinspace pc}}

\def\s{{\rm\thinspace s}}

\newcommand{\halpha}{\Halpha}

\newcommand{\hbeta}{\Hbeta}

\newcommand{\HeIIir}{\ifmmode {\rm He}\,\textsc{ii}\,\lambda8237 \else He\,{\sc ii}$\,\lambda8237$\fi}
\newcommand{\HeIir}{\ifmmode {\rm He}\,\textsc{i}\,\lambda10830 \else He\,{\sc i}$\,\lambda10830$\fi}

\newcommand{\SIII}{\ifmmode \left[{\rm S}\,\textsc{iii}\right]\,\lambda9531 \else [S\,\textsc{ii}]\,$\lambda9531$\fi}

\newcommand {\Lsoftint} {\ifmmode L^{\rm int}_{\mathrm{2-10\ keV}} \else $L^{\rm int}_{\mathrm{2-10\ keV}}$\fi}

\newcommand {\ergpersec} {\ifmmode {\rm erg~s}^{-1} \else erg~s$^{-1}$ \fi}

\def\micron{{\mbox{$\mu{\rm m}$}}}
\def\arcsec{{\mbox{$^{\prime \prime}$}}}
\def\arcmin{{\mbox{$^{\prime}$}}}

\def\arcsec{{\mbox{$^{\prime \prime}$}}}

\def\erg{{\rm\thinspace erg}}

\def\g{{\rm\thinspace g}}

\def\K{{\rm\thinspace K}}

\def\km{{\rm\thinspace km}}
\def\kpc{{\rm\thinspace kpc}}
\def\Lsun{\hbox{$\rm\thinspace L_{\odot}$}}
\def\m{{\rm\thinspace m}}

\def\pc{{\rm\thinspace pc}}

\def\s{{\rm\thinspace s}}

\def\ergps{\hbox{$\erg\s^{-1}\,$}}

\def\kmps{\hbox{$\km\s^{-1}\,$}}

\def\apjl{\hbox{ApJL}}
\def\apjs{\hbox{ApJS}}

\def\aap{\hbox{AAP}}
\def\micron{{\mbox{$\mu{\rm m}$}}}
\def\arcsec{{\mbox{$^{\prime \prime}$}}}
\def\arcmin{{\mbox{$^{\prime}$}}}

\newcommand {\ppxf}{\texttt{pPXF}}
\newcommand {\vorbin}{\texttt{vorbin}}

\newcommand {\xtellcor}{\texttt{xtellcor}}

\newcommand{\nuvr}{\ifmmode {\rm NUV}-r \else NUV-$r$\fi}
\newcommand{\mh}{\ifmmode M_{\rm H_2} \else $M_{\rm H_2}$\fi}
\newcommand{\mhi}{\ifmmode M_{\rm HI} \else $M_{\rm HI}$\fi}

\newcommand{\must}{\ifmmode \mu_{\ast} \else $\mu_{\ast}$\fi}
\newcommand{\hmol}{\ifmmode H_2 \else $H_2$\fi}
\newcommand{\rmol}{\ifmmode R_{\rm mol} \else $R_{\rm mol}$\fi}
\newcommand{\tdep}{\ifmmode t_{\rm dep}({\rm H_2}) \else $t_{\rm dep}({\rm H_2})$\fi}
\newcommand{\tdepHI}{\ifmmode t_{\rm dep}({\rm HI}) \else $t_{\rm dep}({\rm HI})$\fi}
\newcommand{\fgas}{\ifmmode f_{\rm H_2} \else $f_{\rm H_2}$\fi}
\newcommand{\fhi}{\ifmmode f_{\rm HI} \else $f_{\rm HI}$\fi}
\newcommand{\xco}{\ifmmode \alpha_{\rm CO} \else $\alpha_{\rm CO}$\fi}

\newcommand{\SiX}{\ifmmode \left[{\rm Si}\,\textsc{x}\right]\,\lambda14300 \else [Si\,{\sc x}]\,$\lambda14300$\fi}
\newcommand{\SiVI}{\ifmmode \left[{\rm Si}\,\textsc{vi}\right]\,\lambda19640 \else [Si\,{\sc vi}]\,$\lambda19640$\fi}
\newcommand{\SXI}{\ifmmode \left[{\rm S}\,\textsc{xi}\right]\,\lambda19196 \else [S\,{\sc xi}]\,$\lambda19196$\fi}
\newcommand{\SVIII}{\ifmmode \left[{\rm S}\,\textsc{viii}\right]\,\lambda9915 \else [S\,{\sc viii}]\,$\lambda9915$\fi}
\newcommand{\SIX}{\ifmmode \left[{\rm S}\,\textsc{ix}\right]\,\lambda12520 \else [S\,{\sc ix}]\,$\lambda12520$\fi}
\newcommand{\FeXIII}{\ifmmode \left[{\rm Fe}\,\textsc{xiii}\right]\,\lambda10747 \else [Fe\,{\sc xiii}]\,$\lambda10747$\fi}
\newcommand{\SiXI}{\ifmmode \left[{\rm Si}\,\textsc{xi}\right]\,\lambda19320 \else [Si\,{\sc xi}]\,$\lambda19320$\fi}

\def\arcsec{{\mbox{$^{\prime \prime}$}}}

\def\erg{{\rm\thinspace erg}}

\def\g{{\rm\thinspace g}}

\def\K{{\rm\thinspace K}}

\def\km{{\rm\thinspace km}}
\def\kpc{{\rm\thinspace kpc}}
\def\Lsun{\hbox{$\rm\thinspace L_{\odot}$}}
\def\m{{\rm\thinspace m}}

\def\pc{{\rm\thinspace pc}}

\def\s{{\rm\thinspace s}}

\def\ergps{\hbox{$\erg\s^{-1}\,$}}

\def\kmps{\hbox{$\km\s^{-1}\,$}}

\def\apjl{\hbox{ApJL}}
\def\apjs{\hbox{ApJS}}

\def\aap{\hbox{AAP}}
\def\micron{{\mbox{$\mu{\rm m}$}}}
\def\arcsec{{\mbox{$^{\prime \prime}$}}}
\def\arcmin{{\mbox{$^{\prime}$}}}

\received{\today}
\revised{\today}
\accepted{\today}
\submitjournal{ApJL}

\shorttitle{UGC\,4211: A Confirmed Dual AGN at 230 pc Separation}
\shortauthors{Koss et al.}

\graphicspath{{./}{./figures/}}

\begin{document}

\title{UGC\,4211: A Confirmed Dual AGN in the Local Universe at 230 pc Nuclear Separation}

\correspondingauthor{Michael Koss}
\email{mike.koss@eurekasci.com}

\author[0000-0002-7998-9581]{Michael J. Koss}
\affil{Eureka Scientific, 2452 Delmer Street Suite 100, Oakland, CA 94602-3017, USA}

\author[0000-0001-7568-6412]{Ezequiel Treister}
\affil{Instituto de Astrof{\'{\i}}sica, Facultad de F{\'{i}}sica, Pontificia Universidad Cat{\'{o}}lica de Chile, Campus San Joaquín, Av. Vicu{\~{n}}a Mackenna 4860, Macul Santiago, Chile, 7820436} 

\author[0000-0002-2603-2639]{Darshan Kakkad}
\affil{Space Telescope Science Institute, 3700 San Martin Drive, Baltimore, MD 21218, USA}

\author[0000-0002-5557-4007]{J. Andrew Casey-Clyde}
\affiliation{Department of Physics, University of Connecticut, 196 Auditorium Road, U-3046, Storrs, CT 06269-3046, USA}
\affiliation{Center for Computational Astrophysics, Flatiron Institute, 162 Fifth Avenue, New York, NY 10010, USA}

\author[0000-0002-6808-2052]{Taiki Kawamuro}
\affil{RIKEN Cluster for Pioneering Research, 2-1 Hirosawa, Wako, Saitama 351-0198, Japan}

\author[0000-0002-0441-3502]{Jonathan Williams}
\affiliation{Department of Astronomy, University of Maryland, College Park, MD 20742, USA}

\author[0000-0002-1616-1701]{Adi Foord}
\affiliation{Kavli Institute of Particle Astrophysics and Cosmology, Stanford University, Stanford, CA 94305, USA}

\author[0000-0002-3683-7297]{Benny Trakhtenbrot}
\affil{School of Physics and Astronomy, Tel Aviv University, Tel Aviv 69978, Israel}

\author[0000-0002-8686-8737]{Franz E. Bauer}
\affiliation{Instituto de Astrof{\'{\i}}sica, Facultad de F{\'{i}}sica, Pontificia Universidad Cat{\'{o}}lica de Chile, Campus San Joaquín, Av. Vicu{\~{n}}a Mackenna 4860, Macul Santiago, Chile, 7820436} 
\affiliation{Centro de Astroingenier{\'{\i}}a, Facultad de F{\'{i}}sica, Pontificia Universidad Cat{\'{o}}lica de Chile, Campus San Joaquín, Av. Vicu{\~{n}}a Mackenna 4860, Macul Santiago, Chile, 7820436} 
\affiliation{Millennium Institute of Astrophysics, Nuncio Monse{\~{n}}or S{\'{o}}tero Sanz 100, Of 104, Providencia, Santiago, Chile}
\affiliation{Space Science Institute, 4750 Walnut Street, Suite 205, Boulder, Colorado 80301, USA}

\author[0000-0003-3474-1125]{George C. Privon}
\affiliation{National Radio Astronomy Observatory, 520 Edgemont Road, Charlottesville, VA 22903, USA}
\affiliation{Department of Astronomy, University of Florida, P.O. Box 112055, Gainesville, FL 32611, USA}

\author[0000-0001-5231-2645]{Claudio Ricci}
\affil{N\'ucleo de Astronom\'ia de la Facultad de Ingenier\'ia, Universidad Diego Portales, Av. Ej\'ercito Libertador 441, Santiago 22, Chile}
\affil{Kavli Institute for Astronomy and Astrophysics, Peking University, Beijing 100871, People's Republic of China}

\author[0000-0002-7962-5446]{Richard Mushotzky}
\affiliation{Department of Astronomy, University of Maryland, College Park, MD 20742, USA}
\affiliation{Joint Space-Science Institute, University of Maryland, College Park, MD 20742, USA}

 \author[0000-0003-0057-8892]{Loreto Barcos-Munoz}
 \affiliation{National Radio Astronomy Observatory, 520 Edgemont Road, Charlottesville, VA 22903, USA}
 
\author[0000-0002-2183-1087]{Laura Blecha}
\affil{University of Florida, Department of Physics, 2001 Museum Road, Gainesville, FL 32611, USA} 

\author[0000-0002-7898-7664]{Thomas Connor}
\affiliation{Jet Propulsion Laboratory, California Institute of Technology, 4800 Oak Grove Drive, Pasadena, CA 91109, USA}

\author{Fiona Harrison}
\affil{Cahill Center for Astronomy and Astrophysics, California Institute of Technology, Pasadena, CA 91125, USA}

\author[0000-0001-5766-4287]{Tingting Liu}
\affil{Center for Gravitation, Cosmology and Astrophysics, Department of Physics, University of Wisconsin-Milwaukee, Milwaukee, WI 53211, USA}

\author[0000-0002-1292-1451]{Macon Magno}
\affiliation{Department of Physics, Southern Methodist University, Dallas, TX 75205, USA}

\author[0000-0002-4307-1322]{Chiara M. F. Mingarelli}
\affiliation{Department of Physics, University of Connecticut, 196 Auditorium Road, U-3046, Storrs, CT 06269-3046, USA}
\affiliation{Center for Computational Astrophysics, Flatiron Institute, 162 Fifth Ave, New York, NY, 10010, USA}

\author[0000-0002-2713-0628]{Francisco Muller-Sanchez}
\affil{Department of Physics and Materials Science, The University of Memphis, 3720 Alumni Avenue, Memphis, TN 38152, USA}

\author[0000-0002-5037-951X]{Kyuseok Oh}
\affiliation{Korea Astronomy \& Space Science institute, 776, Daedeokdae-ro, Yuseong-gu, Daejeon 34055, Republic of Korea}
\affiliation{Department of Astronomy, Kyoto University, Kitashirakawa-Oiwake-cho, Sakyo-ku, Kyoto 606-8502, Japan}
\affiliation{JSPS Fellow}

\author[0000-0002-2125-4670]{T. Taro Shimizu}
\affil{Max-Planck-Institut f{\"u}r extraterrestrische Physik (MPE), Giessenbachstrasse 1, D-85748 Garching bei M{\"u}unchen, Germany}

\author[0000-0001-5785-7038]{Krista Lynne Smith}
 \affiliation{Department of Physics, Southern Methodist University, Dallas, TX 75205, USA}

\author[0000-0003-2686-9241]{Daniel Stern}
\affil{Jet Propulsion Laboratory, California Institute of Technology, 4800 Oak Grove Drive, MS 169-224, Pasadena, CA 91109, USA}

\author[0000-0001-5649-7798]{Miguel Parra Tello}
\affiliation{Instituto de Astrof{\'{\i}}sica, Facultad de F{\'{i}}sica, Pontificia Universidad Cat{\'{o}}lica de Chile, Campus San Joaquín, Av. Vicu{\~{n}}a Mackenna 4860, Macul Santiago, Chile, 7820436} 

\author[0000-0002-0745-9792]{C. Megan Urry}
\affiliation{Yale Center for Astronomy \& Astrophysics and Department of Physics, Yale University, P.O. Box 208120, New Haven, CT 06520-8120, USA}

\begin{abstract}
  We present multiwavelength high-spatial resolution ($\sim$0\farcs1, 70\,pc) observations  of UGC\,4211 at $z=0.03474$, a late-stage major galaxy merger at the closest nuclear separation yet found in near-IR imaging (0\farcs{32}, $\sim$230\,pc projected separation).  Using Hubble Space Telescope/STIS,  VLT/MUSE+AO, Keck/OSIRIS+AO spectroscopy, and ALMA observations, we show that the spatial distribution, optical and NIR emission lines, and millimeter continuum emission are all consistent with both nuclei being powered by accreting supermassive black holes (SMBHs). Our data, combined with common black hole mass prescriptions, suggests that both SMBHs have similar masses, \lmbh$\sim$8.1 (south) and \lmbh$\sim$8.3 (north), respectively. The projected separation of 230 pc ($\sim$6$\times$ the black hole sphere of influence) represents the closest-separation dual AGN studied to date with multiwavelength resolved spectroscopy and shows the potential of nuclear ($<$50 pc) continuum observations with ALMA to discover hidden growing SMBH pairs. While the exact occurrence rate of close-separation dual AGN is not yet known, it may be surprisingly high, given that UGC\,4211 was found within a small, volume-limited sample of nearby hard X-ray detected AGN.    Observations of dual SMBH binaries in the subkiloparsec regime at the final stages of dynamical friction provide important constraints for future gravitational wave observatories.
\end{abstract}





\section{Introduction} \label{sec:intro}

There is evidence for a strong connection between major galaxy mergers ($<$3:1 stellar mass ratio) and supermassive black hole (SMBH) growth from theoretical models and computational simulations  \citep[e.g.,][and references therein]{2018MNRAS.479.3952B} since they provide a very efficient mechanism to remove angular momentum and drive gas to the nuclear regions as the two black holes are dragged toward each other during the dynamical friction phase. The SMBH pairs can sometimes be seen as dual active galactic nuclei (AGN), which provide a unique signature of merger-driven black hole growth \citep[e.g.,][]{VanWassenhove:2012:L7}.





Previous studies have identified many dozens to hundreds of dual AGN candidates based on several distinct and complementary methods, including optical spectroscopy with emission line ratios \citep{Liu:2011:101}, hard X-ray emission \citep[e.g.,][]{Koss:2011:L42,Koss:2016:L4}, double-peaked narrow emission lines \citep[e.g.,][]{2010ApJ...716..866S}, and most recently astrometry, which is used to identify high-redshift double quasars \citep[e.g.,][and references therein]{2021NatAs...5..569S}. All these methods have caveats and sometimes a significant fraction of false positives when further multiwavelength confirmational studies are performed \citep[e.g.,][]{2012ApJ...745...67F}.

The dawn of gravitational wave (GW) astronomy \citep{Collaboration:2016:061102} and the possible  imminent detection of nHz GWs with pulsar timing arrays \citep[e.g., PTAs;][]{Verbiest:2016:1267} has increased the urgency for solving the long-standing problem of SMBH binary formation timescales.  GW source predictions are largely based on parameterizations of theoretical and empirical galaxy merger rates \citep[e.g.,][]{Buchner19}, and thus carry large systematic uncertainties, reaching orders of magnitude \citep{Bonetti:2018:2599}.   Thus, the study of kiloparsec (kpc) and subkiloparsec (subkpc) dual AGN provides a unique opportunity to study systems with two black holes in the final stage of merging.

However, kpc and subkpc dual AGN are both more rare and more challenging to study than systems at larger separations (e.g. $>$3 kpc). This is likely due to enhanced obscuration in late-stage mergers \citep[e.g.,][]{Koss:2016:85,2021MNRAS.506.5935R}, which are the likely hosts of such sources; the limits of spatial resolution, especially at subkpc scales; the small fraction of radio-bright duals \citep{Burke-Spolaor:2011:2113}, where the emission becomes optically thin;  and/or the inefficiency of optical selection techniques, such as double-peaked narrow emission lines, which suffer from a high rate of false positives \citep[][]{Fu:2011:103}. Based on the observed samples of dual AGN, there has been tantalizing evidence that AGN triggering peaks in advanced-stage mergers where stellar bulge separations are $<$10 kpc \citep[e.g.,][]{Koss:2010:L125,Barrows:2017:129,2018ApJ...856...93F,2021ApJ...923...36S}, consistent with simulations that trace SMBH accretion rate evolution during such mergers \citep[e.g.,][]{Blecha:2018:3056}.  A crucial step forward is to study dual AGN with $0.1-1.0$ kpc separations in nearby galaxies \citep{Steinborn:2016:1013}. Despite intensive observational efforts to search for such subkpc dual AGN \citep[e.g.,][]{Muller-Sanchez:2018:48}, we still do not know how common they are, and we may very well be missing many such systems due to the aforementioned difficulties detecting them. 



While there have been several claims of dual AGN on 100s of pc scales, typically based on a single dataset or diagnostic, subsequent observations have often challenged their dual nature.  Some notable examples include NGC\,3393 \citep{Fabbiano:2011:431}, a third subkpc AGN in NGC\,6240 \citep{Kollatschny:2020:A91}, and  SDSS\,J101022.95+141300.9 \citep{Goulding:2019:L21}, which were later challenged in subsequent studies \citep[e.g.][]{Koss:2015:149,Treister:2020:149,Veres:2021:99}.  Ultimately, it is critical to identify subkpc dual AGN using a multiwavelength analysis to confirm the nature of their nuclei.

High-spatial resolution near infrared (NIR) adaptive optics (AO) observations have provided one of the best methods for confirming dual AGN as the technique can identify multiple stellar bulges using the NIR imaging \citep[e.g.,][]{2011ApJ...735...48S} and potentially probe close separations ($\sim$0\farcs{1}).  This approach was critical to demonstrate that most double-peak \OIII\ AGN are not dual systems \citep[e.g., 98\% of double-peaked \oiii\ emitters,][]{2012ApJ...745...67F}. The largest sample of nearby AGN observed using NIR AO is an imaging study of 96 nearby hard X-ray selected AGN \citep{Koss:2018:214a}.  That study did not focus on dual AGN candidates, but rather on conducting a blind survey of low-redshift AGN ($z<0.075$) detected in the ultrahard X-rays ($>$10 keV) with Swift Burst Alert Telescope (BAT). A search for dual AGN signatures among pairs of NIR nuclei in this sample has the advantage of starting with confirmed advanced mergers, where spatially resolved observations and analysis can be utilized to study each nucleus for AGN activity.


Here we study UGC\,4211 (also known as MCG+02$-$20$-$013 or SWIFT\,J0804.6+1045), the closest-separation dual NIR nuclei found in this NIR AO study. The dual nuclei were identified using segmentation maps with the secondary extended northern nucleus being $\sim4\times$ (1.4 mag) fainter than the southern nucleus in $K^\prime$ AO imaging.   It was previously classified as a  Sy\,2 system at $z=0.03474$ or $\sim$153\,Mpc \citep{Koss:2022:2} based on stellar absorption lines, and first identified as hosting an AGN based on optical spectroscopic follow-up of galaxies with warm far infrared colors \citep{Keel:1988:250}. The AGN was later detected in the hard X-rays as part of the Swift BAT 70 month catalog \citep{Baumgartner:2013:19}.  Previous morphological classifications using the Sloan Digital Sky Survey \citep[SDSS;][]{Koss:2011:57} or deeper Dark Energy Camera Legacy Survey (DECaLS) imaging \citep{Walmsley:2021:3966} did not identify a merger in this galaxy. 
We note that a more distant companion galaxy, at essentially the same redshift and separated by 122\arcsec\ (or 84\,kpc), has been identified in a previous study  \citep[the star forming galaxy SDSS J080440.36+104513.0; see][]{Koss:2012:L22}.  Due to the high equivalent width (EW) of the emission lines in the SDSS spectroscopy, the galaxy was selected as an  E+A (poststarburst) galaxy \citep{2017A&A...597A.134M}.  From a compilation study of BAT AGN \citep{Koss:2021:29}, the host galaxy was found to be relatively massive compared to nearby galaxies hosting AGN ($\log$ (\mstar/\Msun) $=$11.1), with significant molecular gas ($\log (M_{H2}/\Msun)=$9.7) and a high star formation rate ($\log({\rm  SFR}/ \Msun\,{\rm yr}^{-1}){=}0.9$), but still below \citep[e.g., $\log L_{IR}/L_{\sun}$${=}$10.7,][]{Shimizu:2017:3161}  the gas-rich luminous infrared galaxies (LIRGs; $\log L_{IR}/L_{\sun}$$>$$11.0$) among the sample \citep[e.g.,][]{Koss:2014}.   Throughout this study, we adopt $\Omega_m=0.3$, $\Omega_\Lambda=0.7$, and $H_0=70$\,km\,s$^{-1}$\,Mpc$^{-1}$, and a scale of 0\farcs{69}\,kpc$^{-1}$ based on the redshift of the system.  
  



\begin{deluxetable*}{l l l l l l l l}
\tablecaption{Summary of Observations \label{tab:obs}}
\tablehead{
\colhead{Observatory}&\colhead{Instrument}& \colhead{Date}& \colhead{Filter/Mode}& \colhead{Proposal ID}& \colhead{Range}&\colhead{Res}& \colhead{Exp.}\\
\colhead{}&\colhead{}& \colhead{}& \colhead{}& \colhead{}& \colhead{}&\colhead{(\arcsec)}& \colhead{(ks)}
}
\startdata
NuSTAR&&2017-03-11&&BAT Legacy&3--70\,keV&8&20.3\\
Chandra&ACIS&2019-02-08 &&20701055&0.5--8\,keV&0.5&10\\
HST&WFC3&2021-01-21&F225W&16241&2365\,\AA&0.08&1.05\\
HST&STIS&2017-02-05&G430L&14248&2870--5680\,\AA&0.1&1.1\\
HST&STIS&2017-02-05&G750M&14248&6480--7045\,\AA&0.1&1.4\\
VLT&MUSE&2021-11-30 to 2022-1-03&&106.21HW,108.22C1&4800--9300\,\AA&0.09--0.05&6.6\\
HST&ACS&2018-12-28&F814W&15444&8045\,\AA&0.08&0.67\\
Keck&OSIRIS&2021-01-20&Jbb&N156&1.18--1.44\,\micron&0.2&2.4\\
Keck&OSIRIS&2017-11-02&Hbb&C333&1.47--1.80\,\micron&0.1&1.2\\
Keck&OSIRIS&2017-11-02&Kbb&C333&1.97--2.38\,\micron&0.1&3.6\\
ALMA&&2021-10-24, 2022-08-19&&2021.1.01019.S&221--240\,Ghz&$\sim$0.06&2.3\\
JVLA&&2018-12-04&&18B-245&22 Ghz&1.7--1&0.6\\
\enddata
\end{deluxetable*}

\section{Observations and Data Reduction}
\label{sec:obs_data}

Our analysis is based on  multiwavelength observations obtained with multiple facilities.
A summary of the observations including the observation dates, programs, spatial resolution, and exposure times is provided in \autoref{tab:obs},  with more details regarding the observing conditions and data reduction provided in \autoref{new_obs}.  
We utilize new optical observations from the Hubble Space Telescope (HST) Imaging Spectrograph (STIS), optical AO-assisted integral field spectroscopic (IFS) observations from the Multi Unit Spectroscopic Explorer \citep[MUSE;][]{2010SPIE.7735E..08B} instrument in narrow-field mode (NFM) at the Very Large Telescope (VLT), NIR IFS from the OH Suppressing InfraRed Imaging Spectrograph \citep[OSIRIS;][]{Larkin:2006:441} at the W. M. Keck Observatory with the AO system in laser guide star (LGS) mode, and millimeter (mm) observations from the Atacama large Millimeter/submillimeter Array (ALMA). The HST UV data that was a nondetection, archival HST optical imaging data, as well as the NuSTAR, Chandra, and 22 GHz Karl G. Jansky Very Large Array (JVLA) data, which all have insufficient spatial resolution to resolve the two nuclei, are presented in \autoref{chan_vla}.
 




\begin{figure} 
\centering
\includegraphics[width=0.46\columnwidth]{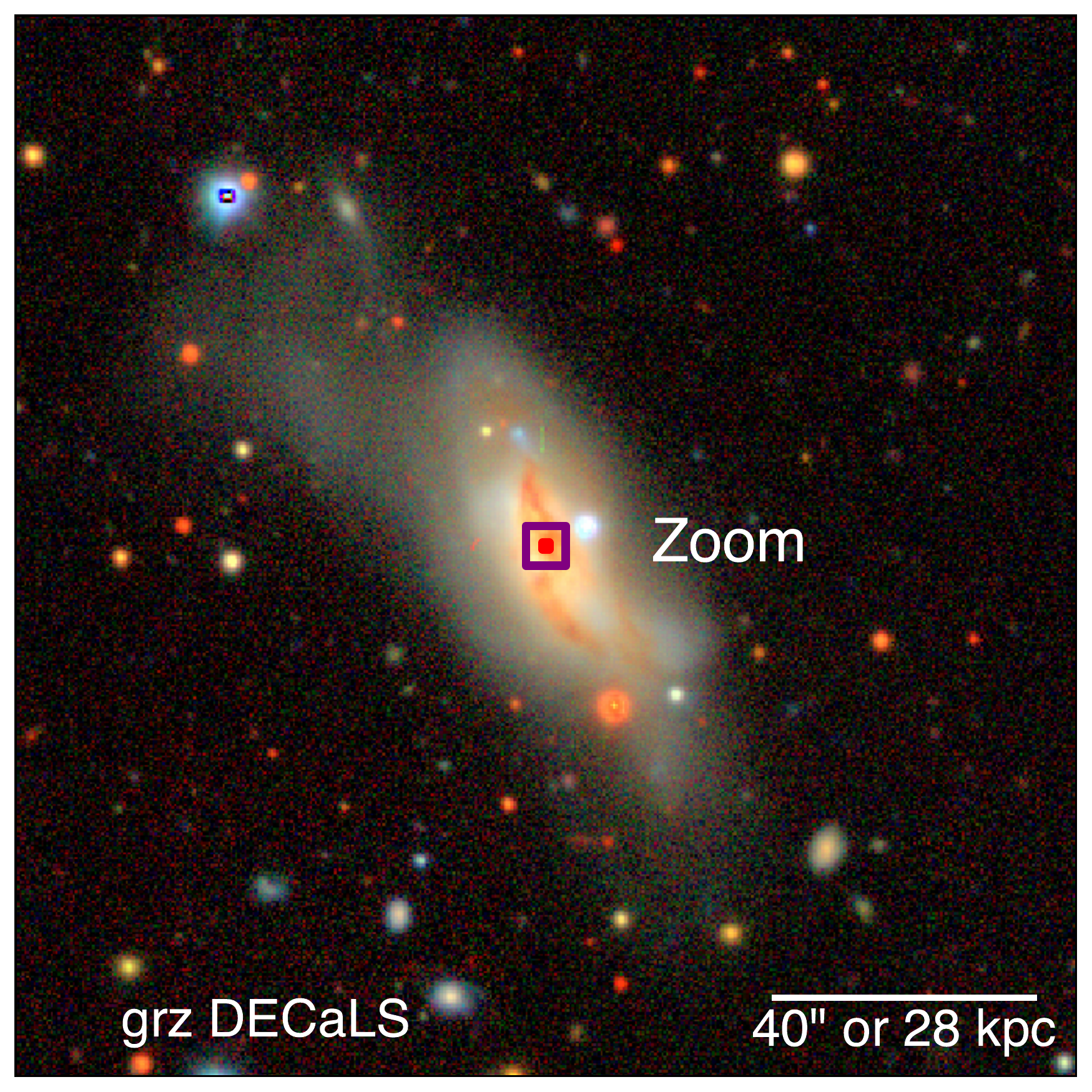}
\includegraphics[width=0.46\columnwidth]{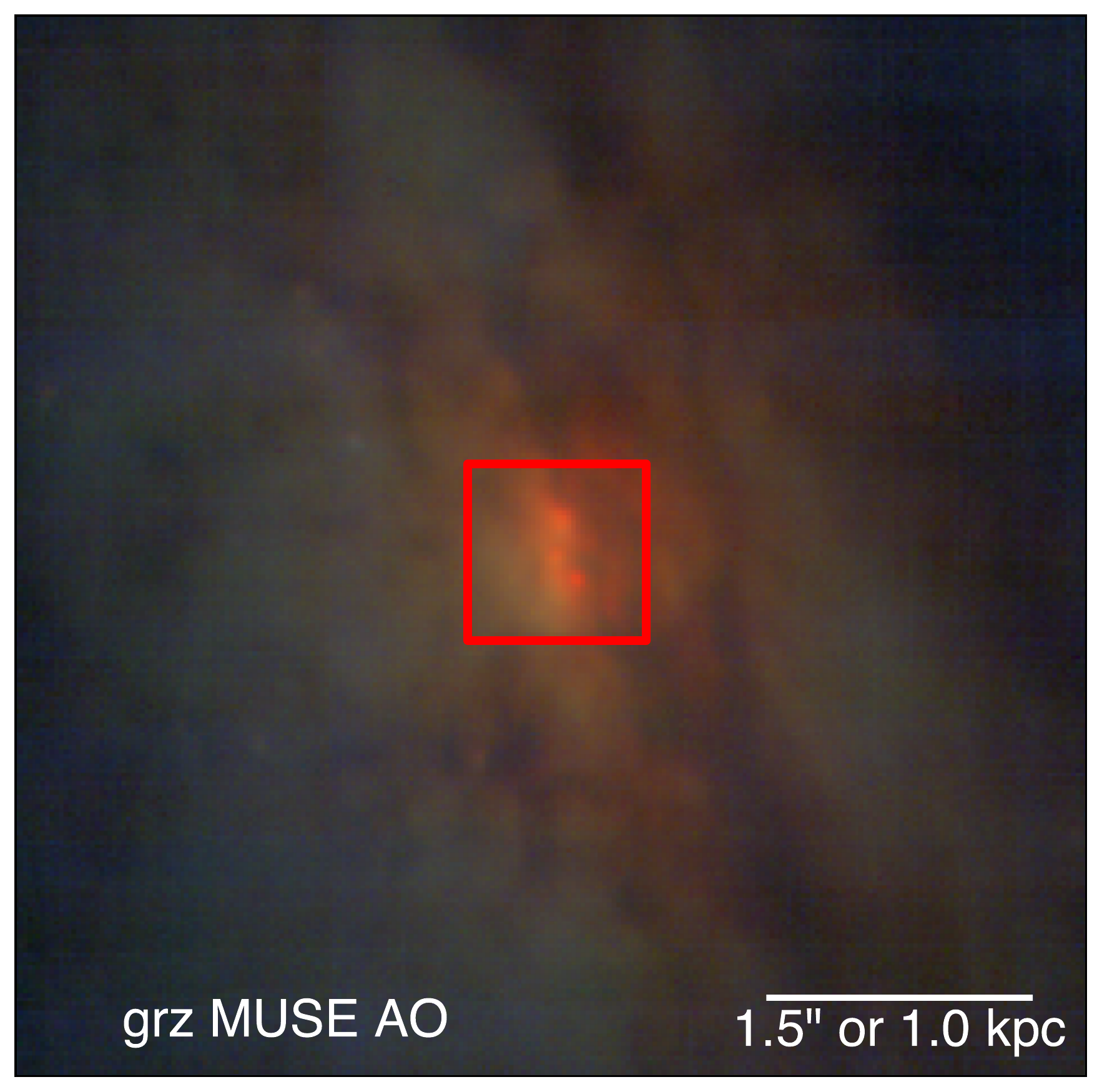}\\
\includegraphics[width=5.6cm]{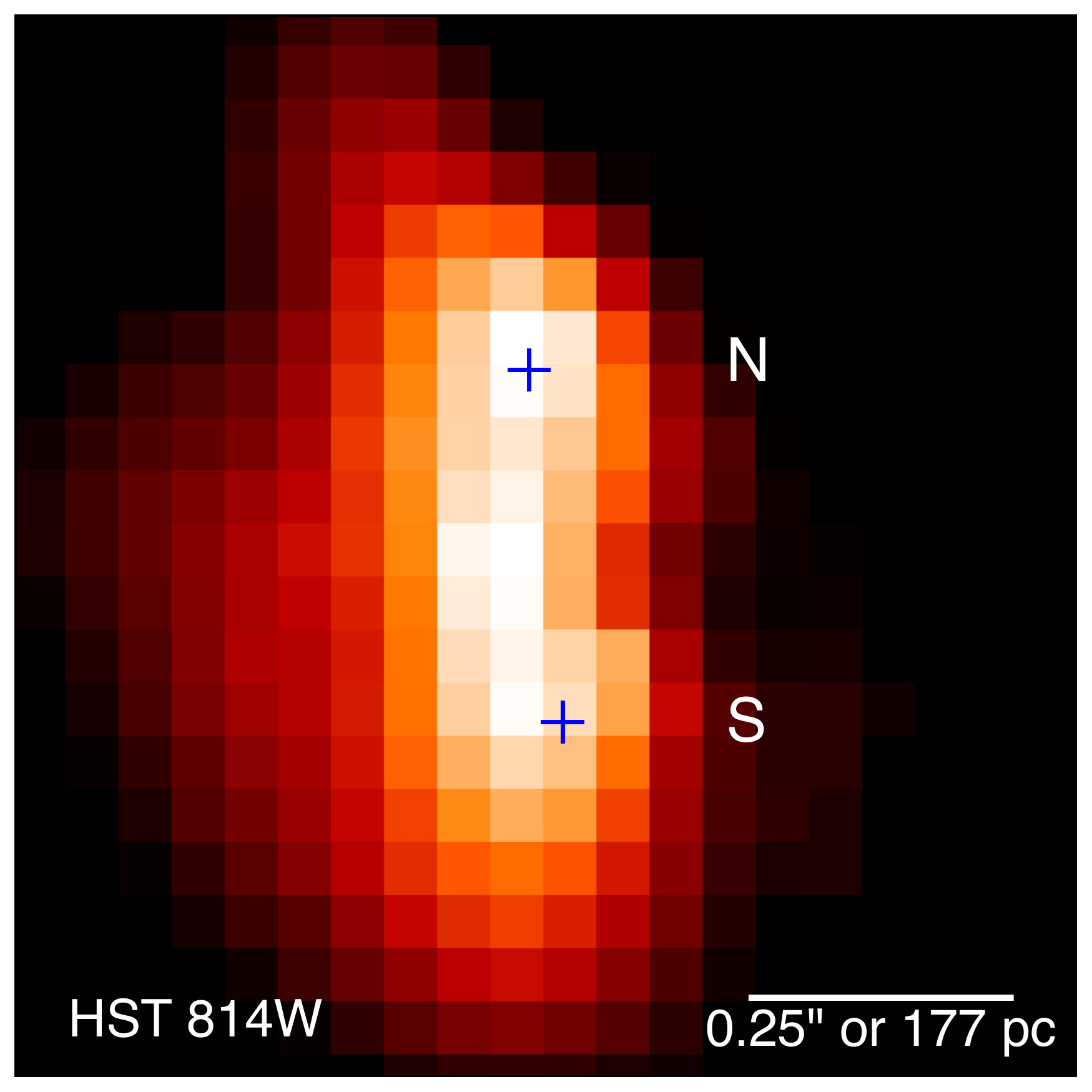}
\includegraphics[width=5.6cm]{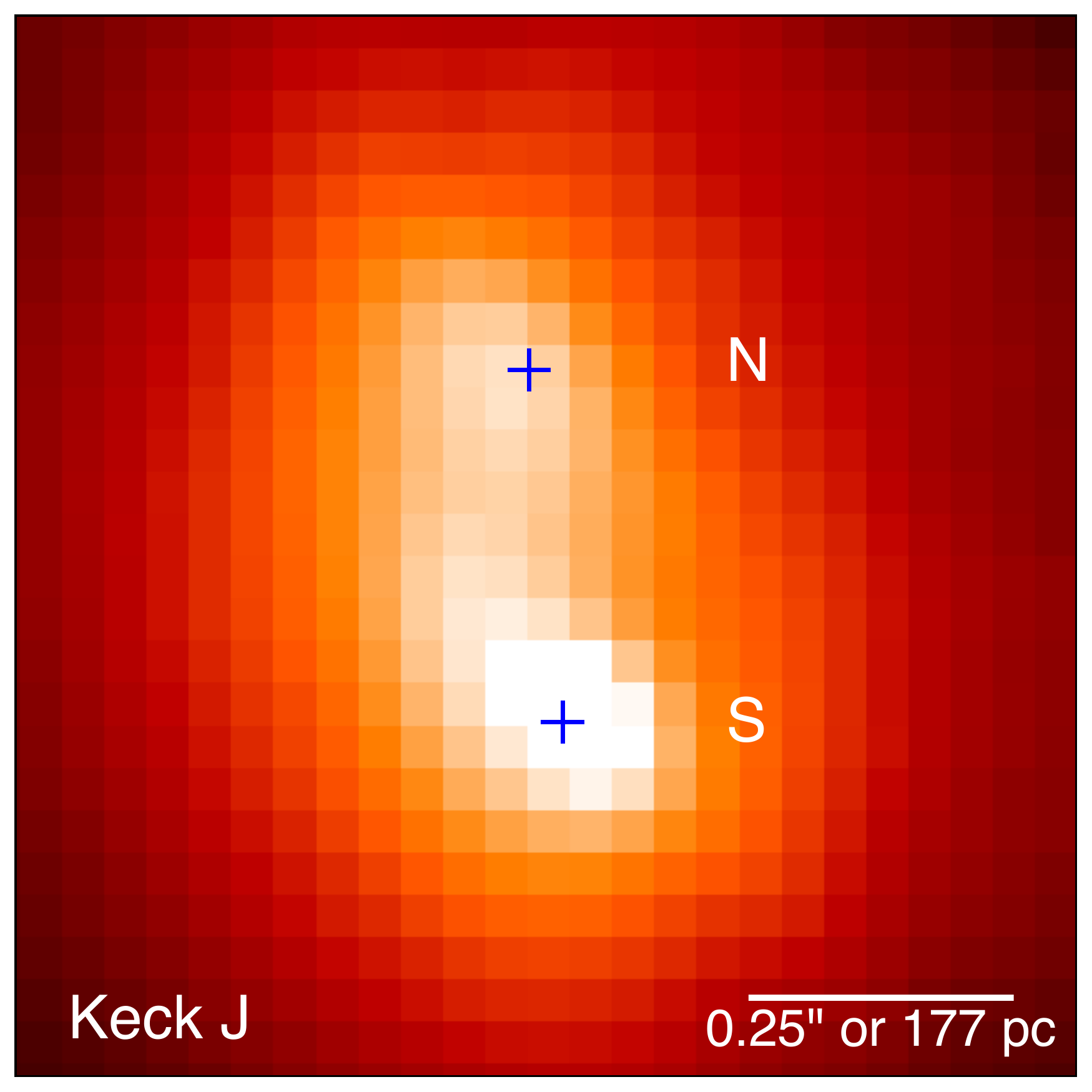}
\includegraphics[width=5.6cm]{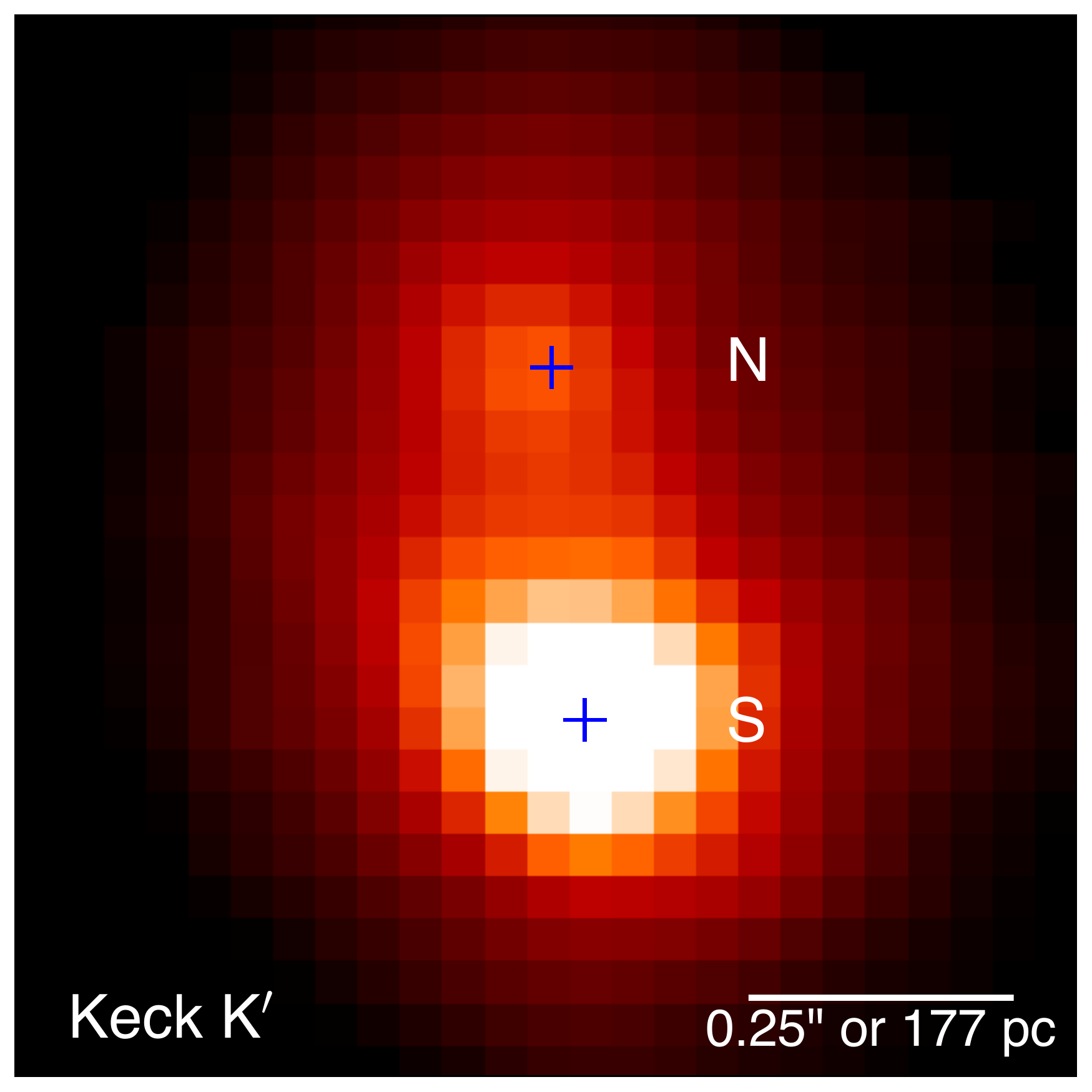}
\includegraphics[width=5.6cm]{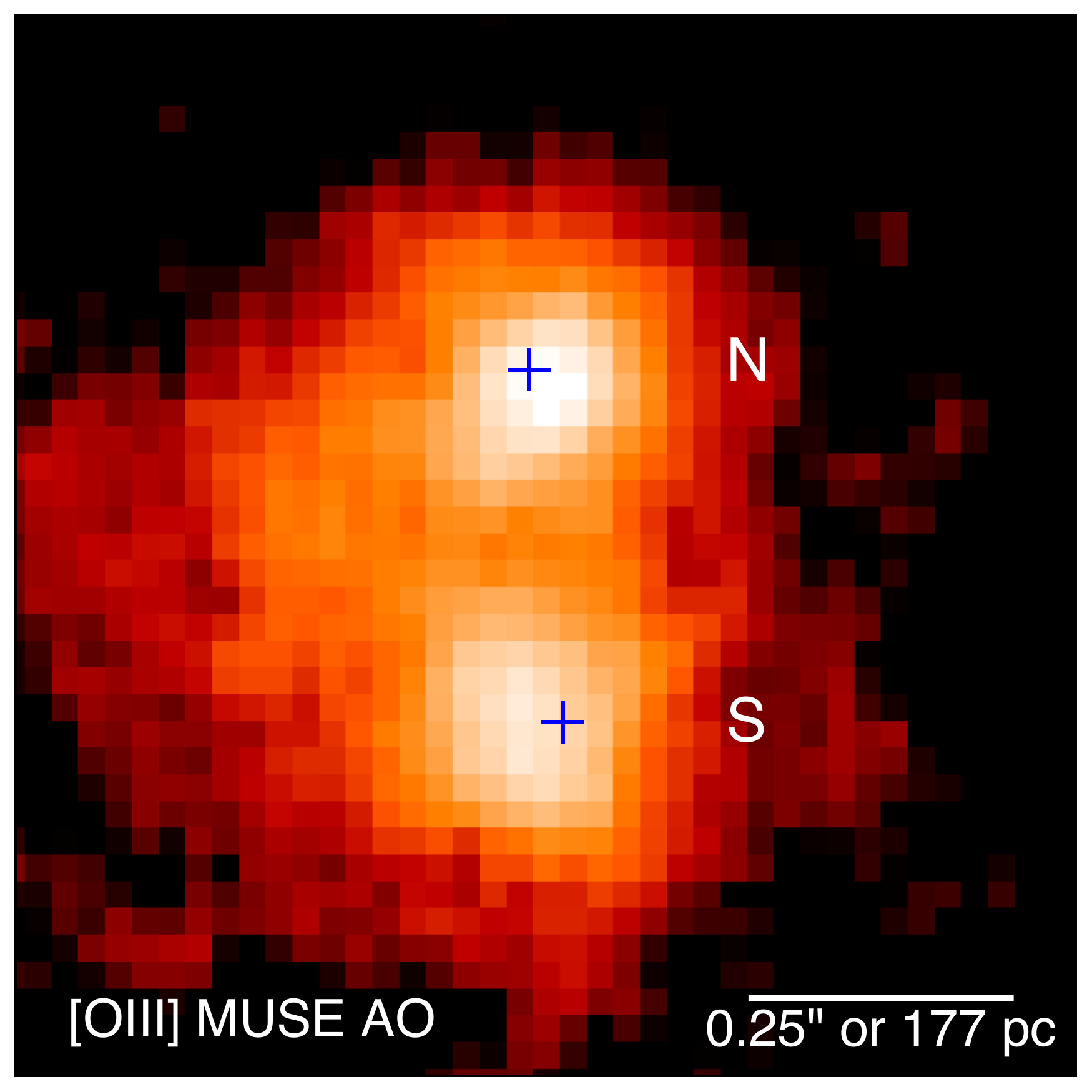}
\includegraphics[width=5.6cm]{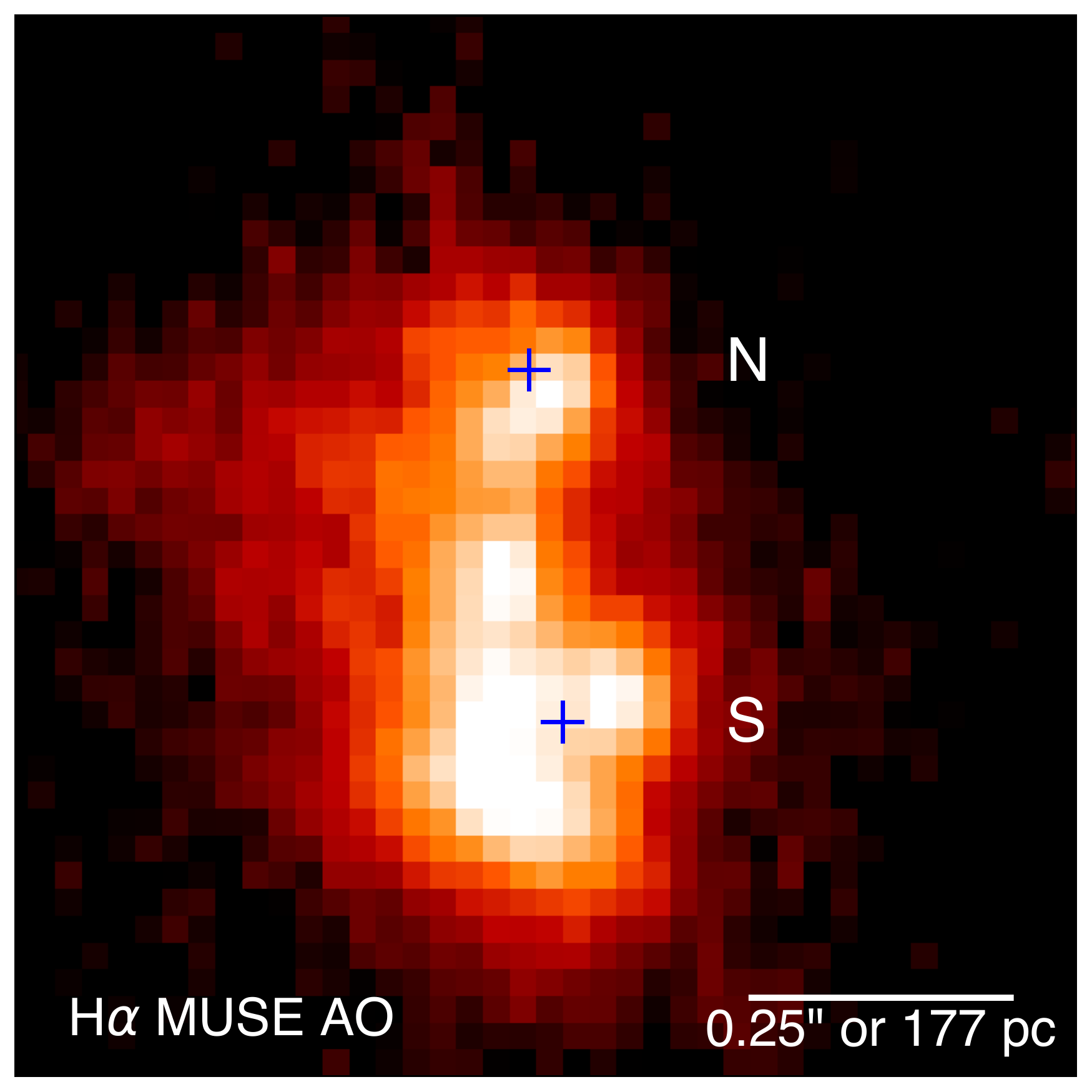}
\includegraphics[width=5.6cm]{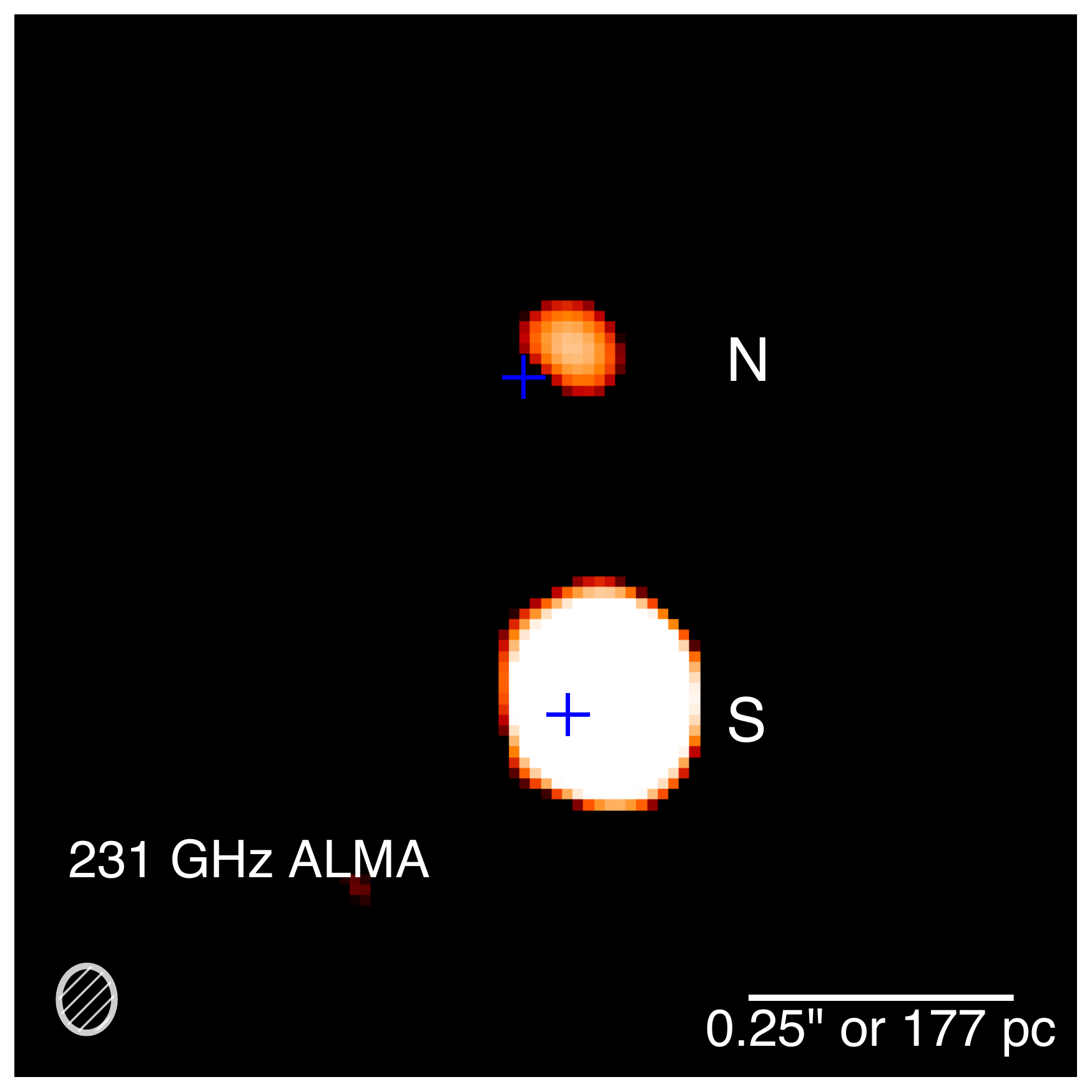}
\caption{Multiwavelength, multiscale images of UGC\,4211.  Top left: 3$\arcmin$-wide $grz$ color image from the DECaLS with an asinh stretch.  The purple square indicates the 6\arcsec\ zoom region used for the upper right panel, while the tiny red square is the 1\arcsec\ zoom region used for our highest-resolution analysis of the dual nuclei.   
Top right: 6\arcsec\ $grz$ color image constructed from our new MUSE AO data cube. Here the red square indicates the 1\arcsec\ zoomed-in circumnuclear region.  
The panels in the second row show the 1$\arcsec$ circumnuclear region in HST F814W (left), NIRC2 AO $J$-band (middle), and NIRC2 AO $K^\prime$ (right) in log stretch. The northern and southern emission components identified in NIR are indicated (N and S, respectively), and are separated by 0\farcs{32}, almost exactly along the north--south direction (PA 7.5\deg\ east of north).   The panels in the {\em bottom row} show emission maps from our MUSE AO data, centered on \oiii\ (left) and \halpha\ (middle), as well as from ALMA continuum emission at $\sim$230\,GHz (right, white hatched circle indicates the beam size.). The positions of the two NIR nuclei are shown with blue crosses.   }
\label{fig:imaging}
\end{figure}



\begin{figure*} 
\centering
\includegraphics[width=\columnwidth]{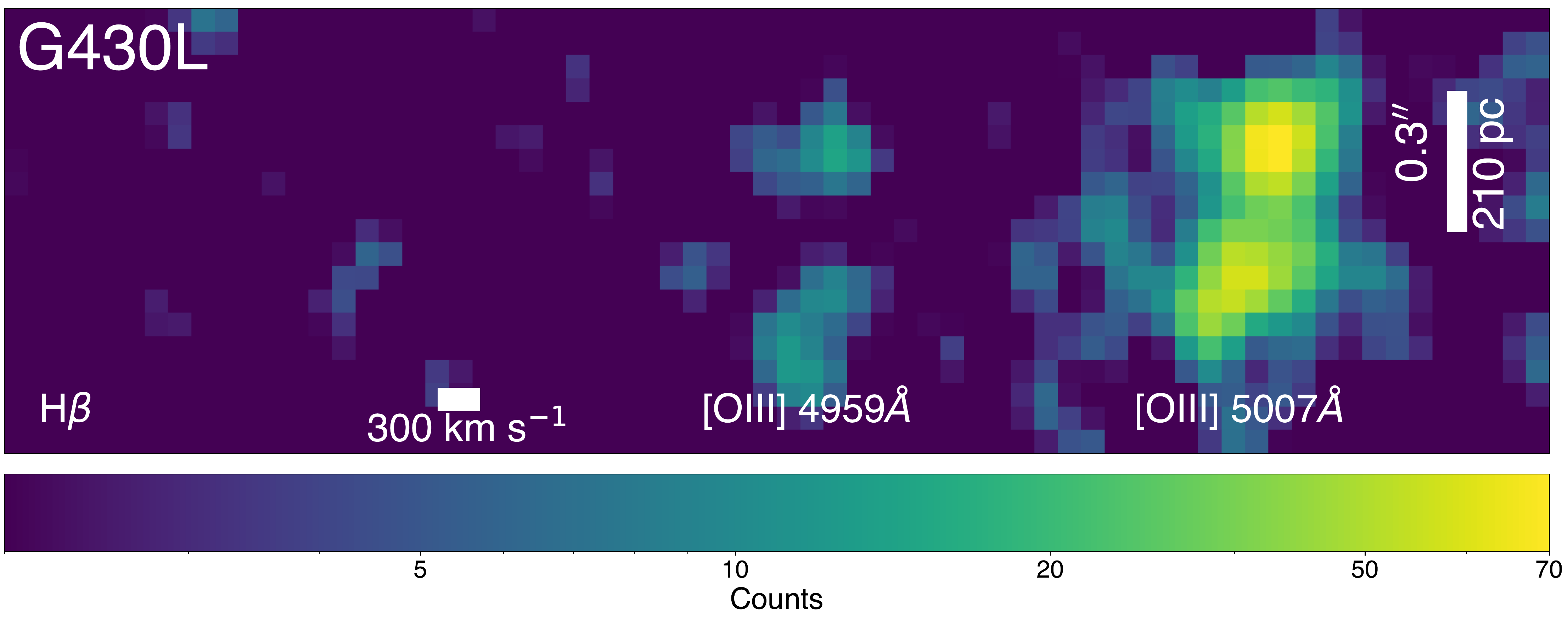}
\includegraphics[width=\columnwidth]{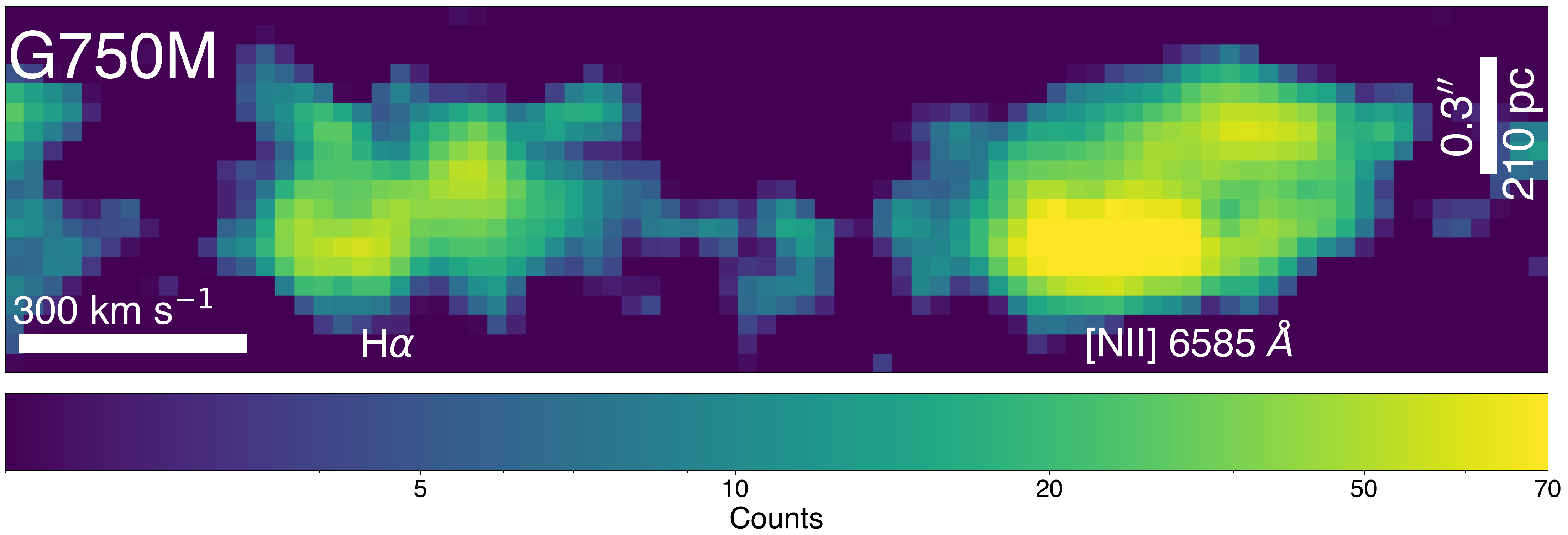}
\includegraphics[width=\textwidth]{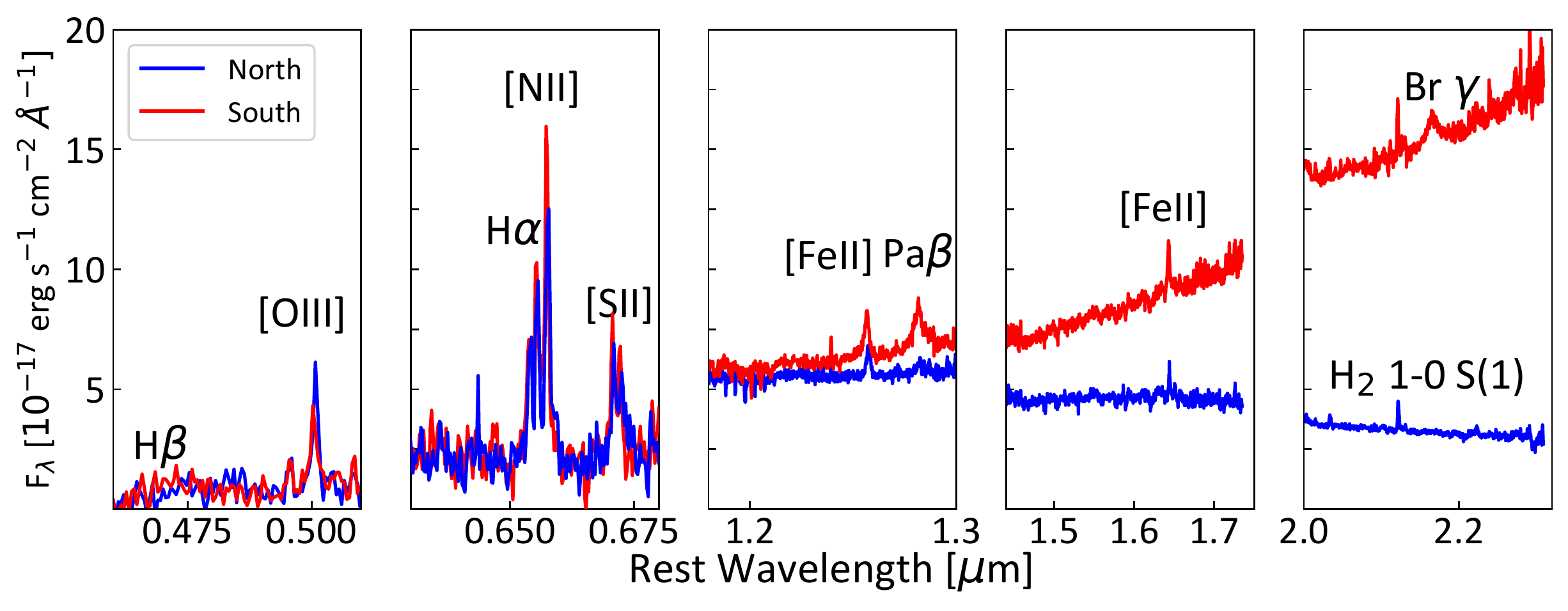}
\caption{HST/STIS spatially resolved 2D spectroscopy of the \oiii\ ({\em top row}) and the \halpha\ spectral region ({\em middle row}).  The spectroscopy is from a 0\farcs{2} long slit, aligned with the two NIR nuclei in the N--S direction.  The vertical axis is spatial (north--south) and the horizontal axis is spectral ($\lambda$ increasing to the right).    {\em Bottom row}: STIS spectra of the northern and southern sources in the \oiii\ and \halpha\ spectral regions (extracted from 0\farcs2 wide apertures; first two panels), along with OSIRIS NIR spectra of the two nuclei (extracted from 0\farcs3 diameter aperture; last three panels).  }
\label{fig:stis}
\end{figure*}

\begin{figure*} 
\centering
\includegraphics[width=5.9cm]{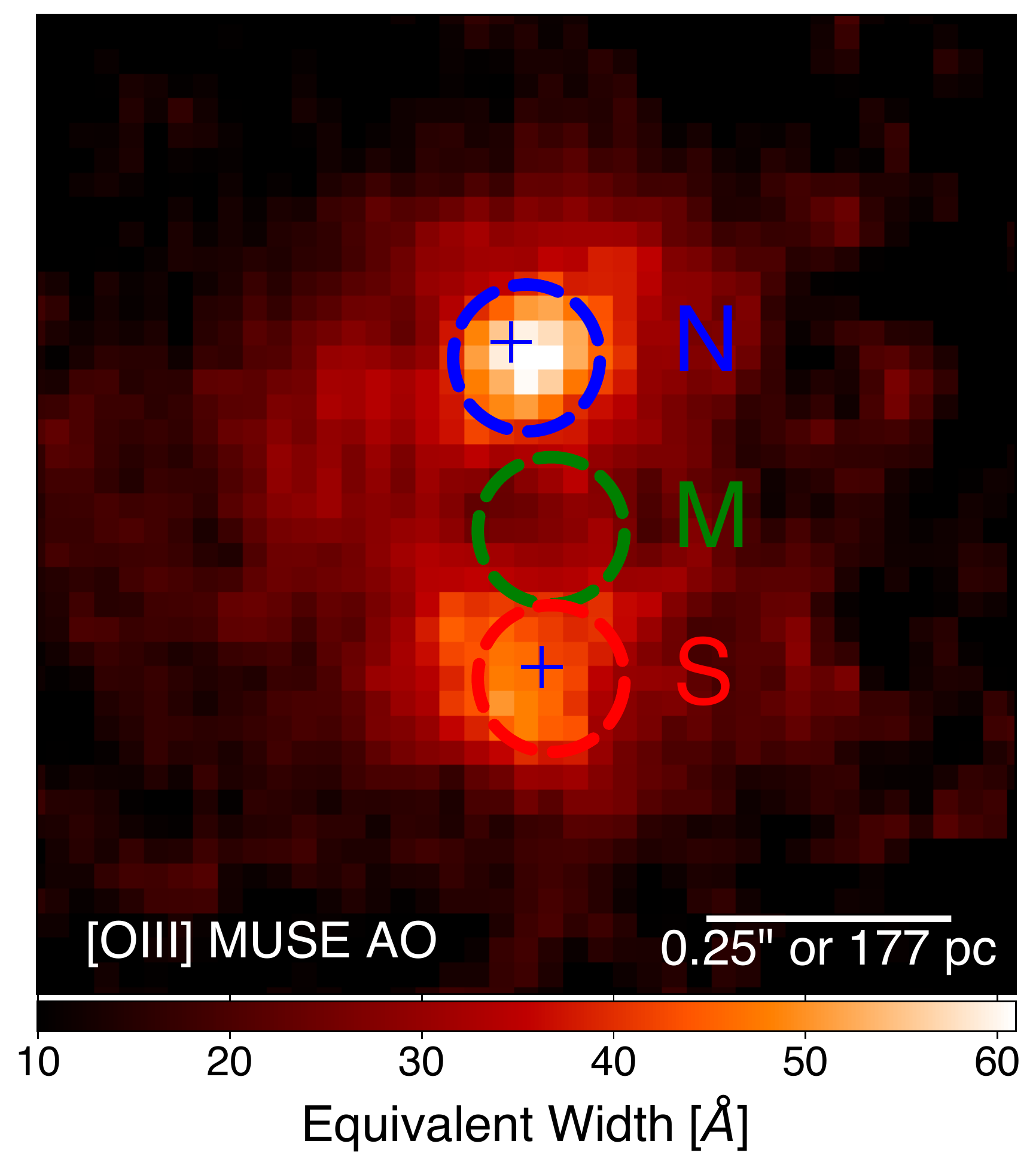}
\includegraphics[width=5.9cm]{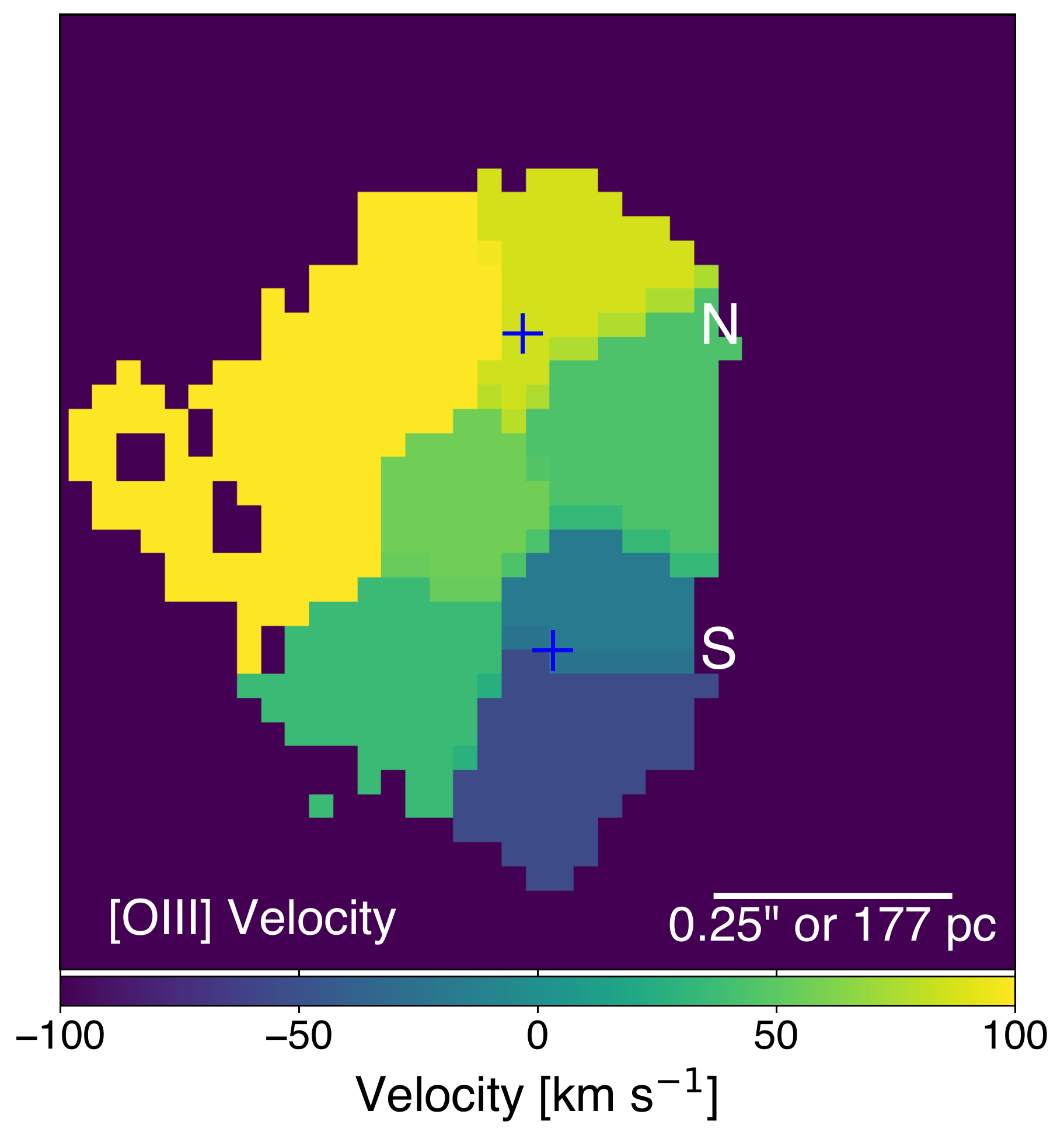}
\includegraphics[width=5.9cm]{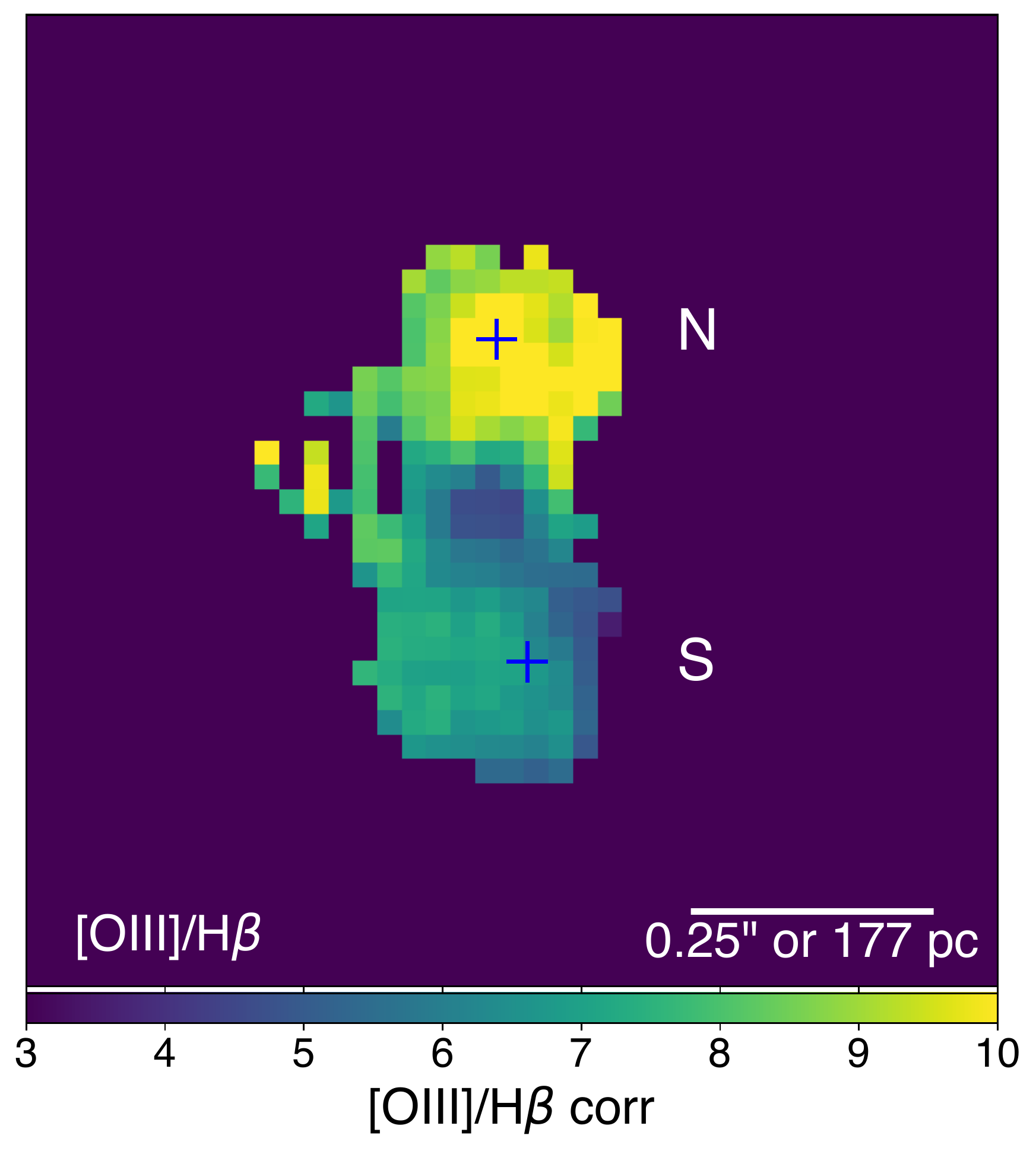}
\includegraphics[width=0.88\columnwidth]{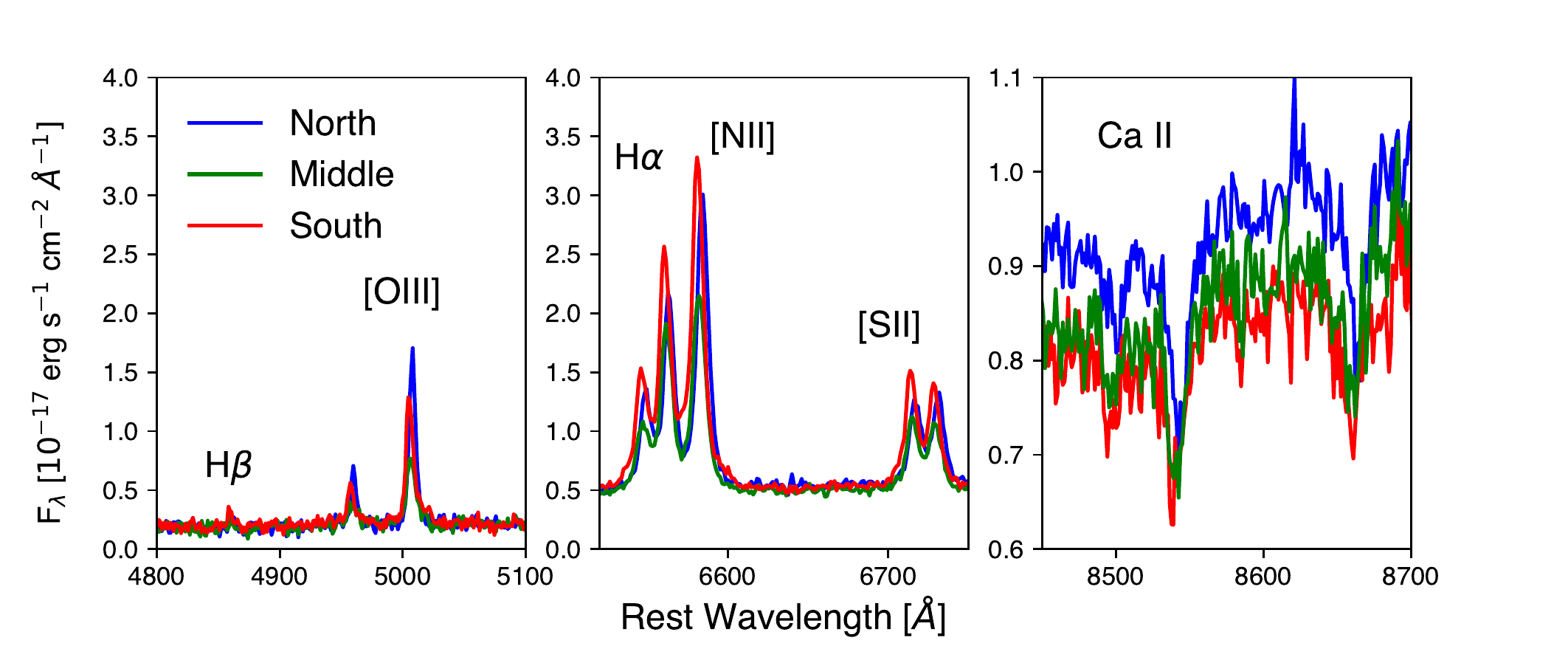}\\
\includegraphics[width=0.88\columnwidth]{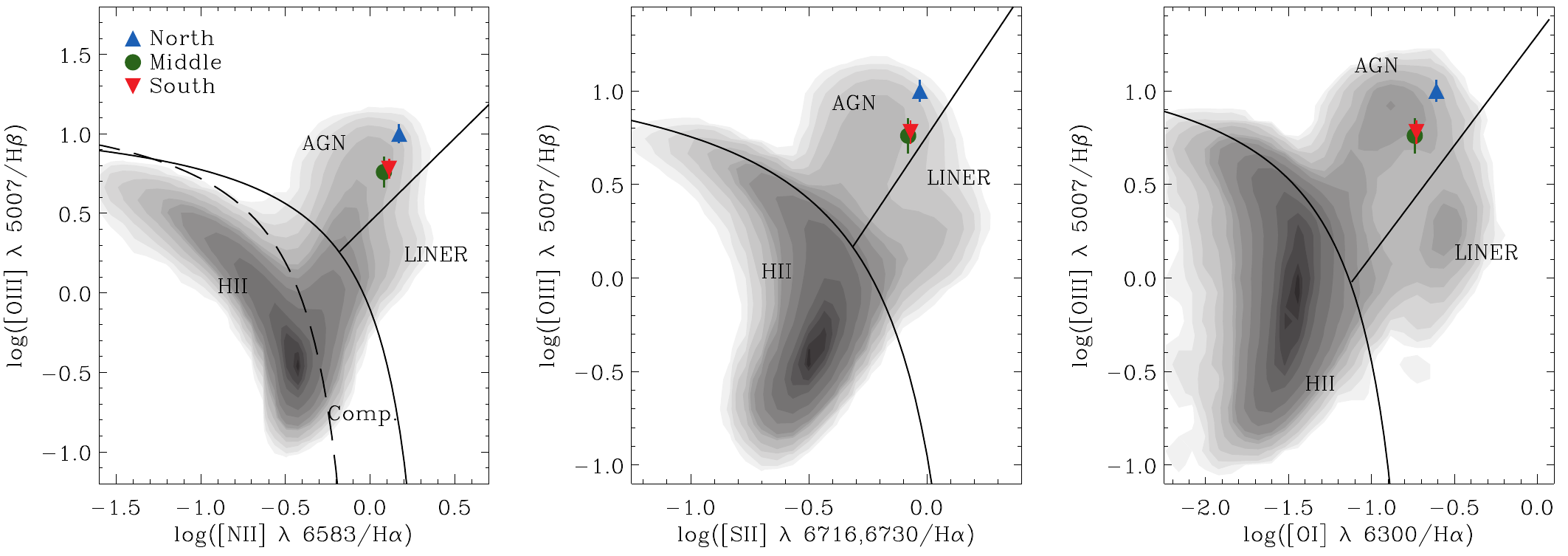}
\caption{ Top left and middle:  VLT/MUSE AO-assisted, 1\arcsec\ wide map of  \oiii\ equivalent width and velocity in the heart of UGC\,4211 scaled with linear stretch.  Dashed circles mark the circular apertures used for 1D spectral extractions (0\farcs{15} diameter) for the northern nucleus (N; blue), the southern nucleus (S; red), and the region halfway between them (M; green).  For the velocity map, Voronoi binning was used to increase S/N to measure the \OIII\ emission line velocity offset (from $z=0.03474$), with excluded low S/N regions in black.  The positions of the two NIR nuclei are shown with blue crosses.  Top right: \OIII/\hbeta\ map.   Since \Hbeta\ is too weak for detection, we used the \Halpha\ emission, assuming a ratio of $\sim$15 as measured in the \halpha/\hbeta\ emission from the three spatial regions, with low S/N regions in black. Second row: spectra of the three spatial regions (north, south, middle), showing the key emission lines (i.e., \hbeta, \oiii, \halpha, \nii, \sii) along with the \Catrip\ absorption. Bottom row: strong line ratios (BPT) diagrams for the three regions.  Gray shaded contours denote the SDSS sample density \citep{Oh:2011:13}.}
\label{fig:muse}
\end{figure*}

\begin{figure*} 
\centering
\includegraphics[width=8cm]{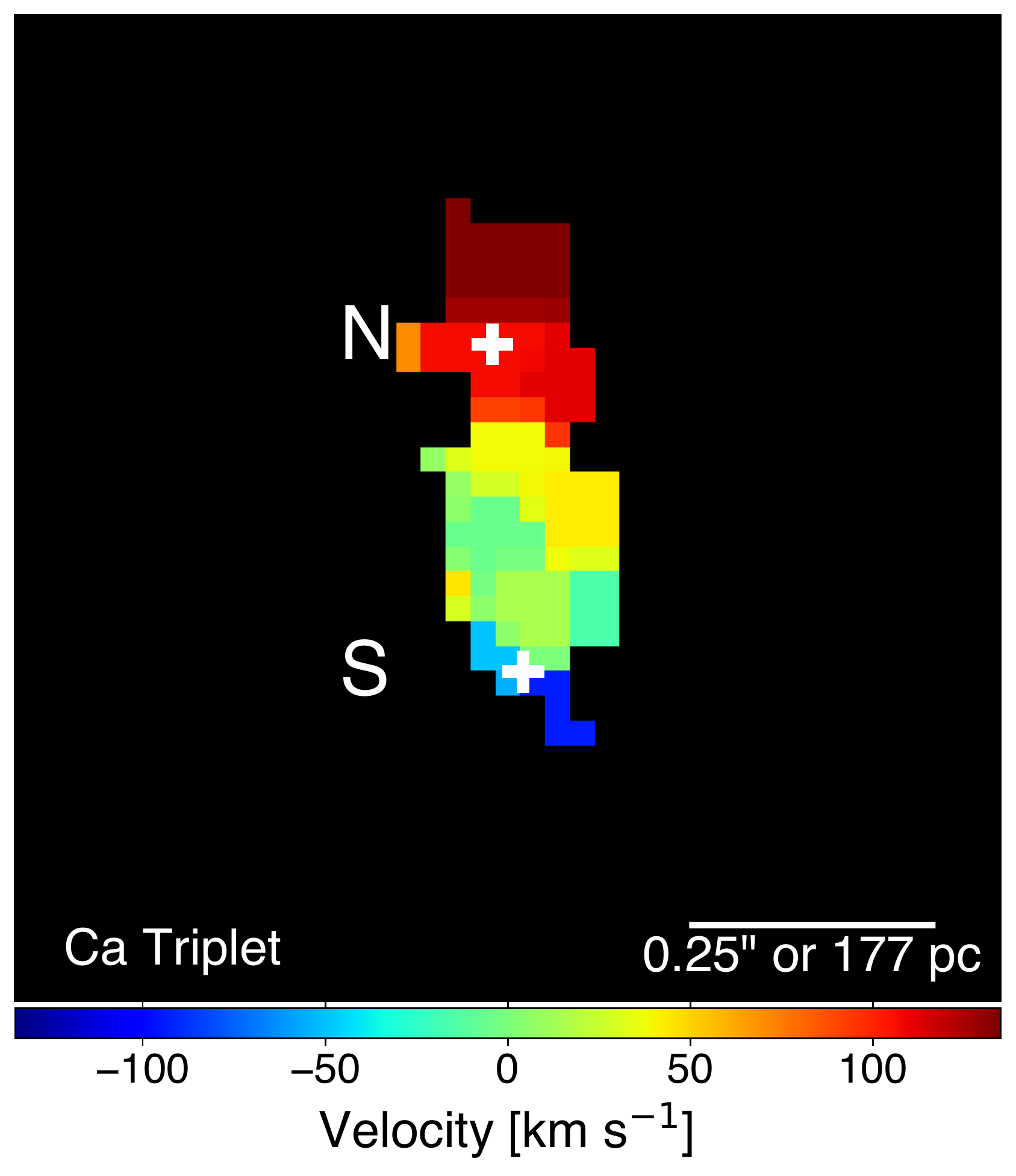}
\includegraphics[width=8cm]{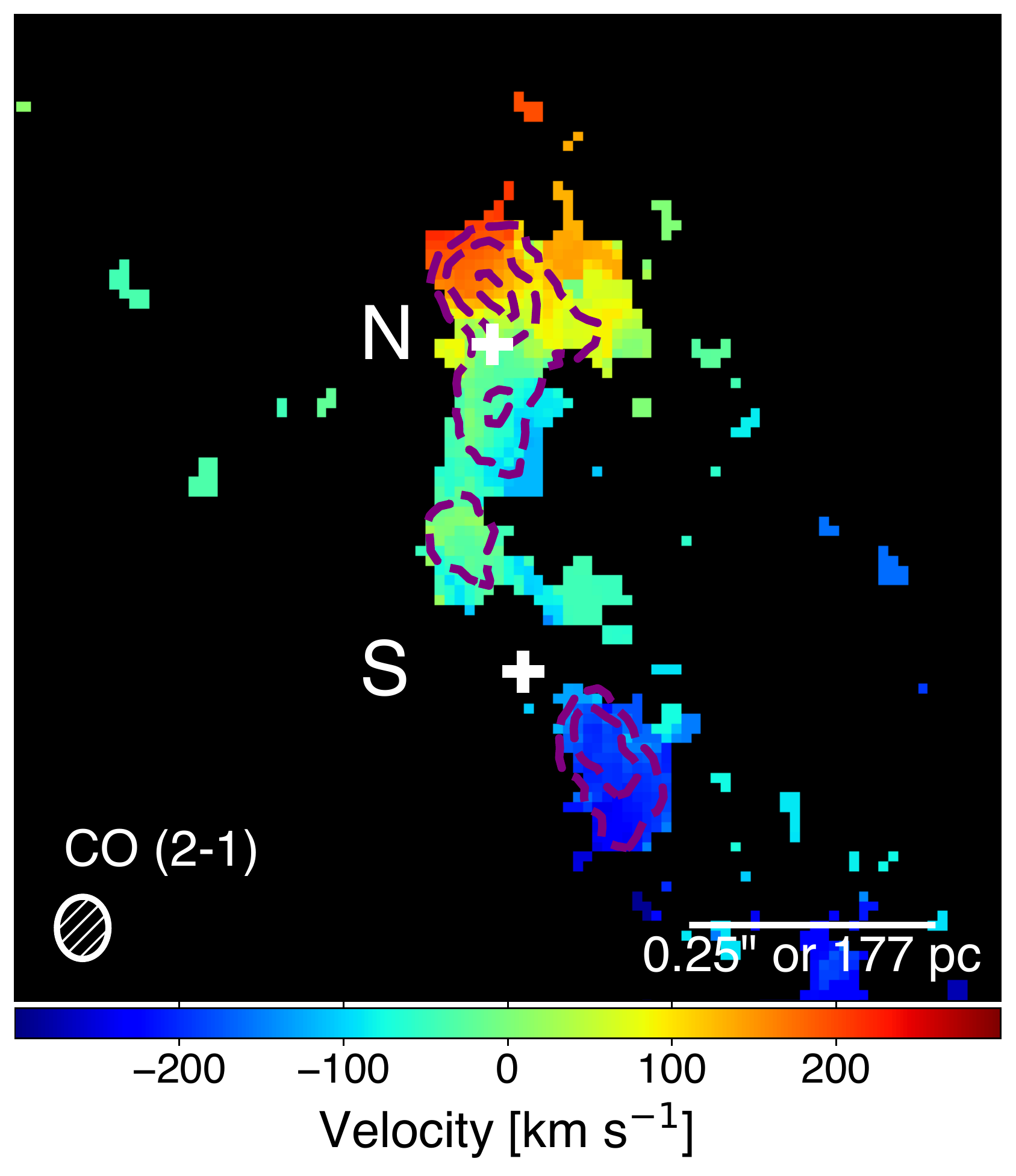}
\caption{Left: VLT/MUSE AO-assisted, 1\arcsec\ wide map of velocity from fitting the CaT stellar absorption region (8300--8900\,\AA) in the heart of UGC\,4211 scaled with linear stretch.    Voronoi binning was used to increase the S/N, with excluded low S/N regions in black.  The positions of the two ALMA continuum sources (which closely match the NIR nuclei) are shown with white crosses. Right: ALMA $^{12}$CO(2-1) 1\arcsec\ wide map showing the distribution of molecular gas velocities.  Purple contours indicate the location of the velocity-integrated $^{12}$CO(2-1) emission. The white hatched circle shows the beam size.}
\label{fig:musecatrip}
\end{figure*}



\section{Analysis and Results}\label{sec:results}
We focus in this Letter on the dual AGN nature of the two NIR nuclei.  In this section we describe our imaging and line emission analysis, along with stellar kinematics and AGN properties.  Additional analysis of the NuSTAR, Chandra, and VLA data are presented in \autoref{chan_vla}. A detailed study of the distribution of the molecular gas observed with ALMA and the ionized gas and stellar populations with VLT/MUSE will be presented in a companion paper (Treister et al., in preparation).

\subsection{Imaging Analysis} \label{sec:imaging}
\autoref{fig:imaging}  shows  the UGC\,4211 system at a range of scales, from the nuclear region (i.e. $<$800 pc or 1\arcsec) at high resolution, to tens of kpc.  On scales of $\sim$30\,\kpc, likely tidal tails can be seen, along with dust lanes extending roughly 10 kpc in the north--south direction. The MUSE/NFM $grz$ pseudoimage shows that the nuclear region hosting the two nuclei is significantly reddened compared to the surrounding (stellar) emission.  

In the central region of the system, both nuclei become prominent in the reddest NIR imaging (i.e., the $K^\prime$ band), consistent with being highly obscured.  In the \OIII\ image from MUSE, two prominent emission areas are seen, while in H$\alpha$ the southern nucleus possibly shows a more complicated structure at small scales ($<$0\farcs{1}) at the center of the nucleus.

The nuclear ALMA 231\,GHz continuum emission ($<$200\,pc), has been found to be a good proxy of the AGN luminosity, with a tight correlation with the 14--150\,\kev\  \citep[0.36 dex;][]{2022ApJ...938...87K}.  This emission is shown in the bottom right panel of \autoref{fig:imaging} from a high resolution $\sim$0\farcs{07} (49\,pc) map, with a brighter southern source detected at a S/N of 116 and fainter northern source at S/N of 6.2 based on the peak flux. Both sources are unresolved at this resolution, and remain unresolved on a higher resolution, $\sim$0\farcs{04} (28\,pc), map obtained using a Briggs robust parameter of -0.5.  Using the \texttt{CASA} \texttt{imfit} task to fit model Gaussians spatially, we find two sources of continuum emission consistent with the northern and southern nuclei (both in separation and position angle). The positions of the two emitters are R.A.=08:04:46.3902, decl.=+10:46:35.9407, for the brighter southern source, with a flux density of $1.705\pm0.037$\,mJy, and R.A.=8:04:46.3921, decl.=+10:46:36.2723 for the fainter northern one, with a flux density of $0.140\pm0.032$\,mJy. This corresponds to a separation of 0\farcs{33}$\pm$0\farcs{01} or 229$\pm$7\,pc at a position angle (PA) of 4.8\deg$\pm$3.   The brighter southern source had a spectral index ($S_{\nu} \propto\nu^{-\alpha}$) of $\alpha=0.02\pm0.26$, which is consistent with other hard X-ray selected AGN  \citep[$\alpha_{\rm mm} = 0.5\pm1.2$;][]{2022ApJ...938...87K}.   The secondary is too faint to derive meaningful constraints on $\alpha$.


Using the brightest pixel in the NIR $K^\prime$ emission for each of the two nuclei, we find a separation of 0\farcs{32}$\pm$0\farcs{03}, with a PA of 7.5\deg$\pm$5\deg.  The NIR positions are R.A.=08:04:46.3917, decl.=+10:46:35.926, for the brighter southern source, with an offset of 0\farcs{03} from the southern ALMA source, and R.A.=8:04:46.3946, decl.=+10:46:36.244, for the fainter northern one, with an offset of 0\farcs{05} from the northern ALMA source. Therefore, the position, separation, and PA of the NIR nuclei closely match the 2 millimeter (mm) sources given the relative astrometric and centroiding errors ($\sim$0\farcs{1}).

In the optical imaging in F814W and the emission lines, however, there is a small shift to the E in the peak emission of the southern nucleus, compared to the $J$ and $K^\prime$ NIR nuclei.  There is a small shift ($\sim$0\farcs{04} to the E) in the \oiii\ emission as well. The F814W and \Halpha\ images, both show some enhanced emission between the two nuclei, but this is not seen in the \oiii\ image.  In summary, we find the northern nucleus to be closely aligned in the optical to the NIR and emission line region, while the optical emission of the southern nucleus shows a small shift (0\farcs{04}) relative to the NIR and mm peaks.

\subsection{Nuclear Emission Line Analysis}
The combination of HST/STIS, VLT/MUSE in AO, and Keck/OSIRIS in AO, provide a high-spatial resolution ($\sim$0\farcs{1}) spectral study of the nuclear region from 0.3 to 2.4\,\micron.  \autoref{fig:stis} (top and middle rows) shows the 2D spatially resolved maps of HST/STIS along the nearly north--south direction, for the brightest emission lines (\oiii, \Halpha, and \nii).  Two bright components in the emission lines are seen, separated by 0\farcs{3} (6 pixels) consistent with the two NIR nuclei. The southern nucleus shows a clear $\sim$150\,\kmps blueshift compared to the northern nucleus. This velocity offset is important as it means that any contamination due to extended emission from the point-spread function (PSF) will not overlap in the two regions.

To examine the potential AGN nature of the nuclei, \autoref{fig:stis} shows the  HST/STIS spectra extracted at the northern and southern nuclei. We extracted a 5 pixel (0\farcs{25}) spectrum along the N--S direction across the nuclei peak \oiii\ emission in HST/STIS which is at similar separation to the two NIR nuclei (0\farcs{3}).  HST/STIS has a slit width of 0\farcs{2} and is aligned in the N--S direction. With Keck/OSIRIS (\autoref{fig:stis}), we extract a spectrum with a 0\farcs{3} diameter aperture around each nuclei (\autoref{sec:osiris_regions}).  

The northern source is brighter in \oiii, while the southern source is brighter in \nii\, consistent with the spatially resolved 2D spectroscopy.  No \hbeta\ is detected in either nucleus.   The [Fe\,{\sc ii}]\,$\lambda1.257$ and [Fe\,{\sc ii}]\,$\lambda1.644$\,\micron\ NIR lines are seen in both nuclei, along with H$_2$ 1-0 S(1) and H$_2$ 1-0 S(2) rovibrational emission lines at 2.03 and 2.12\,\micron, (respectively), which trace warm molecular gas. The southern nucleus shows a sharp brightening in the continuum redward of 1\,\micron, likely due to a contribution from AGN heated dust \citep{Lyu:2018:92},  commonly seen in  broad line AGN.  A hidden broad-line region (BLR) component ($>$1000\,\kms) is also found in the NIR hydrogen recombination lines (e.g. Pa$\beta$, FWHM=1996$\pm$180\,\kms; Br$\gamma$, FWHM=2530$\pm$160\,\kms) of the southern nucleus, but no BLR is found in the northern nucleus. The [Si\,{\sc vi}]\,$\lambda$1.9640  coronal line is not detected in either nucleus, consistent with  the majority of BAT Sy 2 AGN  \citep{Lamperti:2017:540}.

The MUSE \oiii\ EW along with an \OIII/\Hbeta\ ratio maps are shown in  \autoref{fig:muse}.  The \oiii\ EW peaks on the northern nucleus, while the southern nucleus shows weaker \oiii\ EW. The \oiii/\hbeta\ map shows the highest ionization near the northern nucleus consistent with the strong \oiii\ emission seen, then dropping between the two nuclei, and increasing on the southern nucleus.


With the excellent pixel sampling and PSF of VLT/MUSE, we  extract three smaller spectral apertures  (0\farcs{15} diameter) with two centered on the NIR nuclei and an additional region midway between them (\autoref{fig:muse}) where the \oiii\ emission is weaker. This is done because the individual spaxels are too low S/N to perform emission line fitting of the weak \hbeta\ emission.    The extracted MUSE spectra are consistent with the HST/STIS results, with relatively brighter \oiii\ emission from the northern NIR nucleus and \nii\ emission that is brighter in the southern nucleus.  We also find that the middle component shows a half of the \oiii\ flux of the northern component consistent with the \oiii\ STIS map.  In \nii, the southern source is brighter.

In \autoref{fig:muse}, we present the MUSE-based strong line ratios of the key emission regions (the two nuclei and the region between them), namely the \OIII/\Hbeta\ vs.~\NII/\Halpha, \SII/\Halpha, and \OI/\Halpha\ line flux ratios.  Relying on commonly used diagnostics, specifically those revised by \citet{Kewley:2006:961}, we find that all three regions are classified as AGN-(or ``Seyfert"-)powered, rather than H\,{\sc ii}, or composites.  The MUSE data suggest that the northern nucleus is powered by a harder radiation field (i.e., the distance from the `composite' or `H\,{\sc ii}' lines), compared to the southern or middle regions consistent with the \oiii/\hbeta\ map.  Overall, the Balmer decrement for the N, S, and M apertures are consistent with each other with $15.1\pm2.1$, $15.0\pm2.1$, and $16.2\pm3.5$, for the N, S, and M regions, respectively.  A Balmer decrement of $\sim$15 is consistent with some of the highest levels of extinction seen in hard X-ray selected AGN from the BAT sample
\citep[e.g., $>99\%$;][]{Oh:2022:4}.

\subsection{Redshift and Kinematics of the Nuclei}
The emission line and stellar absorption velocities of the northern and southern nuclei are derived from MUSE using the \OIII\ emission and \Catrip\ triplet absorption region (CaT, 8350--8900\,\AA).  A map of the \oiii\ velocity is shown in \autoref{fig:muse}.  The \oiii\ velocity structure follows that found in HST/STIS, with the northern nucleus showing a $\sim$150\,\kmps offset compared to the southern nucleus and a gradual drop in velocity between the two.  Some higher velocities are seen in a region to the east of the northern nucleus.  

A map of the measured nuclear velocities of the stellar absorption lines is shown in \autoref{fig:musecatrip} along with the $^{12}$CO $J$=2--1 transition line velocities from ALMA.\footnote{Hereafter, we simply refer to $^{12}$CO $J$=2--1 as CO.}  Overall the absorption line maps match the emission line distribution, showing a velocity decrease toward the southern nucleus.  The CO velocity map also shows a similar decreasing velocity gradient between the two nuclei, together with a very clear velocity gradient around the position of the northern continuum emitter, where significant CO emission is concentrated.  

We also can compare the velocities from the three MUSE regions (N, S, and M) used for measuring emission lines.  A summary of these velocities is provided in \autoref{tab:musekinem}.  The \oiii\ and absorption line velocities match within error for each of the three apertures and also show a similar offset between the two nuclei, with an offset of 132$\pm$22\,\kmps based on \oiii, and 168$\pm$36\,\kms\ based on the stellar absorption lines in CaT.  The CO velocities also show a similar offset between the two nuclei, though the southern continuum emitter is too weak to measure the gas velocity at its immediate location ($<$0\farcs{15}).

The northern nucleus has somewhat higher CaT velocity dispersion than the southern nucleus (\sigs$=200\pm14$\,\kmps vs. $\sigs=165\pm17$\,\kms, respectively).  In Keck/OSIRIS, the 2.29\,\micron\ stellar velocity dispersion from the CO bandheads of the northern nucleus (\sigs$=204\pm$20\,\kms) are consistent with the CaT region  (the southern nucleus is dominated by the NIR AGN continuum).

\subsection{AGN Bolometric Luminosity, Black Hole Mass, and Eddington ratio}

We estimate the AGN bolometric luminosity, using the tight correlation ($\sim$0.5 dex scatter) between the nuclear peak mm-wave luminosity and the hard X-ray emission  \citep{2022ApJ...938...87K}, giving an absorption-corrected $2-10\,\kev$ luminosity of $\log(\Lsoftint/\ergs)=42.48\pm0.49$ for the northern nucleus and $\log(\Lsoftint/\ergs)=43.98\pm0.47$ for the southern nucleus.  Assuming a fixed correction factor of 20 to convert to bolometric luminosity this yields \lLbol=43.8 and \lLbol=45.3, for the northern and southern nuclei respectively.
The emission from both nuclei is within the range over which the millimeter-to-X-ray relation  was derived ($\log(\Lsoftint/\ergs)=41-44$). In a study of nearby dual AGN detected in the X-ray band \citep{Koss:2012:L22}, the average ratio of X-ray luminosities was $\approx$11, while some dual pairs have X-ray ratios greater than $\approx$1000 (e.g., IRAS\,05589+2828 and UGC\,8327), so the predicted ratio of 32 suggested by their mm emission is not extreme.

An X-ray spectral analysis using Chandra (from 2019) and NuSTAR (from 2017) data assuming a single nucleus suggested a moderate column density of $\log(\NH/\cmii)=22.95\pm0.05$ \citep{Zhao:2021:A57} with an intrinsic luminosity of $\log(\Lsoftint/\ergs)=43.3$,  which is lower than the sum of the ALMA-based estimates by about 0.7 dex.  The separation of the two NIR nuclei ($\sim$0\farcs{3}) is below the limiting resolution of Chandra ($\sim$0\farcs{5}) to possibly resolve the nuclei, and we find no evidence of two sources (see \autoref{chan_vla}). A previous analysis using Swift/XRT (from 2010) and Swift/BAT data \citep[average from 2005--2011,][]{Ricci:2017:17}, had $\log(\Lsoftint/\ergs)=42.7\pm0.2$, 0.6 dex below the later Chandra and NuSTAR data indicating significant X-ray variability.  The Swift/XRT observations also show a $\sim$5.3$\times$ increase in 2-10\,keV count rate between 2010 May ($0.0068\pm0.0014$\,ct\,s$^{-1}$) and 2017 March ($0.036\pm0.005$\,ct\,s$^{-1}$). We do not find evidence (e.g., $<$5\% chance) of variability within NuSTAR data within the observation (less than one day).  For Chandra, an $\chi^2$ test of the probability of constancy within the observation is only 1.8\%, suggesting possible variability. Therefore, it seems probable that the large 0.7 dex offset between the mm-predicted X-ray emission (from 2022) and the measured fluxes (from 2017 and 2019), may be related to source variability.      



For the southern nucleus, hydrogen Pa$\beta$ can be used for black hole mass estimation (Pa\,$\alpha$ is unobservable due to telluric absorption). We use the Pa$\beta$ \mbh\ relation from \citet{Brok:2022:7} yielding $\log(\mbh/\Msun)=8.1$ for the southern nucleus.

We can also roughly estimate the \mbh\ based on $\mbh-\sigs$ relation with \sigs\ from the two NIR nuclei, which is critical for the northern nucleus that has no broad lines for \mbh\ estimation. While velocity dispersions could in principle be affected by the complex stellar dynamics in the two progenitor galaxies during the merger, detailed simulations \citep[e.g.,][]{Stickley:2012:33} suggest that the \sigs\ values are much more likely to fall near the equilibrium value. However, dust attenuation associated with the merger may increase the scatter, though the northern nucleus has the benefit of a $K$-band CO bandhead measurement, which is less affected by dust, and largely agrees with the CaT value. Finally, in NGC\,6240, a similar close-separation dual AGN at $z = 0.0245$ \citep[$\sim$750\,pc;][]{2004AJ....127..239G}, \citet{Medling:2011:32} showed that the \mbh\ measured from directly resolving the sphere of influence of the black hole agreed within the scatter of the \sigs--\mbh\ relation.

Using our data with the \sigs\ relation derived by \citet{Kormendy:2013:511}.  We find $\log(\mbh/\Msun)=8.4$ and $8.1$ for the northern and southern nuclei (respectively). The latter is highly consistent with the broad Paschen line measurement mentioned above given the significant uncertainties beyond the typical intrinsic scatter of 0.3--0.5 dex \citep[e.g.,][and references therein]{Marsden:2020:61a}.   A rough estimate of the BH sphere of influence  $\mathrm{SOI}=33\,\mathrm{pc} \left(\frac{\sigs}{200\,\kmps}\right)^{2.38}$ \citep{Koss:2022:6} yields a r$_{\rm SOI}\sim$40\,pc given the black hole masses. Thus the projected separation is roughly 6$\times$ the black hole SOI (a lower limit given the line-of-sight distance is unknown).  

Using the ALMA continuum to estimate the bolometric luminosity of both sources, and with the black hole masses above from the velocity dispersion for the northern source and broad lines for the southern source, we find Eddington ratios of \lledd=0.002 and \lledd=0.1 for the northern and southern nucleus, respectively. 




\begin{figure}
\centering
\includegraphics[width=0.8\columnwidth]{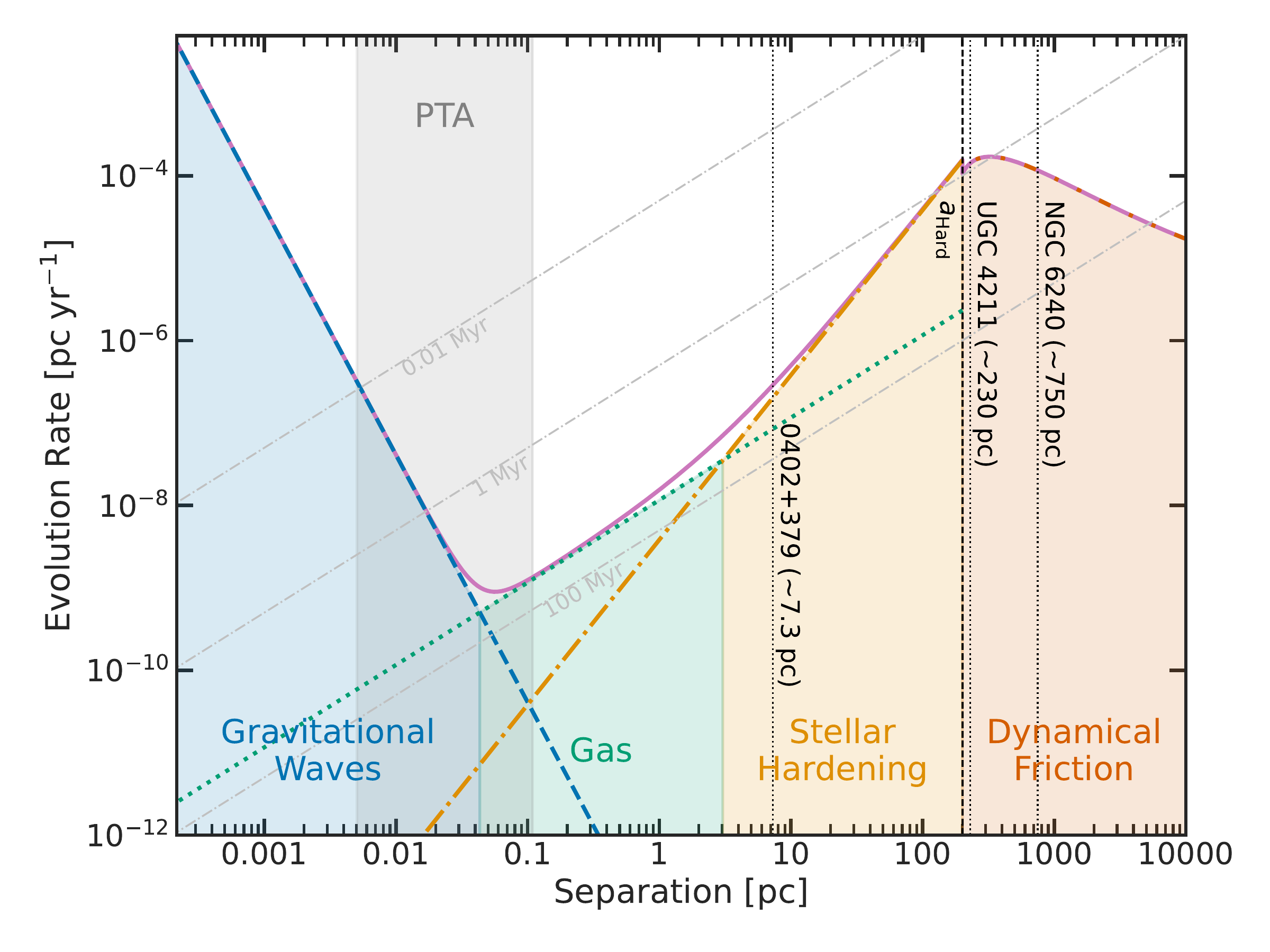}
\caption{Estimate of the future SMBH pair semi-major axis evolution for UGC\,4211 as a function of the semi-major axis. At its current separation (dotted line), UGC\,4211 is likely nearing the end of its dynamical friction phase (DF, orange region). Once it forms a hard binary at $a_{\rm Hard}$ (dashed line) its evolution will be driven by stellar hardening (SH, yellow region), then gas (green region), and finally gravitational waves (GW, blue region) until coalescence. They light gray region shows the semi-major axis scale where UGC\,4211 will emit $\sim$nHz GWs.}
    \label{fig:evol}
\end{figure}

\section{Implications for Gravitational Waves} \label{sec:gw}

Since dual AGN such as UGC\,4211 would be progenitors to SMBH binaries (SMBHBs), we can use the observed properties of this system to approximate key evolutionary timescales.
From largest to smallest separation, the orbital evolution of an SMBH pair is thought to be driven by (1) dynamical friction ($\sim 10~\rm{kpc} - 100~\rm{pc}$); (2) stellar hardening ($\sim 100 - 0.1~\rm{pc}$); and (3) GW emission ($\sim 0.1~\rm{pc} - \rm{coalescence}$; e.g.,  \citealt{izquierdo-villalba_massive_2022}, and references therein).  If sufficient gas is present there may also be a gas-driven phase near $\sim 1 - 0.01~\rm{pc}$. 

In \autoref{fig:evol}, we present estimates of binary hardening rates due to each of these mechanisms following the methodology outlined in \citet{Goulding:2019:L21}.  We have also included the separations of other ``bona fide" dual or binary AGN including NGC\,6240 \citep{2004AJ....127..239G} and 0402+379 \citep{2006ApJ...646...49R}.   These estimated hardening rates were obtained as follows, using the observed properties of UGC\,4211 and assuming $230~\rm{pc}$ separation. UGC\,4211 is nearing the end of its dynamical friction (DF) phase, which drives evolution until the binary hardens with semi-major axis $a_{\rm Hard} \ll G (M_{1} + M_{2}) / \sigs^{2}$ \citep{binney_galactic_2008}, where $M_{1}$ is the mass of the heavier SMBH, $M_{2}$ is the lighter SMBH mass, and $\sigs$ is the stellar velocity dispersion, which we estimate from the virial theorem.  We estimate UGC\,4211 will harden in $\lesssim 1~\rm{Myr}$. Once hard, the SMBH pair sheds energy primarily via stellar three-body interactions, i.e. stellar hardening (SH).  If there are too few stellar interactions at this stage the SMBH pair evolution can stall, taking longer than a Hubble time to reach the GW emission phase \citep{yu_ejection_2003}.  We estimate that, in the absence of efficient gas-driven inspiral, it will take UGC\,4211 $\sim 1~\rm{Gyr}$ to reach GW dominated evolution, but significantly shorter than the merger timescale for `stalled' binaries \citep[e.g.,][]{Kelley:2018:964}.  If enough gas is present it can reduce this time, reaching milliparsec scales in $\sim 200~\rm{Myr}$, by which point they will have formed a gravitationally bound SMBH binary (SMBHB) emitting nHz GWs in the PTA band. We note however, that some studies suggest circumbinary gas disks may not actually be that effective at driving BHs to efficiently merge \citep[e.g.,][]{2019ApJ...871...84M}.   For GW emission, a SBMHB merger with similar black hole masses as UGC\,4211, would be at the edge of LISA's sensitivity \citep[see Fig.~4 in ][]{2021CQGra..38e5009K}, but could be detected up to $z=1$.

\section{Conclusions}  
We find that our multiwavelength observations confirm the presence of two active nuclei in the center of UGC\,4211, separated by 230\,pc (projected) and $\sim$150\,\kms\ (along our line-of-sight) based on the following lines of evidence:
\begin{enumerate}
    \item The detection of NIR broad lines associated with the southern nucleus.
    \item The high-resolution MUSE AO (and HST/STIS) observations, show two separate emission sources of \OIII, each of them unresolved at $\sim$0\farcs{1} resolution. When placed on the BPT diagram, both nuclei are clearly in the Seyfert locus with the northern AGN further in the ionized AGN region.

    \item The copious unresolved nuclear mm emission at the location of the two nuclei, coincident with the unresolved emission-line emitters.  While this could be due to star formation, the fact that they are spatially extremely compact, ~<30 pc, and (at least the south nucleus) consistent with a flat, nonthermal, spectrum, strongly suggests that these signals arise from mm wave emission, as seen in X-ray selected AGN \citep[see discussion in,][]{2022ApJ...938...87K}.  Furthermore, the mm continuum luminosity is consistent with the expected value for bright nearby X-ray selected AGN emission following the \citet{2022ApJ...938...87K} correlation, and is significantly higher than the values expected for star formation processes in such a small size. Furthermore, a non-AGN origin would require an extremely compact and dense star forming region, which at the same time shows a strong AGN-ionized BPT emission line ratios, and is in the center of a NIR nucleus tracing old stellar populations. 
    \item Using the CO(2-1) emission line as a tracer of the molecular gas, the velocity map shows a clear velocity gradient centered on the positions of the northern and southern continuum emitters.  A similar gradient is also seen in the \Catrip\ stellar absorption lines. This demonstrates that both sources are kinematically independent nuclei, while rules out other possibilities, such as a jet knots.
\end{enumerate}


The single AGN scenario, where the narrow emission-line region associated with the northern NIR nucleus is photoionized by the broad-line AGN in the southern nucleus, seems highly unlikely, given the \oiii\ emission peaks on the center of the northern nucleus where the ALMA continuum source also confirms the AGN nature.  The southern nucleus is also highly obscured in the optical-UV, so it is hard to understand how the northern nucleus would be strongly photoionized.   While shocks may sometimes move H\,{\sc II} regions on the BPT diagram it is only into the composite or LINER region on the \NII\ BPT diagram \citep[e.g.,][]{Rich:2011:87}.  Although there are some cases in which a non-AGN can be found in this region, these are mostly extended sources \citep[e.g.,][]{Keel:2012:878, Treister:2018:83,2022ApJ...936...88F}, and not distinct nuclear emitters as it is the case here, which lie directly in the center of two extended NIR regions tracing old stellar populations.

Our analysis and findings clearly demonstrate the benefits of multiwavelength, high-spatial resolution observations ($<$0\farcs{1}).  The ability of ALMA to identify other subkpc dual AGN candidates is also promising, given the large number of high-resolution archival observations of BAT AGN \citep[e.g., $N>100$;][]{2022ApJ...938...87K}, though requiring further investigation.

While the exact occurrence rate of close-separation dual AGN (i.e., $<$300\,pc, like UGC\,4211) is not yet known, it may be surprisingly high, given that UGC\,4211 was found within a small, volume-limited sample of nearby hard X-ray detected AGN \citep[e.g., $z<0.075$;][]{Koss:2018:214a}.  This AGN was specifically found among a population of only 34 luminous obscured AGN (i.e., lacking broad $\hbeta$ and \lbol$>10^{44}$\,\ergps) observed in the NIR at high-spatial resolution (e.g. $<$200\,pc) within this survey and there are five other candidates at $<$3 kpc separation among this sample of 34 (one being NGC 6420, which is a known dual AGN).  While luminous AGN like UGC\,4211 are rare in the nearby universe and require large, multiwavelength observational efforts to confirm their nature, the luminosities are typical of most AGN found in higher redshift surveys \citep[e.g., see Fig. 1 in,][]{Koss:2022:1}.  Among the more numerous nearby low-luminosity AGN, such as optical BPT selected AGN, the frequency is likely significantly lower, consistent with what has been found with dual AGN at larger separations \citep[$>$5\,kpc;][]{Koss:2012:L22} and therefore lower luminosity analogs of UGC\,4211 may be even more difficult to find.

Our observations and analysis of UGC\,4211, combined with an extrapolation of our current knowledge of binary evolution suggest that close SMBHBs in the very nearby universe could be observed through their electromagnetic emission as dual AGN, and detected with future GW facilities such as PTAs and LISA as discrete GW sources. These results also inform simulations of likely SMBHBs hosts. For example, the host morphologies of parsec-scale SMBHBs are thought to be dominated by inactive galaxies, unlike UGC\,4211, with only 0.5--5\%, showing a bolometric luminosity of \lLbol$>43$ based on simulations \citep{2022arXiv220704064I}. In the nearby universe, among massive galaxies, likely SMBHBs hosts are thought to be massive ellipticals, in contrast to the less massive system studied here, which is relatively gas rich.  This underscores the importance of more observations and confirmations of a near-coalescence dual system to complement upcoming GW observations with PTAs and prepare for future GW observatories, such as LISA.




\begin{acknowledgements}
We thank the anonymous referee for very useful comments and suggestions to improve the manuscript. The authors would like to express their appreciation for useful software discussions with Roland Bacon and Yao Yao as well as the observation planning with ESO astronomer Michael Hilker who supported multiple attempts to observe this object.  We acknowledge support from: NASA through ADAP award NNH16CT03C (M.K.),
ANID through Millennium Science Initiative Program-NCN19\_058 (E.T.), and ICN12\_009 (F.E.B), CATA-BASAL - ACE210002 (E.T., F.E.B.) and FB210003 (E.T., F.E.B., C.R.), FONDECYT Regular - 1190818 (E.T., F.E.B.) and 1200495 (F.E.B., E.T.), and Fondecyt Iniciacion 11190831 (C.R.), NSF grants PHY-2020265 (T.L., C.M.F.M.) and AST-2106552 (C.M.F.M.), NSF award AST-1909933 and Cottrell Scholar Award 27553 (L.B.), ADAP award 80NSSC19K1096 (F.M-S.), the European Union's Horizon 2020 research and innovation program (950533) and Israel Science Foundation grant 1849/19 (B.T.), Project No. 2022-1-830-06 from the Korea Astronomy and Space Science Institute (K.O.) and the National Research Foundation of Korea (NRF-2020R1C1C1005462), JSPS KAKENHI grant JP20K14529  and the Special Postdoctoral Researchers Program at RIKEN (T.K.),  and the hospitality of NRAO/NAASC during his sabbatical leave (E.T.). The Flatiron Institute is supported by the Simons Foundation.
\end{acknowledgements}

\appendix

\section{New Observations and Data Reduction} \label{new_obs}
\subsection{HST Spectroscopy} 
\label{sec:hst}

We observed UGC\,4211 using the STIS with a 52\arcsec$\times$0\farcs{2} slit (PI: Koss).   We used the G430L and G750M gratings, which cover 2870-5680\,\AA\ and 6480-7045\,\AA, respectively.  The slit was oriented 5.9\deg\ east of north, along the axis separating the two nuclei (as determined from the NIR image; see \autoref{fig:imaging}).  

The calibrated 2D STIS spectra were directly downloaded from the HST archive.  We used the {\tt stistools} software (version 1.3) to remove cosmic rays and combine images. We replaced hot pixels above 4$\sigma$ with median values. We then manually extracted 1D spectra from 5 pixel (0\farcs{25}) width regions centered on the northern and southern nuclei, which are present in the 2D spectra and separated by 6 pixels (0\farcs{3}), using the {\tt x1d} task (see \autoref{fig:stis}).

\subsection{Optical IFS Spectroscopy}
AO-assisted integral field spectroscopic (IFS) observations were performed using MUSE \citep[][]{2010SPIE.7735E..08B} instrument in NFM at the Very VLT as part of programs studying subkpc mergers (PI: Treister).  The NFM covers a field of 7\farcs{5} by 7\farcs{5} on the sky, sampled by 0\farcs{025} spaxels.  Calibration and data reduction were done using the ESO VLT/MUSE pipeline in the ESO {\tt Reflex} environment \citep{Freudling:2013:A96}. Sky-subtracted individual 600s frames were coadded and manually aligned using the \OIII\ emission before stacking using {\tt QFitsView}.  Assuming that the northern nucleus is spatially unresolved, we measure a FWHM of 0\farcs{09} at \oiii. While there are no sources that we can assume unresolved spatially, according to the MUSE user manual and commissioning data\footnote{\url{https://www.eso.org/sci/facilities/paranal/instruments/muse/doc/ESO-261650_MUSE_User_Manual.pdf}} we expect the resolution at $\sim$9000\AA~to be $\sim$0\farcs{05}.

We use the {\tt MUSE python data analysis framework} \citep[{\tt MPDAF}][]{2016ascl.soft11003B} to extract cubes and perform a 3 pixel median filter due to improve S/N.  To measure the equivalent width, line emission, and BPT ratio, we performed a first-order polynomial fit to nearby line-free regions to measure and subtract the continuum on a spaxel-by-spaxel basis.

%


%

\subsection{NIR IFU Spectroscopy}
UGC\,4211 was observed using OSIRIS.  We used the 0\farcs{05} scale with an field of view (FOV) of 0\farcs{8} ×3\farcs{2} in the classical object-sky-object dithering pattern, with an exposure of 600s and a sky offset of 20\arcsec\ in each of the $Jbb$, $Hbb$, and $Kbb$ filters.  Due to the rectangular nature of the FOV, the galaxy was observed in the N--S direction (i.e., PA=0\deg), approximately corresponding to the alignment of the two nuclei.  The data were reduced using the OSIRIS data reduction pipeline (version 4.2) to preform dark-frame subtraction, crosstalk removal, sky subtraction, rectification, data cube assembly, and the wavelength solution manually refined based on OH lines.  The spectra were telluric corrected and flux calibrated using the software \xtellcor\ \citep{Cushing:2004:362} using the A0V star HD 65158. Based on fitting the BLR in the southern source with a Gaussian, the PSF FWHM is 0\farcs{2} and 0\farcs{1}, in $Jbb$ and $Kbb$, respectively.   

\subsection{ALMA}
UGC\,4211 was observed by ALMA on 2021 October 24 and 2022 August 19 in band 6 ($\approx$230\,GHz; program ID 2021.1.01019.S, PI: Treister), aimed to study the molecular gas contents of nearby major galaxy mergers in the BAT AGN sample. The observations were made combining C-5 with C-8 configurations at two different epochs, reaching baselines up to $\sim$8\,km, yielding a minimum beam size of 0\farcs{06}, for a total of 31 minutes on source in C-8 with 46 12\,m antennae, and for 7 minutes on target with 44 12 m antennae in the C-5 configuration with baselines ranging from 15\,m to 1.3\,kms. Data reduction was carried out using the ALMA pipeline v2021.2.0.128 based on Common Astronomy Software Applications package \citep[CASA, v.6.2.1.7;][]{McMullin:2007:127}. As it is usually done, four spectral windows were defined, two covering the $^{12}$CO(2-1) emission line at an observed frequency of 222.78\,GHz with a velocity range of $\pm$1000\,\kms, and two in the surrounding continuum. The continuum map analyzed here was computed coadding emission in line-free regions in four ALMA spectral windows ranging from 221 to 240\,GHz, while a cube covering the $^{12}$CO(2-1) emission line was generated with a spectral width of 15\,\kms, and a Briggs robust parameter of 0.5, resulting in a beam size of 0\farcs{061}$\times$0\farcs{073}. Additional details about the CO map will be provided in a companion paper (Treister et al., in preparation).

\subsection{Emission and absorption line fitting}
We use {\tt PySpecKit} to fit the optical emission lines from HST/STIS, the MUSE \OIII\ map, and the NIR emission lines from Keck/OSIRIS, which uses a Levenberg--Marquardt algorithm for spectral fitting \citep[version 1.0.2;][]{Ginsburg:2011:1109.001} with Gaussians following the approach of \citet{Koss:2017:74}.  For the MUSE data from the three spatial regions corresponding to the two nuclei and a region between them (N, S, and M), we include galaxy template and emission line fitting to appropriately measure the \hbeta\ emission lines following the approach of \citet{Oh:2022:4}.  

For velocity dispersion measurements, we follow  \citep{Koss:2017:74,Koss:2022:6}, using the penalized PiXel Fitting (\ppxf) software \citep[version 7.4.3;][]{Cappellari:2017:798} to measure stellar kinematics and the central stellar velocity dispersion (\sigs). We used the X-Shooter Spectral Library (XSL, specifically DR2; \citealt{Gonneau:2020:A133}).  

Unbiased measurements of stellar kinematics require a minimum S/N, so for an adaptive spatial-binning scheme we use the Voronoi algorithm as implemented in \vorbin\ \citep[version 3.1.5;][]{2003MNRAS.342..345C}, requiring a S/N of 35 for each bin.  For measurements of the \OIII\ emission line velocity, we also use voronoi binning before fitting, requiring a S/N of 10.


\subsection{Relative Astrometric Alignment}
Due to the small separation of the two nuclei, we have aligned the different wavelength images.  The broad emission line region traced in the NIR in the Keck/OSIRIS pseudoimage is consistent with the center of the southern Keck/NIRC2 nucleus.  Using Gaia DR1 data the Keck/NIRC2 AO and HST/F814W images have numerous stars that have been aligned with an expected astrometric error of 0\farcs{1}.  To align the MUSE data to the larger HST F814W data, we use {\tt MPDAF}, to generate an image at the same spectral region and weighting as the F814W and align the MUSE cube using the northern nucleus.  

\section{Additional Data} \label{chan_vla}
Here we discuss the data and analysis of additional UV, optical, and NIR imaging as well as X-ray, and radio data for this system. This includes Chandra, NuSTAR,  and 22\,Ghz VLA data, which all have insufficient spatial resolution to resolve the two nuclei, and were thus not discussed in detail in the main text. 

\subsection{High Resolution Imaging with HST and Keck} \label{uvdata}

UGC 4211 was imaged in the optical and the UV regimes with HST.  All the reduced, drizzled HST imaging data were obtained from the Barbara A. Mikulski Archive for Space Telescopes (MAST) at the Space Telescope Science Institute. The specific HST observations analyzed can be accessed via \dataset[10.17909/cjjm-wr56]{https://doi.org/10.17909/cjjm-wr56}.  UGC\,4211 was imaged with the Wide-Field Camera for Surveys 3 (WFC3) F225W UV filter. The galaxy was also observed with the Advanced Camera for Surveys (ACS) Wide-Field Camera (WFC), on board the HST, in the F814W band as part of a snapshot gap filler program studying 70-month Swift BAT AGN from the BASS DR1 Survey \citep{Koss:2017:74}, described in \cite{Kim:2021:40}.   Guide stars with Gaia DR1 positions were used, yielding a typical image alignment error of 0\farcs{1}.   In the F225W UV HST imaging, no host galaxy emission is detected, which is consistent with considerable dust attenuation in the galaxy center and across the galaxy.
 
   The Near Infrared Camera 2 (NIRC2)+AO imaging in the $J$ and $K^\prime$ bands (0\farcs{04} pix$^{-1}$) of nearby ($z<0.075$) hard X-ray selected AGN from the Swift BAT is described in \citet{Koss:2018:214a}. For both the optical and NIR imaging, guide stars with Gaia DR1 positions were used, yielding a typical image alignment error of 0\farcs{1}. 


\subsection{X-ray Data} \label{sec:xray_obs}
Chandra observed UGC\,4211 on axis in a cool attitude program study BAT AGN (PI: Koss).  Standard reductions with {\tt CIAO, v4.14} using the {\tt chandra$\_$repro} task were done. Subpixel event repositioning (at 0\farcs{25} pix$^{-1}$) was applied to improve the resolution of the image beyond the native 0\farcs{5} pix$^{-1}$ sampling.    The reported astrometric accuracy of Chandra is only $\sim$0\farcs{71} at the 95\% level based on the Chandra Source Catalog.  We use the Chandra X-ray spectra of the source to estimate the PSF using Chandra Ray Tracer (ChaRT) v2.  We fit the X-ray emission using the PSF model and a 2D Gaussian to estimate the centroid position and possible extended emission.  To create a lightcurve, we binned to 500s bins.

When fitting the Chandra data with the PSF and a 2D Gaussian, we find evidence of a extended emission (FWHM=$1.82^{+0.93}_{-0.38}$\,\arcsec).  However, there is no constraints on the ellipticity or position angle. We then used {\tt BAYMAX} \citep[Bayesian AnalYsis of Multiple AGN in X-rays][]{2019ApJ...877...17F} to search for the presence of two X-ray point sources. BAYMAX calculates the likelihood (Bayes factor, calculated in natural log space, hereafter $\log{\mathcal{BF_{2/1}}}$) that Chandra observations are composed of two versus one point source.  We analyze the 0.5-8 keV counts within a 20\arcsec x 20\arcsec (40 x 40 sky pixel) box centered on UGC\,4211. We run BAYMAX on the data within this field of view twice, using different prior distributions on the locations of the primary and secondary X-ray point source. We first allow $\mu_{\mathrm{primary}}$ and $\mu_{\mathrm{secondary}}$ to be anywhere within this field of view (represented by $x$, $y$, priors that are uniform distributions across the full extent of the sky $x$, $y$ range). We find that the data do not strongly support the dual point-source hypothesis, with $\log{\mathcal{BF_{2/1}}}=-0.8 \pm 1.5$. We then rerun our analysis constraining the $\mu_{\mathrm{primary}}$ and $\mu_{\mathrm{secondary}}$ via $x$,$y$ priors that are uniform and constrained to a 1 x 1\arcsec (2 x 2 sky pixel) box centered on the observed locations of each resolved IR stellar core. Although the resultant $\log{\mathcal{BF_{2/1}}}$ is marginally higher, the data still do not significantly support the dual point-source hypothesis at the 95\% confidence interval, with $\log{\mathcal{BF_{2/1}}}=1.5 \pm 1.9$.

On average, for separations below 0\farcs{35}, {\tt{BAYMAX}} will not necessarily be sensitive to detecting dual AGN. Analyzing a suite of dual AGN simulations across a range of separations and count ratios with {\tt{BAYMAX}}, it was previously found that with $>$700 0.5--8\,keV counts (UGC\,4211 has 1163 cts) {\tt{BAYMAX}} is, on average, sensitive to correctly identifying dual AGN at separations 0\farcs{30}$<r<$0\farcs{35} with count ratios $f\ge0.8$ (assuming similar X-ray spectral shapes for the primary and secondary AGN; \citealt{2019ApJ...877...17F}). Thus, in the probable case of the dual AGN in UGC\,4211 having a count ratio less than 0.8, or a secondary AGN that has high levels of nuclear obscuration, we do not expect to find strong evidence for a dual X-ray point source. Future, deeper X-ray observations may aid in pushing our sensitivity to lower count ratios.

NuSTAR data was also used to look for source variability.  The data was processed to extract spectra and light curves using \texttt{NUPRODUCTS} from the \texttt{NuSTARDAS} (version 1.9.7) software package and CALDB (version 20220608). For spectral extraction, we used circular regions 50\arcsec\ in radius centered on the point-source peak.   A background spectrum was extracted from a polygonal region surrounding both sources on the same chip.  The light curves were background corrected using \texttt{lcmath} and were split into the 3-10 keV and 10-40 keV range. We then combined background-corrected light curves from Focal Plane module A and Focal Plane Module B using \texttt{lcmath}.  

To search for the hard X-ray variability within observations in Chandra and NuSTAR, we use \texttt{lcstats} task from the \texttt{XRONOS} (version 6.0) package to analyze the light curves.

\subsection{JVLA} \label{sec:vla_obs}
We observed UGC\,4211 with the JVLA in the $K$ band with C-array configuration giving 1\arcsec\ spatial resolution as part of our 22~GHz radio survey of the BAT AGN \citep{Smith:2020:4216}. This was observed on  2018 December 4 (PI: Smith). The total on-source integration time was 9 minutes and 10 s, yielding a 1$\sigma$ sensitivity of 22\,$\mu$Jy per beam. The source’s peak brightness of 2.259\,mJy beam$^{-1}$ enables us to perform self-calibration using CASA’s \texttt{gaincal} command, which calibrates the antenna-based gains as a function of time. We impose a solution interval of 180 s, and apply the these temporal gains to the data set using \texttt{applycal}. 

\begin{figure*} 
\centering
\includegraphics[width=12cm]{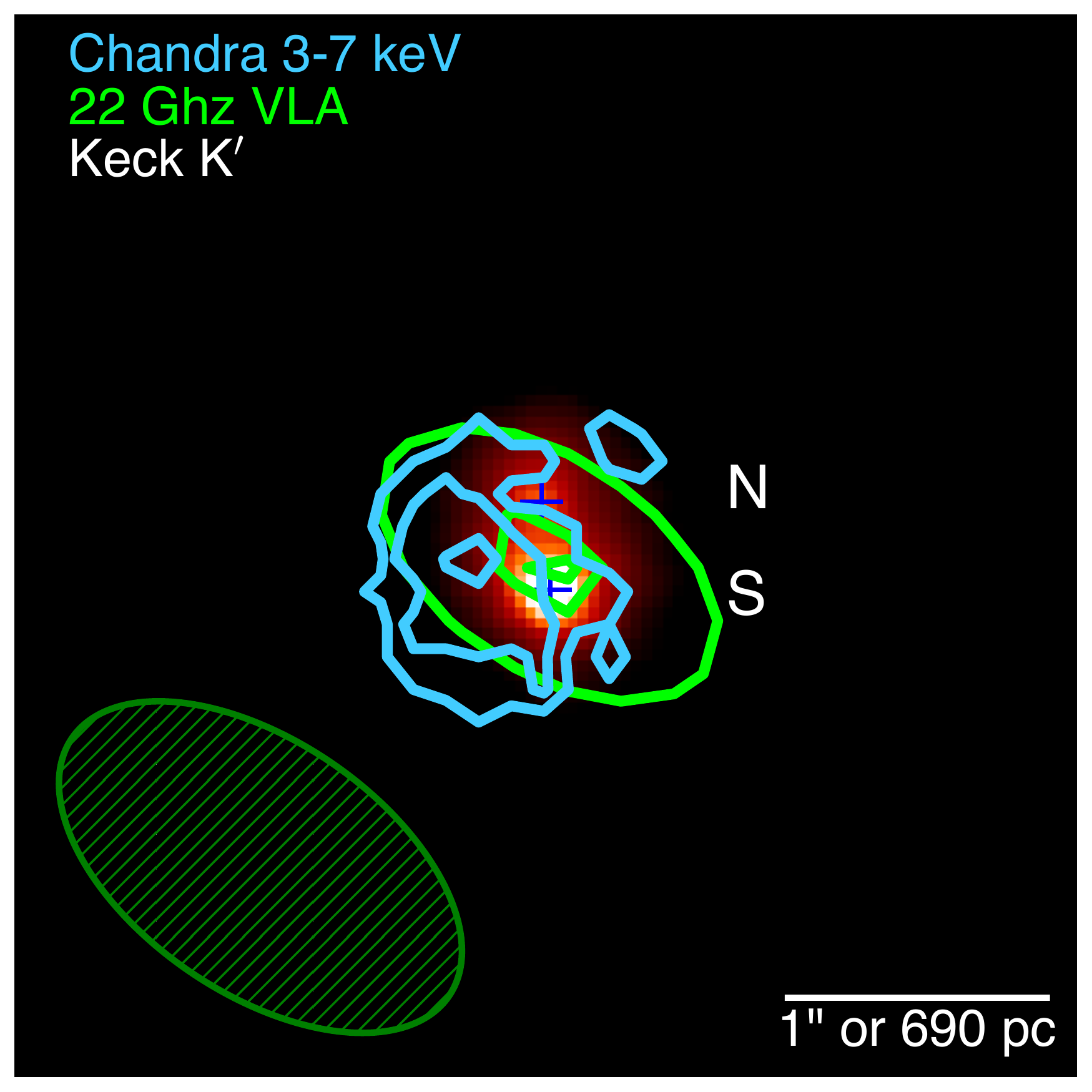}
\caption{NIRC2 AO $K^\prime$ 4\arcsec\ image of UGC\,4211 in log stretch overlaid with Chandra 3-7 keV (light blue) emission and 22\,Ghz VLA emission (green).  In Chandra, contours represent 5, 10, and 30 counts, respectively.  The green hatched circle indicates the VLA beam size.  The positions of the two NIR nuclei are shown with blue crosses.  For both Chandra (FWHM$\sim$0\farcs{5}) and the VLA ($\sim$1\arcsec) the resolution is insufficient to resolve the two sources.  The offset of Chandra is likely due to astrometric errors in the alignment of the images ($\sim$0\farcs{3}), even after using Gaia as a reference.     }
\label{fig:chanvla}
\end{figure*}

Using \texttt{CASA} task \texttt{imfit} the centroid position of the single detection is R.A.=08:04:46.391, decl.=+10:46:36.01 with a flux of 3.74$\pm$0.017\,mJy within 6\arcsec\ and 3.2741$\pm$0.0072\,mJy.  The absolute astrometric accuracy for the VLA $<$0\farcs{01}  The VLA detection is 0\farcs{07} at position angle of 12.6\deg$\pm$4 from the brighter southern ALMA source (\autoref{fig:chanvla}).  This places it along the line between the the northern and southern ALMA sources, but closer to the brighter southern mm source detected with ALMA.  It is however, unclear whether this middle position is due solely to the unresolved 22\,Ghz emission from the two nuclei, or from some other emission. 

\section{Emission and Absorption Line Redshifts} \label{sec:musevel}
A summary of emission and absorption line redshifts from MUSE is provided in \autoref{tab:musekinem}.

\begin{deluxetable*}{l c c}
\tabletypesize{\large}
\tablecaption{Summary of MUSE Line Redshifts \label{tab:musekinem}}
\tablehead{
\colhead{Aperture}&\colhead{\OIII}& \colhead{\Catrip}\\
\colhead{}&\colhead{(\kms)}& \colhead{(\kms)}}
\startdata
MUSE north& 96$\pm$14 &105$\pm$22\\
MUSE middle& 26$\pm$19 &24$\pm$26\\
MUSE south& -36$\pm$17  &-63$\pm$28\\
\enddata
\tablecomments{Summary of of emission and absorption line redshifts for the three 0\farcs{15} diameter circular regions (from north, south, and middle)  extracted from MUSE corresponding to the two nuclei and a region between them. Velocity offsets  are from the central redshift ($z=0.03474$).}
\end{deluxetable*}

\facilities{ALMA, CXO, HST (STIS, ACS, WFC3), Keck:I (OSIRIS), Keck:II (NIRC2), NuSTAR, Swift (BAT and XRT), VLA,  VLT:Yepun}

\section{OSIRIS Spectra} \label{sec:osiris_regions}
\begin{figure*} 
\centering
\includegraphics[width=14cm]{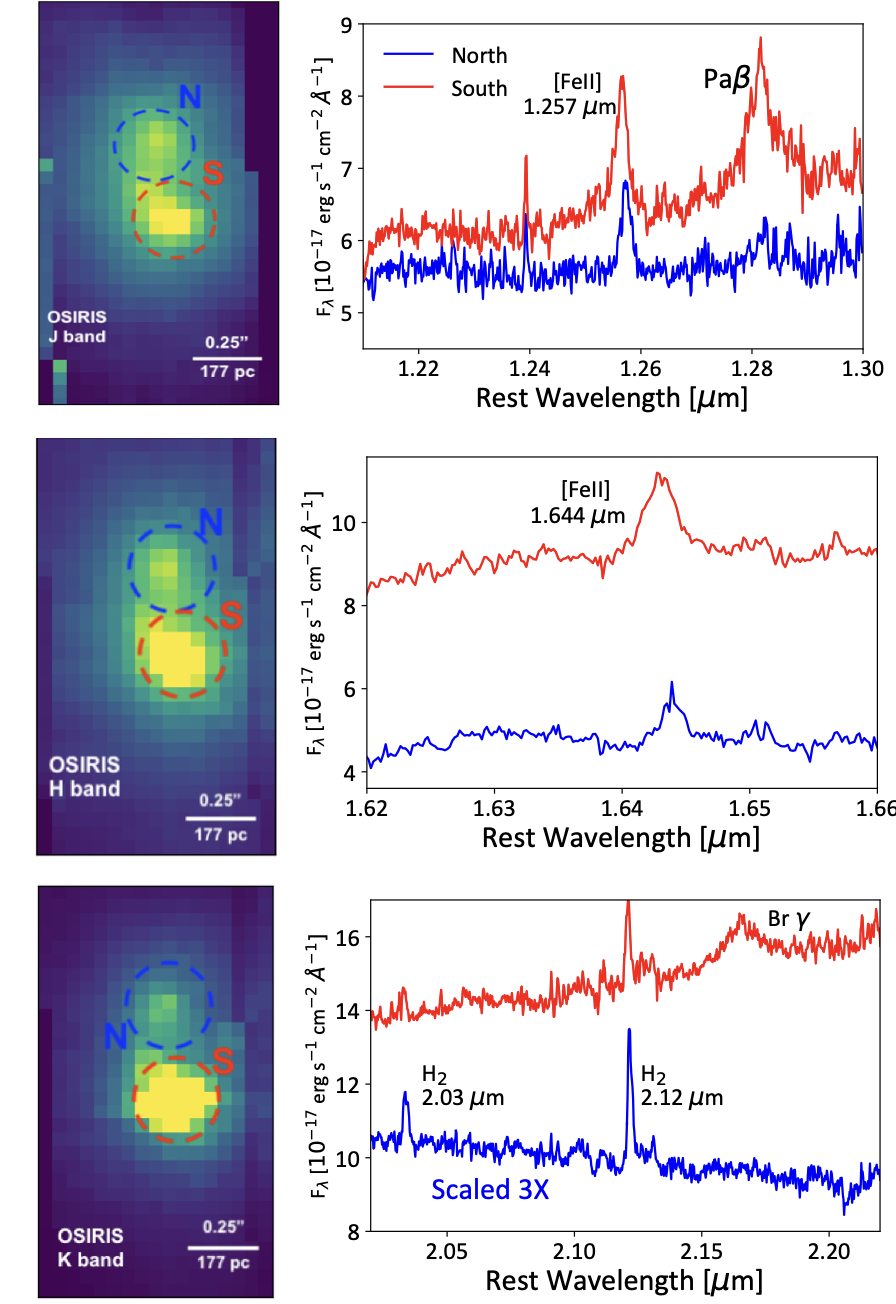}
\caption{Left Column:  Keck/OSIRIS image in the $J$, $H$, and $K$ band. Dashed circles mark the circular apertures used for 1D spectral extraction apertures (0\farcs{3} diameter) for the northern nucleus (blue) and southern nucleus (red).   Right column: spectra of the two regions (north and south), showing the bright emission lines in the region ([Fe {\sc ii}], warm H$_2$, Paschen $\beta$, and Bracket $\gamma$).  For the $K$-band spectra, we have scaled the northern source emission by a factor of 3 to enable easier comparison with the southern nucleus.  }
\label{fig:osiris}
\end{figure*}

For clarity, the extraction regions, and associated spectra are shown in \autoref{fig:osiris}.

\software{APLpy \citep{Robitaille:2012:1208.017},
    astropy \citep{Collaboration:2013:A33},  
          Matplotlib \citep{Hunter:2007:90}, 
          jupyter notebook,
          Numpy \citep{vanderWalt:2011:22},
          MPDAF \citep{2016ascl.soft11003B},
          pPXF \citep{Cappellari:2017:798},
          PySpecKit \citep{Ginsburg:2011:1109.001},
          vorbin \citep{2003MNRAS.342..345C}
          }

\bibliography{ms.bbl}{}

\begin{thebibliography}{}
\expandafter\ifx\csname natexlab\endcsname\relax\def\natexlab#1{#1}\fi
\providecommand{\url}[1]{\href{#1}{#1}}
\providecommand{\dodoi}[1]{doi:~\href{http://doi.org/#1}{\nolinkurl{#1}}}
\providecommand{\doeprint}[1]{\href{http://ascl.net/#1}{\nolinkurl{http://ascl.net/#1}}}
\providecommand{\doarXiv}[1]{\href{https://arxiv.org/abs/#1}{\nolinkurl{https://arxiv.org/abs/#1}}}

\bibitem[{{Bacon} {et~al.}(2016){Bacon}, {Piqueras}, {Conseil}, {Richard}, \&
  {Shepherd}}]{2016ascl.soft11003B}
{Bacon}, R., {Piqueras}, L., {Conseil}, S., {Richard}, J., \& {Shepherd}, M.
  2016, {MPDAF: MUSE Python Data Analysis Framework}, Astrophysics Source Code
  Library, record ascl:1611.003.
\newblock \doeprint{1611.003}

\bibitem[{{Bacon} {et~al.}(2010){Bacon}, {Accardo}, {Adjali}, {Anwand},
  {Bauer}, {Biswas}, {Blaizot}, {Boudon}, {Brau-Nogue}, {Brinchmann},
  {Caillier}, {Capoani}, {Carollo}, {Contini}, {Couderc}, {Daguis{\'e}},
  {Deiries}, {Delabre}, {Dreizler}, {Dubois}, {Dupieux}, {Dupuy}, {Emsellem},
  \& {Fechner}}]{2010SPIE.7735E..08B}
{Bacon}, R., {Accardo}, M., {Adjali}, L., {et~al.} 2010, in Society of
  Photo-Optical Instrumentation Engineers (SPIE) Conference Series, Vol. 7735,
  Ground-based and Airborne Instrumentation for Astronomy III, ed. I.~S.
  {McLean}, S.~K. {Ramsay}, \& H.~{Takami}, 773508, \dodoi{10.1117/12.856027}

\bibitem[{Barrows {et~al.}(2017)Barrows, Comerford, Greene, \&
  Pooley}]{Barrows:2017:129}
Barrows, R.~S., Comerford, J.~M., Greene, J.~E., \& Pooley, D. 2017, \apj, 838,
  129, \dodoi{10.3847/1538-4357/aa64d9}

\bibitem[{Baumgartner {et~al.}(2013)Baumgartner, Tueller, Markwardt, Skinner,
  Barthelmy, Mushotzky, Evans, \& Gehrels}]{Baumgartner:2013:19}
Baumgartner, W.~H., Tueller, J., Markwardt, C.~B., {et~al.} 2013, \apjs, 207,
  19, \dodoi{10.1088/0067-0049/207/2/19}

\bibitem[{{Binney} \& {Tremaine}(2008)}]{binney_galactic_2008}
{Binney}, J., \& {Tremaine}, S. 2008, {Galactic Dynamics: Second Edition}

\bibitem[{Blecha {et~al.}(2018)Blecha, Snyder, Satyapal, \&
  Ellison}]{Blecha:2018:3056}
Blecha, L., Snyder, G.~F., Satyapal, S., \& Ellison, S.~L. 2018, \mnras, 478,
  3056, \dodoi{10.1093/mnras/sty1274}

\bibitem[{{Blumenthal} \& {Barnes}(2018)}]{2018MNRAS.479.3952B}
{Blumenthal}, K.~A., \& {Barnes}, J.~E. 2018, \mnras, 479, 3952,
  \dodoi{10.1093/mnras/sty1605}

\bibitem[{Bonetti {et~al.}(2018)Bonetti, Sesana, Barausse, \&
  Haardt}]{Bonetti:2018:2599}
Bonetti, M., Sesana, A., Barausse, E., \& Haardt, F. 2018, \mnras, 477, 2599,
  \dodoi{10.1093/mnras/sty874}

\bibitem[{Brok {et~al.}(2022)Brok, Koss, Trakhtenbrot, Stern, Cantalupo,
  Lamperti, Ricci, Ricci, Oh, Bauer, Riffel, Rodríguez-Ardila, Bär, Harrison,
  Ichikawa, Mejía-Restrepo, Mushotzky, Powell, Boissay-Malaquin, Stalevski,
  Treister, Urry, \& Veilleux}]{Brok:2022:7}
Brok, J. S.~d., Koss, M.~J., Trakhtenbrot, B., {et~al.} 2022, \apjs, 261, 7,
  \dodoi{10.3847/1538-4365/ac5b66}

\bibitem[{{Buchner} {et~al.}(2019){Buchner}, {Treister}, {Bauer}, {Sartori}, \&
  {Schawinski}}]{Buchner19}
{Buchner}, J., {Treister}, E., {Bauer}, F.~E., {Sartori}, L.~F., \&
  {Schawinski}, K. 2019, \apj, 874, 117, \dodoi{10.3847/1538-4357/aafd32}

\bibitem[{Burke-Spolaor(2011)}]{Burke-Spolaor:2011:2113}
Burke-Spolaor, S. 2011, \mnras, 410, 2113,
  \dodoi{10.1111/j.1365-2966.2010.17586.x}

\bibitem[{Cappellari(2017)}]{Cappellari:2017:798}
Cappellari, M. 2017, \mnras, 466, 798, \dodoi{10.1093/mnras/stw3020}

\bibitem[{{Cappellari} \& {Copin}(2003)}]{2003MNRAS.342..345C}
{Cappellari}, M., \& {Copin}, Y. 2003, \mnras, 342, 345,
  \dodoi{10.1046/j.1365-8711.2003.06541.x}

\bibitem[{Collaboration {et~al.}(2013)Collaboration, Robitaille, Tollerud,
  Greenfield, Droettboom, Bray, Aldcroft, Davis, Ginsburg, Price-Whelan,
  Kerzendorf, Conley, Crighton, Barbary, Muna, Ferguson, Grollier, Parikh,
  Nair, Unther, Deil, Woillez, Conseil, Kramer, Turner, Singer, Fox, Weaver,
  Zabalza, Edwards, Azalee~Bostroem, Burke, Casey, Crawford, Dencheva, Ely,
  Jenness, Labrie, Lim, Pierfederici, Pontzen, Ptak, Refsdal, Servillat, \&
  Streicher}]{Collaboration:2013:A33}
Collaboration, A., Robitaille, T.~P., Tollerud, E.~J., {et~al.} 2013, \aap,
  558, A33, \dodoi{10.1051/0004-6361/201322068}

\bibitem[{Cushing {et~al.}(2004)Cushing, Vacca, \& Rayner}]{Cushing:2004:362}
Cushing, M.~C., Vacca, W.~D., \& Rayner, J.~T. 2004, PASP, 116, 362,
  \dodoi{10.1086/382907}

\bibitem[{Fabbiano {et~al.}(2011)Fabbiano, Wang, Elvis, \&
  Risaliti}]{Fabbiano:2011:431}
Fabbiano, G., Wang, J., Elvis, M., \& Risaliti, G. 2011, \nat, 477, 431,
  \dodoi{10.1038/nature10364}

\bibitem[{{Finlez} {et~al.}(2022){Finlez}, {Treister}, {Bauer}, {Keel}, {Koss},
  {Nagar}, {Sartori}, {Maksym}, {Venturi}, {Tub{\'\i}n}, \&
  {Harvey}}]{2022ApJ...936...88F}
{Finlez}, C., {Treister}, E., {Bauer}, F., {et~al.} 2022, \apj, 936, 88,
  \dodoi{10.3847/1538-4357/ac854e}

\bibitem[{{Foord} {et~al.}(2019){Foord}, {G{\"u}ltekin}, {Reynolds},
  {Hodges-Kluck}, {Cackett}, {Comerford}, {King}, {Miller}, \&
  {Runnoe}}]{2019ApJ...877...17F}
{Foord}, A., {G{\"u}ltekin}, K., {Reynolds}, M.~T., {et~al.} 2019, \apj, 877,
  17, \dodoi{10.3847/1538-4357/ab18a3}

\bibitem[{Freudling {et~al.}(2013)Freudling, Romaniello, Bramich, Ballester,
  Forchi, García-Dabló, Moehler, \& Neeser}]{Freudling:2013:A96}
Freudling, W., Romaniello, M., Bramich, D.~M., {et~al.} 2013, \aap, 559, A96,
  \dodoi{10.1051/0004-6361/201322494}

\bibitem[{Fu {et~al.}(2011)Fu, Myers, Djorgovski, \& Yan}]{Fu:2011:103}
Fu, H., Myers, A.~D., Djorgovski, S.~G., \& Yan, L. 2011, \apj, 733, 103,
  \dodoi{10.1088/0004-637X/733/2/103}

\bibitem[{{Fu} {et~al.}(2012){Fu}, {Yan}, {Myers}, {Stockton}, {Djorgovski},
  {Aldering}, \& {Rich}}]{2012ApJ...745...67F}
{Fu}, H., {Yan}, L., {Myers}, A.~D., {et~al.} 2012, \apj, 745, 67,
  \dodoi{10.1088/0004-637X/745/1/67}

\bibitem[{{Fu} {et~al.}(2018){Fu}, {Steffen}, {Gross}, {Dai}, {Isbell}, {Lin},
  {Wake}, {Xue}, {Bizyaev}, \& {Pan}}]{2018ApJ...856...93F}
{Fu}, H., {Steffen}, J.~L., {Gross}, A.~C., {et~al.} 2018, \apj, 856, 93,
  \dodoi{10.3847/1538-4357/aab364}

\bibitem[{{Gallimore} \& {Beswick}(2004)}]{2004AJ....127..239G}
{Gallimore}, J.~F., \& {Beswick}, R. 2004, \aj, 127, 239,
  \dodoi{10.1086/379959}

\bibitem[{Ginsburg \& Mirocha(2011)}]{Ginsburg:2011:1109.001}
Ginsburg, A., \& Mirocha, J. 2011, Astrophysics Source Code Library, 1109.001.
\newblock
  \url{http://adsabs.harvard.edu/cgi-bin/nph-data_query?bibcode=2011ascl.soft09001G&link_type=EJOURNAL}

\bibitem[{Gonneau {et~al.}(2020)Gonneau, Lyubenova, Lançon, Trager, Peletier,
  Arentsen, Chen, Coelho, Dries, Falcón-Barroso, Prugniel, Sánchez-Blázquez,
  Vazdekis, \& Verro}]{Gonneau:2020:A133}
Gonneau, A., Lyubenova, M., Lançon, A., {et~al.} 2020, \aap, 634, A133,
  \dodoi{10.1051/0004-6361/201936825}

\bibitem[{Goulding {et~al.}(2019)Goulding, Pardo, Greene, Mingarelli, Nyland,
  \& Strauss}]{Goulding:2019:L21}
Goulding, A.~D., Pardo, K., Greene, J.~E., {et~al.} 2019, \apjl, 879, L21,
  \dodoi{10.3847/2041-8213/ab2a14}

\bibitem[{Hunter(2007)}]{Hunter:2007:90}
Hunter, J.~D. 2007, Computing in Science and Engineering, 9, 90,
  \dodoi{10.1109/MCSE.2007.55}

\bibitem[{{Izquierdo-Villalba}
  {et~al.}(2022{\natexlab{a}}){Izquierdo-Villalba}, {Sesana}, {Bonoli}, \&
  {Colpi}}]{izquierdo-villalba_massive_2022}
{Izquierdo-Villalba}, D., {Sesana}, A., {Bonoli}, S., \& {Colpi}, M.
  2022{\natexlab{a}}, \mnras, 509, 3488, \dodoi{10.1093/mnras/stab3239}

\bibitem[{{Izquierdo-Villalba}
  {et~al.}(2022{\natexlab{b}}){Izquierdo-Villalba}, {Sesana}, \&
  {Colpi}}]{2022arXiv220704064I}
{Izquierdo-Villalba}, D., {Sesana}, A., \& {Colpi}, M. 2022{\natexlab{b}},
  arXiv e-prints, arXiv:2207.04064.
\newblock \doarXiv{2207.04064}

\bibitem[{{Kaiser} \& {McWilliams}(2021)}]{2021CQGra..38e5009K}
{Kaiser}, A.~R., \& {McWilliams}, S.~T. 2021, Classical and Quantum Gravity,
  38, 055009, \dodoi{10.1088/1361-6382/abd4f6}

\bibitem[{{Kawamuro} {et~al.}(2022){Kawamuro}, {Ricci}, {Imanishi},
  {Mushotzky}, {Izumi}, {Ricci}, {Bauer}, {Koss}, {Trakhtenbrot}, {Ichikawa},
  {Rojas}, {Smith}, {Shimizu}, {Oh}, {den Brok}, {Baba}, {Balokovi{\'c}},
  {Chang}, {Kakkad}, {Pfeifle}, {Privon}, {Temple}, {Ueda}, {Harrison},
  {Powell}, {Stern}, {Urry}, \& {Sanders}}]{2022ApJ...938...87K}
{Kawamuro}, T., {Ricci}, C., {Imanishi}, M., {et~al.} 2022, \apj, 938, 87,
  \dodoi{10.3847/1538-4357/ac8794}

\bibitem[{Keel {et~al.}(1988)Keel, de~Grijp, \& Miley}]{Keel:1988:250}
Keel, W.~C., de~Grijp, M. H.~K., \& Miley, G.~K. 1988, Astronomy and
  Astrophysics, Vol. 203, p. 250-254 (1988), 203, 250.
\newblock
  \url{https://ui.adsabs.harvard.edu/abs/1988A%26A...203..250K/abstract}

\bibitem[{Keel {et~al.}(2012)Keel, Chojnowski, Bennert, Schawinski, Lintott,
  Lynn, Pancoast, Harris, Nierenberg, Sonnenfeld, \& Proctor}]{Keel:2012:878}
Keel, W.~C., Chojnowski, S.~D., Bennert, V.~N., {et~al.} 2012, \mnras, 420,
  878, \dodoi{10.1111/j.1365-2966.2011.20101.x}

\bibitem[{Kelley {et~al.}(2018)Kelley, Blecha, Hernquist, Sesana, \&
  Taylor}]{Kelley:2018:964}
Kelley, L.~Z., Blecha, L., Hernquist, L., Sesana, A., \& Taylor, S.~R. 2018,
  \mnras, 477, 964, \dodoi{10.1093/mnras/sty689}

\bibitem[{Kewley {et~al.}(2006)Kewley, Groves, Kauffmann, \&
  Heckman}]{Kewley:2006:961}
Kewley, L.~J., Groves, B., Kauffmann, G., \& Heckman, T. 2006, \mnras, 372,
  961, \dodoi{10.1111/j.1365-2966.2006.10859.x}

\bibitem[{Kim {et~al.}(2021)Kim, Barth, Ho, \& Son}]{Kim:2021:40}
Kim, M., Barth, A.~J., Ho, L.~C., \& Son, S. 2021, \apjs, 256, 40,
  \dodoi{10.3847/1538-4365/ac133e}

\bibitem[{Kollatschny {et~al.}(2020)Kollatschny, Grupe, Parker, Ochmann,
  Schartel, Herwig, Komossa, Romero-Colmenero, \&
  Santos-Lleo}]{Kollatschny:2020:A91}
Kollatschny, W., Grupe, D., Parker, M.~L., {et~al.} 2020, \aap, 638, A91,
  \dodoi{10.1051/0004-6361/202037897}

\bibitem[{Kormendy \& Ho(2013)}]{Kormendy:2013:511}
Kormendy, J., \& Ho, L.~C. 2013, \araa, 51, 511,
  \dodoi{10.1146/annurev-astro-082708-101811}

\bibitem[{Koss {et~al.}(2014)Koss, Blecha, Mushotzky, Veilleux, Hung, Man, \&
  Li}]{Koss:2014}
Koss, M., Blecha, L., Mushotzky, R., {et~al.} 2014, AAS, 223.
\newblock \url{http://adsabs.harvard.edu/abs/2014AAS...22325120K}

\bibitem[{Koss {et~al.}(2012)Koss, Mushotzky, Treister, Veilleux, Vasudevan, \&
  Trippe}]{Koss:2012:L22}
Koss, M., Mushotzky, R., Treister, E., {et~al.} 2012, \apj, 746, L22,
  \dodoi{10.1088/2041-8205/746/2/L22}

\bibitem[{Koss {et~al.}(2010)Koss, Mushotzky, Veilleux, \&
  Winter}]{Koss:2010:L125}
Koss, M., Mushotzky, R., Veilleux, S., \& Winter, L. 2010, \apjl, 716, L125,
  \dodoi{10.1088/2041-8205/716/2/L125}

\bibitem[{Koss {et~al.}(2011{\natexlab{a}})Koss, Mushotzky, Veilleux, Winter,
  Baumgartner, Tueller, Gehrels, \& Valencic}]{Koss:2011:57}
Koss, M., Mushotzky, R., Veilleux, S., {et~al.} 2011{\natexlab{a}}, \apj, 739,
  57, \dodoi{10.1088/0004-637X/739/2/57}

\bibitem[{Koss {et~al.}(2011{\natexlab{b}})Koss, Mushotzky, Treister, Veilleux,
  Vasudevan, Miller, Sanders, Schawinski, \& Trippe}]{Koss:2011:L42}
Koss, M., Mushotzky, R., Treister, E., {et~al.} 2011{\natexlab{b}}, \apjl, 735,
  L42, \dodoi{10.1088/2041-8205/735/2/L42}

\bibitem[{Koss {et~al.}(2017)Koss, Trakhtenbrot, Ricci, Lamperti, Oh, Berney,
  Schawinski, Baloković, Baronchelli, Crenshaw, Fischer, Gehrels, Harrison,
  Hashimoto, Hogg, Ichikawa, Masetti, Mushotzky, Sartori, Stern, Treister,
  Ueda, Veilleux, \& Winter}]{Koss:2017:74}
Koss, M., Trakhtenbrot, B., Ricci, C., {et~al.} 2017, \apj, 850, 74,
  \dodoi{10.3847/1538-4357/aa8ec9}

\bibitem[{Koss {et~al.}(2015)Koss, Romero-Cañizales, Baronchelli, Teng,
  Baloković, Puccetti, Bauer, Arévalo, Assef, Ballantyne, Brandt, Brightman,
  Comastri, Gandhi, Harrison, Luo, Schawinski, Stern, \&
  Treister}]{Koss:2015:149}
Koss, M.~J., Romero-Cañizales, C., Baronchelli, L., {et~al.} 2015, \apj, 807,
  149, \dodoi{10.1088/0004-637X/807/2/149}

\bibitem[{Koss {et~al.}(2016{\natexlab{a}})Koss, Glidden, Baloković, Stern,
  Lamperti, Assef, Bauer, Ballantyne, Boggs, Craig, Farrah, Fürst, Gandhi,
  Gehrels, Hailey, Harrison, Markwardt, Masini, Ricci, Treister, Walton, \&
  Zhang}]{Koss:2016:L4}
Koss, M.~J., Glidden, A., Baloković, M., {et~al.} 2016{\natexlab{a}}, \apjl,
  824, L4, \dodoi{10.3847/2041-8205/824/1/L4}

\bibitem[{Koss {et~al.}(2016{\natexlab{b}})Koss, Assef, Baloković, Stern,
  Gandhi, Lamperti, Alexander, Ballantyne, Bauer, Berney, Brandt, Comastri,
  Gehrels, Harrison, Lansbury, Markwardt, Ricci, Rivers, Schawinski,
  Trakhtenbrot, Treister, \& Urry}]{Koss:2016:85}
Koss, M.~J., Assef, R., Baloković, M., {et~al.} 2016{\natexlab{b}}, \apj, 825,
  85, \dodoi{10.3847/0004-637X/825/2/85}

\bibitem[{Koss {et~al.}(2018)Koss, Blecha, Bernhard, Hung, Lu, Trakthenbrot,
  Treister, Weigel, Sartori, Mushotzky, Schawinski, Ricci, Veilleux, \&
  Sanders}]{Koss:2018:214a}
Koss, M.~J., Blecha, L., Bernhard, P., {et~al.} 2018, \nat, 563, 214.
\newblock \url{https://doi.org/10.1038/s41586-018-0652-7}

\bibitem[{Koss {et~al.}(2021)Koss, Strittmatter, Lamperti, Shimizu,
  Trakhtenbrot, Saintonge, Treister, Cicone, Mushotzky, Oh, Ricci, Stern,
  Ananna, Bauer, Privon, Bär, De~Breuck, Harrison, Ichikawa, Powell, Rosario,
  Sanders, Schawinski, Shao, Megan~Urry, \& Veilleux}]{Koss:2021:29}
Koss, M.~J., Strittmatter, B., Lamperti, I., {et~al.} 2021, \apjs, 252, 29,
  \dodoi{10.3847/1538-4365/abcbfe}

\bibitem[{Koss {et~al.}(2022{\natexlab{a}})Koss, Ricci, Trakhtenbrot, Oh, Brok,
  Mejía-Restrepo, Stern, Privon, Treister, Powell, Mushotzky, Bauer, Ananna,
  Baloković, Bär, Becker, Bessiere, Burtscher, Caglar, Congiu, Evans,
  Harrison, Heida, Ichikawa, Kamraj, Lamperti, Pacucci, Ricci, Riffel, Rojas,
  Schawinski, Temple, Urry, Veilleux, \& Williams}]{Koss:2022:2}
Koss, M.~J., Ricci, C., Trakhtenbrot, B., {et~al.} 2022{\natexlab{a}}, \apjs,
  261, 2, \dodoi{10.3847/1538-4365/ac6c05}

\bibitem[{Koss {et~al.}(2022{\natexlab{b}})Koss, Trakhtenbrot, Ricci, Oh,
  Bauer, Stern, Caglar, Brok, Mushotzky, Ricci, Mejía-Restrepo, Lamperti,
  Treister, Bär, Harrison, Powell, Privon, Riffel, Rojas, Schawinski, \&
  Urry}]{Koss:2022:6}
Koss, M.~J., Trakhtenbrot, B., Ricci, C., {et~al.} 2022{\natexlab{b}}, \apjs,
  261, 6, \dodoi{10.3847/1538-4365/ac650b}

\bibitem[{Koss {et~al.}(2022{\natexlab{c}})Koss, Trakhtenbrot, Ricci, Bauer,
  Treister, Mushotzky, Urry, Ananna, Baloković, Brok, Cenko, Harrison,
  Ichikawa, Lamperti, Lein, Mejía-Restrepo, Oh, Pacucci, Pfeifle, Powell,
  Privon, Ricci, Salvato, Schawinski, Shimizu, Smith, \& Stern}]{Koss:2022:1}
---. 2022{\natexlab{c}}, \apjs, 261, 1, \dodoi{10.3847/1538-4365/ac6c8f}

\bibitem[{Lamperti {et~al.}(2017)Lamperti, Koss, Trakhtenbrot, Schawinski,
  Ricci, Oh, Landt, Riffel, Rodríguez-Ardila, Gehrels, Harrison, Masetti,
  Mushotzky, Treister, Ueda, \& Veilleux}]{Lamperti:2017:540}
Lamperti, I., Koss, M., Trakhtenbrot, B., {et~al.} 2017, \mnras, 467, 540,
  \dodoi{10.1093/mnras/stx055}

\bibitem[{Larkin {et~al.}(2006)Larkin, Barczys, Krabbe, Adkins, Aliado, Amico,
  Brims, Campbell, Canfield, Gasaway, Honey, Iserlohe, Johnson, Kress,
  LaFreniere, Lyke, Magnone, Magnone, McElwain, Moon, Quirrenbach, Skulason,
  Song, Spencer, Weiss, \& Wright}]{Larkin:2006:441}
Larkin, J., Barczys, M., Krabbe, A., {et~al.} 2006, in Ground-based and
  {Airborne} {Instrumentation} for {Astronomy}, Vol. 6269 (SPIE), 441--445,
  \dodoi{10.1117/12.672061}

\bibitem[{{LIGO Scientific Collaboration and Virgo Collaboration}
  {et~al.}(2016){LIGO Scientific Collaboration and Virgo Collaboration},
  Abbott, Abbott, Abbott, Abernathy, Acernese, Ackley, Adams, Adams, Addesso,
  Adhikari, Adya, Affeldt, Agathos, Agatsuma, Aggarwal, Aguiar, Aiello, Ain,
  Ajith, Allen, Allocca, Altin, Anderson, Anderson, Arai, Arain, Araya,
  Arceneaux, Areeda, Arnaud, Arun, Ascenzi, Ashton, Ast, Aston, Astone,
  Aufmuth, Aulbert, Babak, Bacon, Bader, Baker, Baldaccini, Ballardin, Ballmer,
  Barayoga, Barclay, Barish, Barker, Barone, Barr, Barsotti, Barsuglia, Barta,
  Bartlett, Barton, Bartos, Bassiri, Basti, Batch, Baune, Bavigadda, Bazzan,
  Behnke, Bejger, Belczynski, Bell, Bell, Berger, Bergman, Bergmann, Berry,
  Bersanetti, Bertolini, Betzwieser, Bhagwat, Bhandare, Bilenko, Billingsley,
  Birch, Birney, Birnholtz, Biscans, Bisht, Bitossi, Biwer, Bizouard,
  Blackburn, Blair, Blair, Blair, Bloemen, Bock, Bodiya, Boer, Bogaert, Bogan,
  Bohe, Bojtos, Bond, Bondu, Bonnand, Boom, Bork, Boschi, Bose, Bouffanais,
  Bozzi, Bradaschia, Brady, Braginsky, Branchesi, Brau, Briant, Brillet,
  Brinkmann, Brisson, Brockill, Brooks, Brown, Brown, Brown, Buchanan, Buikema,
  Bulik, Bulten, Buonanno, Buskulic, Buy, Byer, Cabero, Cadonati, Cagnoli,
  Cahillane, Bustillo, Callister, Calloni, Camp, Cannon, Cao, Capano, Capocasa,
  Carbognani, Caride, Diaz, Casentini, Caudill, Cavaglià, Cavalier, Cavalieri,
  Cella, Cepeda, Baiardi, Cerretani, Cesarini, Chakraborty, Chalermsongsak,
  Chamberlin, Chan, Chao, Charlton, Chassande-Mottin, Chen, Chen, Cheng,
  Chincarini, Chiummo, Cho, Cho, Chow, Christensen, Chu, Chua, Chung, Ciani,
  Clara, Clark, Cleva, Coccia, Cohadon, Colla, Collette, Cominsky, Constancio,
  Conte, Conti, Cook, Corbitt, Cornish, Corsi, Cortese, Costa, Coughlin,
  Coughlin, Coulon, Countryman, Couvares, Cowan, Coward, Cowart, Coyne, Coyne,
  Craig, Creighton, Creighton, Cripe, Crowder, Cruise, Cumming, Cunningham,
  Cuoco, Canton, Danilishin, D’Antonio, Danzmann, Darman, Da~Silva~Costa,
  Dattilo, Dave, Daveloza, Davier, Davies, Daw, Day, De, DeBra, Debreczeni,
  Degallaix, De~Laurentis, Deléglise, Del~Pozzo, Denker, Dent, Dereli,
  Dergachev, DeRosa, De~Rosa, DeSalvo, Dhurandhar, Díaz, Di~Fiore,
  Di~Giovanni, Di~Lieto, Di~Pace, Di~Palma, Di~Virgilio, Dojcinoski, Dolique,
  Donovan, Dooley, Doravari, Douglas, Downes, Drago, Drever, Driggers, Du,
  Ducrot, Dwyer, Edo, Edwards, Effler, Eggenstein, Ehrens, Eichholz,
  Eikenberry, Engels, Essick, Etzel, Evans, Evans, Everett, Factourovich,
  Fafone, Fair, Fairhurst, Fan, Fang, Farinon, Farr, Farr, Favata, Fays,
  Fehrmann, Fejer, Feldbaum, Ferrante, Ferreira, Ferrini, Fidecaro, Finn,
  Fiori, Fiorucci, Fisher, Flaminio, Fletcher, Fong, Fournier, Franco, Frasca,
  Frasconi, Frede, Frei, Freise, Frey, Frey, Fricke, Fritschel, Frolov, Fulda,
  Fyffe, Gabbard, Gair, Gammaitoni, Gaonkar, Garufi, Gatto, Gaur, Gehrels,
  Gemme, Gendre, Genin, Gennai, George, Gergely, Germain, Ghosh, Ghosh, Ghosh,
  Giaime, Giardina, Giazotto, Gill, Glaefke, Gleason, Goetz, Goetz, Gondan,
  González, Castro, Gopakumar, Gordon, Gorodetsky, Gossan, Gosselin, Gouaty,
  Graef, Graff, Granata, Grant, Gras, Gray, Greco, Green, Greenhalgh, Groot,
  Grote, Grunewald, Guidi, Guo, Gupta, Gupta, Gushwa, Gustafson, Gustafson,
  Hacker, Hall, Hall, Hammond, Haney, Hanke, Hanks, Hanna, Hannam, Hanson,
  Hardwick, Harms, Harry, Harry, Hart, Hartman, Haster, Haughian, Healy,
  Heefner, Heidmann, Heintze, Heinzel, Heitmann, Hello, Hemming, Hendry, Heng,
  Hennig, Heptonstall, Heurs, Hild, Hoak, Hodge, Hofman, Hollitt, Holt, Holz,
  Hopkins, Hosken, Hough, Houston, Howell, Hu, Huang, Huerta, Huet, Hughey,
  Husa, Huttner, Huynh-Dinh, Idrisy, Indik, Ingram, Inta, Isa, Isac, Isi,
  Islas, Isogai, Iyer, Izumi, Jacobson, Jacqmin, Jang, Jani, Jaranowski,
  Jawahar, Jiménez-Forteza, Johnson, Johnson-McDaniel, Jones, Jones, Jonker,
  Ju, Haris, Kalaghatgi, Kalogera, Kandhasamy, Kang, Kanner, Karki, Kasprzack,
  Katsavounidis, Katzman, Kaufer, Kaur, Kawabe, Kawazoe, Kéfélian, Kehl,
  Keitel, Kelley, Kells, Kennedy, Keppel, Key, Khalaidovski, Khalili, Khan,
  Khan, Khan, Khazanov, Kijbunchoo, Kim, Kim, Kim, Kim, Kim, Kim, King, King,
  Kinzel, Kissel, Kleybolte, Klimenko, Koehlenbeck, Kokeyama, Koley,
  Kondrashov, Kontos, Koranda, Korobko, Korth, Kowalska, Kozak, Kringel,
  Krishnan, Królak, Krueger, Kuehn, Kumar, Kumar, Kuo, Kutynia, Kwee, Lackey,
  Landry, Lange, Lantz, Lasky, Lazzarini, Lazzaro, Leaci, Leavey, Lebigot, Lee,
  Lee, Lee, Lee, Lenon, Leonardi, Leong, Leroy, Letendre, Levin, Levine, Li,
  Libson, Littenberg, Lockerbie, Logue, Lombardi, London, Lord, Lorenzini,
  Loriette, Lormand, Losurdo, Lough, Lousto, Lovelace, Lück, Lundgren, Luo,
  Lynch, Ma, MacDonald, Machenschalk, MacInnis, Macleod, Magaña-Sandoval,
  Magee, Mageswaran, Majorana, Maksimovic, Malvezzi, Man, Mandel, Mandic,
  Mangano, Mansell, Manske, Mantovani, Marchesoni, Marion, Márka, Márka,
  Markosyan, Maros, Martelli, Martellini, Martin, Martin, Martynov, Marx,
  Mason, Masserot, Massinger, Masso-Reid, Matichard, Matone, Mavalvala,
  Mazumder, Mazzolo, McCarthy, McClelland, McCormick, McGuire, McIntyre,
  McIver, McManus, McWilliams, Meacher, Meadors, Meidam, Melatos, Mendell,
  Mendoza-Gandara, Mercer, Merilh, Merzougui, Meshkov, Messenger, Messick,
  Meyers, Mezzani, Miao, Michel, Middleton, Mikhailov, Milano, Miller,
  Millhouse, Minenkov, Ming, Mirshekari, Mishra, Mitra, Mitrofanov,
  Mitselmakher, Mittleman, Moggi, Mohan, Mohapatra, Montani, Moore, Moore,
  Moraru, Moreno, Morriss, Mossavi, Mours, Mow-Lowry, Mueller, Mueller, Muir,
  Mukherjee, Mukherjee, Mukherjee, Mukund, Mullavey, Munch, Murphy, Murray,
  Mytidis, Nardecchia, Naticchioni, Nayak, Necula, Nedkova, Nelemans, Neri,
  Neunzert, Newton, Nguyen, Nielsen, Nissanke, Nitz, Nocera, Nolting,
  Normandin, Nuttall, Oberling, Ochsner, O’Dell, Oelker, Ogin, Oh, Oh, Ohme,
  Oliver, Oppermann, Oram, O’Reilly, O’Shaughnessy, Ott, Ottaway, Ottens,
  Overmier, Owen, Pai, Pai, Palamos, Palashov, Palomba, Pal-Singh, Pan, Pan,
  Pankow, Pannarale, Pant, Paoletti, Paoli, Papa, Paris, Parker, Pascucci,
  Pasqualetti, Passaquieti, Passuello, Patricelli, Patrick, Pearlstone,
  Pedraza, Pedurand, Pekowsky, Pele, Penn, Perreca, Pfeiffer, Phelps, Piccinni,
  Pichot, Pickenpack, Piergiovanni, Pierro, Pillant, Pinard, Pinto, Pitkin,
  Poeld, Poggiani, Popolizio, Post, Powell, Prasad, Predoi, Premachandra,
  Prestegard, Price, Prijatelj, Principe, Privitera, Prix, Prodi, Prokhorov,
  Puncken, Punturo, Puppo, Pürrer, Qi, Qin, Quetschke, Quintero,
  Quitzow-James, Raab, Rabeling, Radkins, Raffai, Raja, Rakhmanov, Ramet,
  Rapagnani, Raymond, Razzano, Re, Read, Reed, Regimbau, Rei, Reid, Reitze,
  Rew, Reyes, Ricci, Riles, Robertson, Robie, Robinet, Rocchi, Rolland,
  Rollins, Roma, Romano, Romano, Romanov, Romie, Rosińska, Rowan, Rüdiger,
  Ruggi, Ryan, Sachdev, Sadecki, Sadeghian, Salconi, Saleem, Salemi, Samajdar,
  Sammut, Sampson, Sanchez, Sandberg, Sandeen, Sanders, Sanders, Sassolas,
  Sathyaprakash, Saulson, Sauter, Savage, Sawadsky, Schale, Schilling, Schmidt,
  Schmidt, Schnabel, Schofield, Schönbeck, Schreiber, Schuette, Schutz, Scott,
  Scott, Sellers, Sengupta, Sentenac, Sequino, Sergeev, Serna, Setyawati,
  Sevigny, Shaddock, Shaffer, Shah, Shahriar, Shaltev, Shao, Shapiro, Shawhan,
  Sheperd, Shoemaker, Shoemaker, Siellez, Siemens, Sigg, Silva, Simakov,
  Singer, Singer, Singh, Singh, Singhal, Sintes, Slagmolen, Smith, Smith,
  Smith, Smith, Son, Sorazu, Sorrentino, Souradeep, Srivastava, Staley,
  Steinke, Steinlechner, Steinlechner, Steinmeyer, Stephens, Stevenson, Stone,
  Strain, Straniero, Stratta, Strauss, Strigin, Sturani, Stuver, Summerscales,
  Sun, Sutton, Swinkels, Szczepańczyk, Tacca, Talukder, Tanner, Tápai,
  Tarabrin, Taracchini, Taylor, Theeg, Thirugnanasambandam, Thomas, Thomas,
  Thomas, Thorne, Thorne, Thrane, Tiwari, Tiwari, Tokmakov, Tomlinson, Tonelli,
  Torres, Torrie, Töyrä, Travasso, Traylor, Trifirò, Tringali, Trozzo, Tse,
  Turconi, Tuyenbayev, Ugolini, Unnikrishnan, Urban, Usman, Vahlbruch, Vajente,
  Valdes, Vallisneri, van Bakel, van Beuzekom, van~den Brand, Van Den~Broeck,
  Vander-Hyde, van~der Schaaf, van Heijningen, van Veggel, Vardaro, Vass,
  Vasúth, Vaulin, Vecchio, Vedovato, Veitch, Veitch, Venkateswara, Verkindt,
  Vetrano, Viceré, Vinciguerra, Vine, Vinet, Vitale, Vo, Vocca, Vorvick, Voss,
  Vousden, Vyatchanin, Wade, Wade, Wade, Waldman, Walker, Wallace, Walsh, Wang,
  Wang, Wang, Wang, Wang, Ward, Ward, Warner, Was, Weaver, Wei, Weinert,
  Weinstein, Weiss, Welborn, Wen, Weßels, Westphal, Wette, Whelan, Whitcomb,
  White, Whiting, Wiesner, Wilkinson, Willems, Williams, Williams, Williamson,
  Willis, Willke, Wimmer, Winkelmann, Winkler, Wipf, Wiseman, Wittel, Woan,
  Worden, Wright, Wu, Yablon, Yakushin, Yam, Yamamoto, Yancey, Yap, Yu, Yvert,
  Zadrożny, Zangrando, Zanolin, Zendri, Zevin, Zhang, Zhang, Zhang, Zhang,
  Zhao, Zhou, Zhou, Zhu, Zucker, Zuraw, \& Zweizig}]{Collaboration:2016:061102}
{LIGO Scientific Collaboration and Virgo Collaboration}, Abbott, B., Abbott,
  R., {et~al.} 2016, Phys. Rev. Lett., 116, 061102,
  \dodoi{10.1103/PhysRevLett.116.061102}

\bibitem[{Liu {et~al.}(2011)Liu, Shen, Strauss, \& Hao}]{Liu:2011:101}
Liu, X., Shen, Y., Strauss, M.~A., \& Hao, L. 2011, \apj, 737, 101,
  \dodoi{10.1088/0004-637X/737/2/101}

\bibitem[{Lyu \& Rieke(2018)}]{Lyu:2018:92}
Lyu, J., \& Rieke, G.~H. 2018, \apj, 866, 92, \dodoi{10.3847/1538-4357/aae075}

\bibitem[{Marsden {et~al.}(2020)Marsden, Shankar, Ginolfi, \&
  Zubovas}]{Marsden:2020:61a}
Marsden, C., Shankar, F., Ginolfi, M., \& Zubovas, K. 2020, Frontiers in
  Physics, 8, 61, \dodoi{10.3389/fphy.2020.00061}

\bibitem[{McMullin {et~al.}(2007)McMullin, Waters, Schiebel, Young, \&
  Golap}]{McMullin:2007:127}
McMullin, J.~P., Waters, B., Schiebel, D., Young, W., \& Golap, K. 2007, 376,
  127.
\newblock \url{https://ui.adsabs.harvard.edu/abs/2007ASPC..376..127M}

\bibitem[{Medling {et~al.}(2011)Medling, Ammons, Max, Davies, Engel, \&
  Canalizo}]{Medling:2011:32}
Medling, A.~M., Ammons, S.~M., Max, C.~E., {et~al.} 2011, \apj, 743, 32,
  \dodoi{10.1088/0004-637X/743/1/32}

\bibitem[{{Meusinger} {et~al.}(2017){Meusinger}, {Br{\"u}necke}, {Schalldach},
  \& {in der Au}}]{2017A&A...597A.134M}
{Meusinger}, H., {Br{\"u}necke}, J., {Schalldach}, P., \& {in der Au}, A. 2017,
  \aap, 597, A134, \dodoi{10.1051/0004-6361/201629139}

\bibitem[{Muller-Sanchez {et~al.}(2018)Muller-Sanchez, Hicks, Malkan, Davies,
  Yu, Shaver, \& Davis}]{Muller-Sanchez:2018:48}
Muller-Sanchez, F., Hicks, E. K.~S., Malkan, M., {et~al.} 2018, \apj, 858, 48,
  \dodoi{10.3847/1538-4357/aab9ad}

\bibitem[{{Munoz} {et~al.}(2019){Munoz}, {Miranda}, \&
  {Lai}}]{2019ApJ...871...84M}
{Munoz}, D.~J., {Miranda}, R., \& {Lai}, D. 2019, \apj, 871, 84,
  \dodoi{10.3847/1538-4357/aaf867}

\bibitem[{Oh {et~al.}(2011)Oh, Sarzi, Schawinski, \& Yi}]{Oh:2011:13}
Oh, K., Sarzi, M., Schawinski, K., \& Yi, S.~K. 2011, \apjs, 195, 13,
  \dodoi{10.1088/0067-0049/195/2/13}

\bibitem[{Oh {et~al.}(2022)Oh, Koss, Ueda, Stern, Ricci, Trakhtenbrot, Powell,
  Brok, Lamperti, Mushotzky, Ricci, Bär, Rojas, Ichikawa, Riffel, Treister,
  Harrison, Urry, Bauer, \& Schawinski}]{Oh:2022:4}
Oh, K., Koss, M.~J., Ueda, Y., {et~al.} 2022, \apjs, 261, 4,
  \dodoi{10.3847/1538-4365/ac5b68}

\bibitem[{Ricci {et~al.}(2017)Ricci, Trakhtenbrot, Koss, Ueda, Del~Vecchio,
  Treister, Schawinski, Paltani, Oh, Lamperti, Berney, Gandhi, Ichikawa, Bauer,
  Ho, Asmus, Beckmann, Soldi, Baloković, Gehrels, \&
  Markwardt}]{Ricci:2017:17}
Ricci, C., Trakhtenbrot, B., Koss, M.~J., {et~al.} 2017, \apjs, 233, 17,
  \dodoi{10.3847/1538-4365/aa96ad}

\bibitem[{{Ricci} {et~al.}(2021){Ricci}, {Privon}, {Pfeifle}, {Armus},
  {Iwasawa}, {Torres-Alb{\`a}}, {Satyapal}, {Bauer}, {Treister}, {Ho}, {Aalto},
  {Ar{\'e}valo}, {Barcos-Mu{\~n}oz}, {Charmandaris}, {Diaz-Santos}, {Evans},
  {Gao}, {Inami}, {Koss}, {Lansbury}, {Linden}, {Medling}, {Sanders}, {Song},
  {Stern}, {U}, {Ueda}, \& {Yamada}}]{2021MNRAS.506.5935R}
{Ricci}, C., {Privon}, G.~C., {Pfeifle}, R.~W., {et~al.} 2021, \mnras, 506,
  5935, \dodoi{10.1093/mnras/stab2052}

\bibitem[{Rich {et~al.}(2011)Rich, Kewley, \& Dopita}]{Rich:2011:87}
Rich, J.~A., Kewley, L.~J., \& Dopita, M.~A. 2011, \apj, 734, 87,
  \dodoi{10.1088/0004-637X/734/2/87}

\bibitem[{Robitaille \& Bressert(2012)}]{Robitaille:2012:1208.017}
Robitaille, T., \& Bressert, E. 2012, Astrophysics Source Code Library,
  1208.017.
\newblock
  \url{http://adsabs.harvard.edu/cgi-bin/nph-data_query?bibcode=2012ascl.soft08017R&link_type=EJOURNAL}

\bibitem[{{Rodriguez} {et~al.}(2006){Rodriguez}, {Taylor}, {Zavala}, {Peck},
  {Pollack}, \& {Romani}}]{2006ApJ...646...49R}
{Rodriguez}, C., {Taylor}, G.~B., {Zavala}, R.~T., {et~al.} 2006, \apj, 646,
  49, \dodoi{10.1086/504825}

\bibitem[{{Shen} {et~al.}(2011){Shen}, {Liu}, {Greene}, \&
  {Strauss}}]{2011ApJ...735...48S}
{Shen}, Y., {Liu}, X., {Greene}, J.~E., \& {Strauss}, M.~A. 2011, \apj, 735,
  48, \dodoi{10.1088/0004-637X/735/1/48}

\bibitem[{{Shen} {et~al.}(2021){Shen}, {Chen}, {Hwang}, {Liu}, {Zakamska},
  {Oguri}, {Li}, {Lazio}, \& {Breiding}}]{2021NatAs...5..569S}
{Shen}, Y., {Chen}, Y.-C., {Hwang}, H.-C., {et~al.} 2021, Nature Astronomy, 5,
  569, \dodoi{10.1038/s41550-021-01323-1}

\bibitem[{Shimizu {et~al.}(2017)Shimizu, Mushotzky, Meléndez, Koss, Barger, \&
  Cowie}]{Shimizu:2017:3161}
Shimizu, T.~T., Mushotzky, R.~F., Meléndez, M., {et~al.} 2017, \mnras, 466,
  3161, \dodoi{10.1093/mnras/stw3268}

\bibitem[{{Smith} {et~al.}(2010){Smith}, {Shields}, {Bonning}, {McMullen},
  {Rosario}, \& {Salviander}}]{2010ApJ...716..866S}
{Smith}, K.~L., {Shields}, G.~A., {Bonning}, E.~W., {et~al.} 2010, \apj, 716,
  866, \dodoi{10.1088/0004-637X/716/1/866}

\bibitem[{Smith {et~al.}(2020)Smith, Mushotzky, Koss, Trakhtenbrot, Ricci,
  Wong, Bauer, Ricci, Vogel, Stern, Powell, Urry, Harrison, Mejía-Restrepo,
  Oh, Baek, \& Chung}]{Smith:2020:4216}
Smith, K.~L., Mushotzky, R.~F., Koss, M., {et~al.} 2020, \mnras, 492, 4216,
  \dodoi{10.1093/mnras/stz3608}

\bibitem[{Steinborn {et~al.}(2016)Steinborn, Dolag, Comerford, Hirschmann,
  Remus, \& Teklu}]{Steinborn:2016:1013}
Steinborn, L.~K., Dolag, K., Comerford, J.~M., {et~al.} 2016, \mnras, 458,
  1013, \dodoi{10.1093/mnras/stw316}

\bibitem[{{Stemo} {et~al.}(2021){Stemo}, {Comerford}, {Barrows}, {Stern},
  {Assef}, {Griffith}, \& {Schechter}}]{2021ApJ...923...36S}
{Stemo}, A., {Comerford}, J.~M., {Barrows}, R.~S., {et~al.} 2021, \apj, 923,
  36, \dodoi{10.3847/1538-4357/ac0bbf}

\bibitem[{Stickley \& Canalizo(2012)}]{Stickley:2012:33}
Stickley, N.~R., \& Canalizo, G. 2012, \apj, 747, 33,
  \dodoi{10.1088/0004-637X/747/1/33}

\bibitem[{Treister {et~al.}(2018)Treister, Privon, Sartori, Nagar, Bauer,
  Schawinski, Messias, Ricci, U, Casey, Comerford, Muller-Sanchez, Evans,
  Finlez, Koss, Sanders, \& Urry}]{Treister:2018:83}
Treister, E., Privon, G.~C., Sartori, L.~F., {et~al.} 2018, \apj, 854, 83,
  \dodoi{10.3847/1538-4357/aaa963}

\bibitem[{Treister {et~al.}(2020)Treister, Messias, Privon, Nagar, Medling, U,
  Bauer, Cicone, Muñoz, Evans, Muller-Sanchez, Comerford, Armus, Chang, Koss,
  Venturi, Schawinski, Casey, Urry, Sanders, Scoville, \&
  Sheth}]{Treister:2020:149}
Treister, E., Messias, H., Privon, G.~C., {et~al.} 2020, \apj, 890, 149,
  \dodoi{10.3847/1538-4357/ab6b28}

\bibitem[{van~der Walt {et~al.}(2011)van~der Walt, Colbert, \&
  Varoquaux}]{vanderWalt:2011:22}
van~der Walt, S., Colbert, S.~C., \& Varoquaux, G. 2011, Computing in Science
  and Engineering, 13, 22, \dodoi{10.1109/MCSE.2011.37}

\bibitem[{Van~Wassenhove {et~al.}(2012)Van~Wassenhove, Volonteri, Mayer, Dotti,
  Bellovary, \& Callegari}]{VanWassenhove:2012:L7}
Van~Wassenhove, S., Volonteri, M., Mayer, L., {et~al.} 2012, \apjl, 748, L7,
  \dodoi{10.1088/2041-8205/748/1/L7}

\bibitem[{Verbiest {et~al.}(2016)Verbiest, Lentati, Hobbs, van Haasteren,
  Demorest, Janssen, Wang, Desvignes, Caballero, Keith, Champion, Arzoumanian,
  Babak, Bassa, Bhat, Brazier, Brem, Burgay, Burke-Spolaor, Chamberlin,
  Chatterjee, Christy, Cognard, Cordes, Dai, Dolch, Ellis, Ferdman, Fonseca,
  Gair, Garver-Daniels, Gentile, Gonzalez, Graikou, Guillemot, Hessels, Jones,
  Karuppusamy, Kerr, Kramer, Lam, Lasky, Lassus, Lazarus, Lazio, Lee, Levin,
  Liu, Lynch, Lyne, Mckee, McLaughlin, McWilliams, Madison, Manchester,
  Mingarelli, Nice, Osłowski, Palliyaguru, Pennucci, Perera, Perrodin,
  Possenti, Petiteau, Ransom, Reardon, Rosado, Sanidas, Sesana, Shaifullah,
  Shannon, Siemens, Simon, Smits, Spiewak, Stairs, Stappers, Stinebring,
  Stovall, Swiggum, Taylor, Theureau, Tiburzi, Toomey, Vallisneri, van Straten,
  Vecchio, Wang, Wen, You, Zhu, \& Zhu}]{Verbiest:2016:1267}
Verbiest, J. P.~W., Lentati, L., Hobbs, G., {et~al.} 2016, \mnras, 458, 1267,
  \dodoi{10.1093/mnras/stw347}

\bibitem[{Veres {et~al.}(2021)Veres, Gabányi, Frey, Paragi, Kun, \&
  An}]{Veres:2021:99}
Veres, P.~M., Gabányi, K.~E., Frey, S., {et~al.} 2021, \apj, 922, 99,
  \dodoi{10.3847/1538-4357/ac307d}

\bibitem[{Walmsley {et~al.}(2021)Walmsley, Lintott, Géron, Kruk, Krawczyk,
  Willett, Bamford, Kelvin, Fortson, Gal, Keel, Masters, Mehta, Simmons,
  Smethurst, Smith, Baeten, \& Macmillan}]{Walmsley:2021:3966}
Walmsley, M., Lintott, C., Géron, T., {et~al.} 2021, \mnras, 509, 3966,
  \dodoi{10.1093/mnras/stab2093}

\bibitem[{Yu \& Tremaine(2003)}]{yu_ejection_2003}
Yu, Q., \& Tremaine, S. 2003, The Astrophysical Journal, 599, 1129,
  \dodoi{10.1086/379546}

\bibitem[{Zhao {et~al.}(2021)Zhao, Marchesi, Ajello, Cole, Hu, Silver, \&
  Torres-Albà}]{Zhao:2021:A57}
Zhao, X., Marchesi, S., Ajello, M., {et~al.} 2021, Astronomy \&amp;
  Astrophysics, Volume 650, id.A57, 14, 650, A57,
  \dodoi{10.1051/0004-6361/202140297}

\end{thebibliography}
\bibliographystyle{aasjournal}

\end{document}